\documentclass[12pt,preprint]{aastex}
\newcommand{\nraoblurb}{The National Radio Astronomy Observatory is
a facility of the National Science Foundation operated under cooperative
agreement by Associated Universities, Inc.}

\newcommand{\degree}{\ensuremath{\,^\circ}}

\newcommand{\myr}{\ensuremath{\,{\rm Myr}}}
\newcommand{\gyr}{\ensuremath{\,{\rm Gyr}}}

\newcommand{\mhz}{\ensuremath{\,{\rm MHz}}}
\newcommand{\ghz}{\ensuremath{\,{\rm GHz}}}

\newcommand{\K}{\ensuremath{\,{\rm K}}}
\newcommand{\mk}{\ensuremath{\,{\rm mK}}}

\newcommand{\kpc}{\ensuremath{\,{\rm kpc}}}

\newcommand{\kms}{\ensuremath{\,{\rm km\, s^{-1}}}}

\newcommand{\s}{\,s}

\newcommand{\dexkpc}{\ensuremath{\rm \,dex\,kpc^{-1}}}
\newcommand{\Kkpc}{\ensuremath{\rm \,K\,kpc^{-1}}}
\newcommand{\te}{\ensuremath{T_{\rm e}}}

\newcommand{\rgal}{\ensuremath{{R_{\rm gal}}}}
\newcommand{\rsun}{\ensuremath{{R_{\rm Sun}}}}
\newcommand{\hy}[2]{${\rm H}#1#2\alpha$}

\newcommand{\hi}{H\,{\sc i}}

\newcommand{\hii}{H\,{\sc ii}}

\newcommand{\htwoco}{{H$_{2}$CO}}
\newcommand{\hiea}{H~{\sc i}~E/A}
\newcommand{\hisa}{H~{\sc i}~SA}

\newcommand\urltilda{\kern -.15em\lower .7ex\hbox{\~{}}\kern .04em}

\begin{document}


\title{Azimuthal Metallicity Structure in the Milky Way Disk}

\author{Dana S. Balser\altaffilmark{1}, Trey V. Wenger\altaffilmark{2,1},
L. D. Anderson\altaffilmark{3}, \& T. M. Bania\altaffilmark{4}}

\altaffiltext{1}{National Radio Astronomy Observatory, 520 Edgemont Rd., 
Charlottesville, VA 22903, USA.}
\altaffiltext{2}{Astronomy Department, University of Virginia, 
P.O. Box 400325, Charlottesville VA 22904-4325, USA.}
\altaffiltext{3}{Department of Physics, West Virginia University, Morgantown, WV 26506, USA.}
\altaffiltext{4}{Institute for Astrophysical Research, Department of Astronomy,
Boston University, 725 Commonwealth Avenue, Boston MA 02215, USA.}

\begin{abstract}

  Elemental abundance patterns in the Galactic disk constrain theories
  of the formation and evolution of the Milky Way.  \hii\ region
  abundances are the result of billions of years of chemical
  evolution. We made radio recombination line and continuum
  measurements of 21 \hii\ regions located between Galactic azimuth
  $Az = 90$\degree$-130$\degree, a previously unexplored region.  We
  derive the plasma electron temperatures using the line-to-continuum
  ratios and use them as proxies for the nebular [O/H] abundances,
  because in thermal equilibrium the abundance of the coolants (O, N,
  and other heavy elements) in the ionized gas sets the electron
  temperature, with high abundances producing low temperatures.
  Combining these data with our previous work produces a sample of 90
  \hii\ regions with high quality electron temperature determinations.
  We derive kinematic distances in a self-consistent way for the
  entire sample.  The radial gradient in [O/H] is $-0.082 \pm\
  0.014$\dexkpc\ for $Az = 90$\degree$-130$\degree, about a factor of
  two higher than the average value between $Az =
  0$\degree$-60$\degree.  Monte Carlo simulations show that the
  azimuthal structure we reported for $Az = 0$\degree$-60$\degree\ is
  not significant because kinematic distance uncertainties can be as
  high as 50\% in this region.  Nonetheless, the flatter radial
  gradients between $Az = 0$\degree$-60$\degree\ compared with $Az =
  90$\degree$-130$\degree, are significant within the uncertainty. We
  suggest that this may be due to radial mixing from the Galactic Bar
  whose major axis is aligned toward $Az \sim\ 30$\degree.

\end{abstract}


\section{Introduction}\label{sec:intro}

Measurements of the elemental abundance distribution in the Galactic
disk provide important constraints to the formation and evolution of
the Milky Way \citep[e.g.,][]{pagel97}.  \citet{searle71} was the
first to detect radial gradients in nearby galaxies using
collisionally excited lines (CELs) of oxygen, nitrogen, and sulfur
from \hii\ regions.  Observations of CELs in the Milky Way disk are
more sensitive, but they are affected by dust and it is difficult to
determine distances which are necessary to derive radial abundance
gradients.  Nevertheless, radial gradients have been detected in the
Galactic disk and interpreted as higher processing of metals by stars
in the central regions of the Milky Way \citep[e.g.,][]{henry99,
  tosi00}.

The primary metallicity tracers in the Galactic disk are open
clusters \citep[e.g.,][]{janes79, twarog97, friel02, young12,
  frinchaboy13}, Cepheids \citep[e.g.,][]{caputo01, lemasle13,
korotin14}, OB stars \citep[e.g.,][]{fitzsimmons90, rolleston00,
daflon04}, red giant stars \citep[e.g.,][]{hayden13, boeche14,
bovy14}, planetary nebulae \citep[e.g.,][]{maciel03, pottasch06,
stanghellini10, henry10}, and \hii\ regions
\citep[e.g.,][]{peimbert78, shaver83, afflerbach97, deharveng00,
  rudolph06, esteban13}.  Typically, Fe/H is used as a proxy for
metallicity in stars using absorption lines, whereas O/H is the main
metallicity diagnostic in the interstellar medium (ISM) using emission
lines \citep{henry99}.  Other abundance ratios (e.g., $\alpha$/Fe)
provide additional information \citep{hayden13}.  Each tracer has
advantages and disadvantages.  

Here we explore the metallicity distribution of the Milky Way using
\hii\ regions.  They are created by the hydrogen ionizing photons from
massive OB stars and can be detected at infrared (IR) and radio
wavelengths throughout the Galactic disk.  Since \hii\ regions are
young ($< 10$\myr), their abundances provide a measure of the nuclear
processing of many stellar generations.  \hii\ regions with maser
parallax measurements have accurate distances that are critical for
exploring Galactic azimuthal metallicity structure.  Optical and IR
CELs are the primary diagnostics of metallicity in \hii\ regions, but
more recently optical recombination lines (ORLs) have been used
\citep[e.g.,][]{esteban05}.  Radio recombination line (RRL) and
thermal continuum emission provide an accurate measure of the nebular
electron temperature which has been used to indirectly probe
metallicity \citep{churchwell75, churchwell78, mezger79, lichten79,
  wilson79, wink83, afflerbach96, quireza06, balser11}.  Such radio
diagnostics of \hii\ regions yield an extinction free tracer to map
the metallicity distribution across the entire Milky Way disk.


\section{Observations and Data Reduction}\label{sec:obs}

\begin{figure}
\includegraphics[angle=360,scale=0.8]{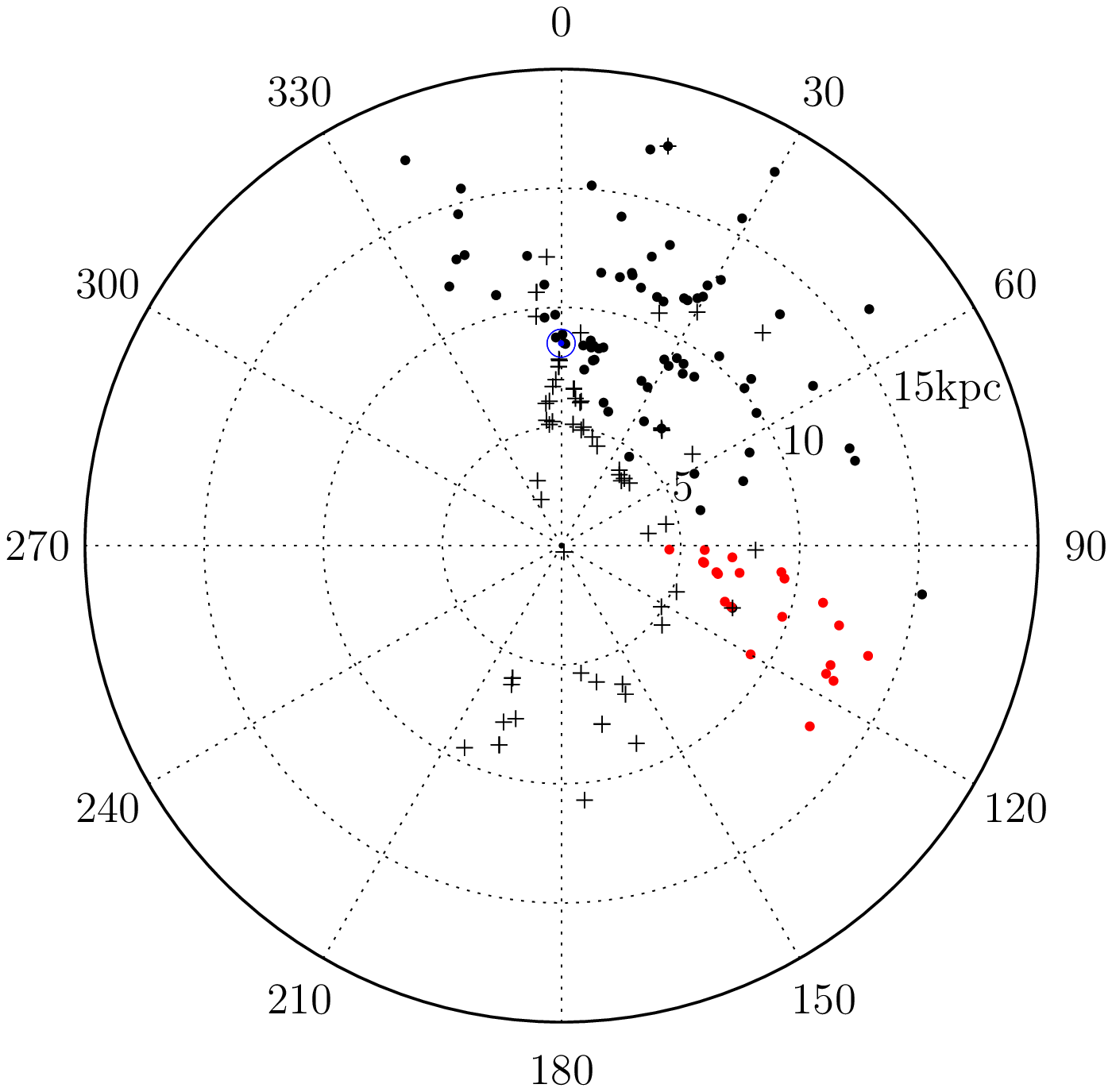} 
\caption{Galactic distribution of \hii\ regions in the Green Bank
  sample.  Plotted are the Galactic location ($Az, R_{\rm gal}$) of
  \hii\ regions from \citet{quireza06} (black crosses) and
  \citet{balser11} (black circles).  The HRDS \hii\ regions studied
  here are shown as red circles.  The blue circle marks the location
  of the Sun 8.5\kpc\ from the Galactic Center, which lies at the plot
  origin.}
\label{fig:dist}
\end{figure}

Here we expand on our previous efforts of using RRL and continuum
emission from \hii\ regions throughout the Galactic Plane to derive
electron temperatures that are used as a proxy for metallicity.
Figure~\ref{fig:dist} shows the Galactic distribution of \hii\ regions
with derived electron temperatures using data from the National Radio
Astronomy Observatory (NRAO)\footnote{\nraoblurb} 140 Foot telescope
\citep[][black crosses; hereafter the ``140 Foot Sample'']{quireza06}
and the Green Bank Telescope (GBT) \citep[][black circles; hereafter
the ``GBT Sample'']{balser11}.

The \hii\ Region Discovery Survey (HRDS) has significantly expanded
the census of \hii\ regions in the Galaxy using the GBT
\citep{bania10, anderson11, anderson12}.  Here we use nebulae from
this larger sample to study the metallicity distribution in the Milky
Way.  Since the measurement uncertainty in deriving the \hii\ region
electron temperature from single-dish radio data is dominated by
continuum measurements, we first performed a continuum survey of 143
HRDS nebulae to select the best sources.  We chose the relatively
unexplored Galactic azimuth\footnote{Here we define azimuth as zero in
  the direction of the Sun as viewed from the Galactic Center,
  increasing clockwise or as the Sun orbits the Galactic Center.}
range between $Az = 90$\degree$-130$\degree.  The continuum
observations were made in excellent weather conditions using the GBT
dynamic scheduling system (see below for details about continuum
observations).

We select 21 of the best continuum sources based on the following
criteria. (1) Sources having a continuum intensity signal-to-noise
ratio greater than 10.  (2) Since we are probing trans-Galactic paths,
source confusion can be a significant problem.  We chose sources where
the \hii\ region was either isolated or where any blending of
components could be well-fit by Gaussian profiles.  (3) We select
sources with either flat RRL spectral baselines or with baseline
structure that could be well-fit by a low order polynomial function.
(4) Since we seek to explore metallicity structure, we chose sources
that span a wide range of Galactic radii over the specified azimuth
range.  These HRDS sources are shown as red circles in
Figure~\ref{fig:dist}.  Hereafter, all the \hii\ regions shown in
Figure~\ref{fig:dist} will be called the Green Bank Sample.  The Green
Bank sample contains 90 nebulae with accurate electron temperatures.

We used the GBT to observe the RRL and continuum emission toward the
21 HRDS \hii\ regions in the sample that did not yet have data of
sufficient quality to derive their metallicities.  The observations
spanned 10 distinct observing sessions between 2011 March 25 and 2012
April 24.  The observing procedures and data reduction were similar to
that of \citet{balser11}.  We briefly describe them below.

Total power, position-switched spectra were taken with the GBT
auto-correlation spectrometer (ACS).  We first observed a reference
(Off) position for 6 minutes and then, tracking the same sky path,
observed the target position (On) for 6 minutes, for a total time of
12 minutes.  The ACS was tuned to 8 distinct frequencies, each with a
bandwidth of 50\mhz.  Since the X-band receiver samples two orthogonal
circular polarizations, each total power pair contains 16 independent
spectra.  The 8 spectral windows included seven Hn$\alpha$ RRLs
(\hy87-\hy93) covering 8-10\ghz\ with half-power beam-widths (HPBWs)
from 73 to 90 arcsec.  There are eight Hn$\alpha$ RRLs within the
X-band receiver, but the \hy86\ RRL is confused with higher order
RRLs.  Therefore, we tuned the last spectral window to the H114$\beta$
and H130$\gamma$ RRLs.  The center rest frequencies include: 8045.6,
8300.0, 8584.8, 8665.3, 8877.0, 9183.0, 9505.0, and 9812.0 MHz.

We made continuum observations at two circular orthogonal
polarizations with the Digital Continuum Receiver (DCR) at a frequency
of 8556\mhz, near the \hy91\ RRL, with a bandwidth of 320\mhz.  We
scanned the telescope at a rate of $80\,{\rm arcmin}\,{\rm
  minute}^{-1}$ in R.A. and then in Decl. for 60\s.  The data were
sampled by the DCR at 0.1\s.  We performed offset pointing and focus
observations on a calibrator every two hours to ensure that we were
centered on the \hii\ region.  Each continuum scan is twice the size
of that used in \citet{balser11} to better establish the zero level in
these directions with more complex continuum emission structure.  The
line-to-continuum ratio is used to derive the electron temperature.
Therefore, we observed the continuum before each set of spectral line
observations for each source so that fluctuations in the weather or
telescope gain between the continuum and spectral line observations
would be minimized.

We reduced and analyzed the RRL and continuum data following the
procedures in \citet{balser11}.  Six Hn$\alpha$ spectra were averaged
to increase the signal-to-noise ratio.  The \hy90\ RRL was not
included because a nearby, higher order RRL effected the baseline fit.
We re-sampled the velocity scale to that of the \hy87\ RRL, and
shifted all spectra to match the Doppler tracked \hy89\ RRL.  Since
the GBT's beam varies by 20\% across the X-band receiver, we assume
Gaussian source brightness distributions and HPBWs to scale the RRL
intensities to the \hy91\ RRL \citep[see][]{balser11}.

\begin{figure}
\includegraphics[angle=0,scale=0.55]{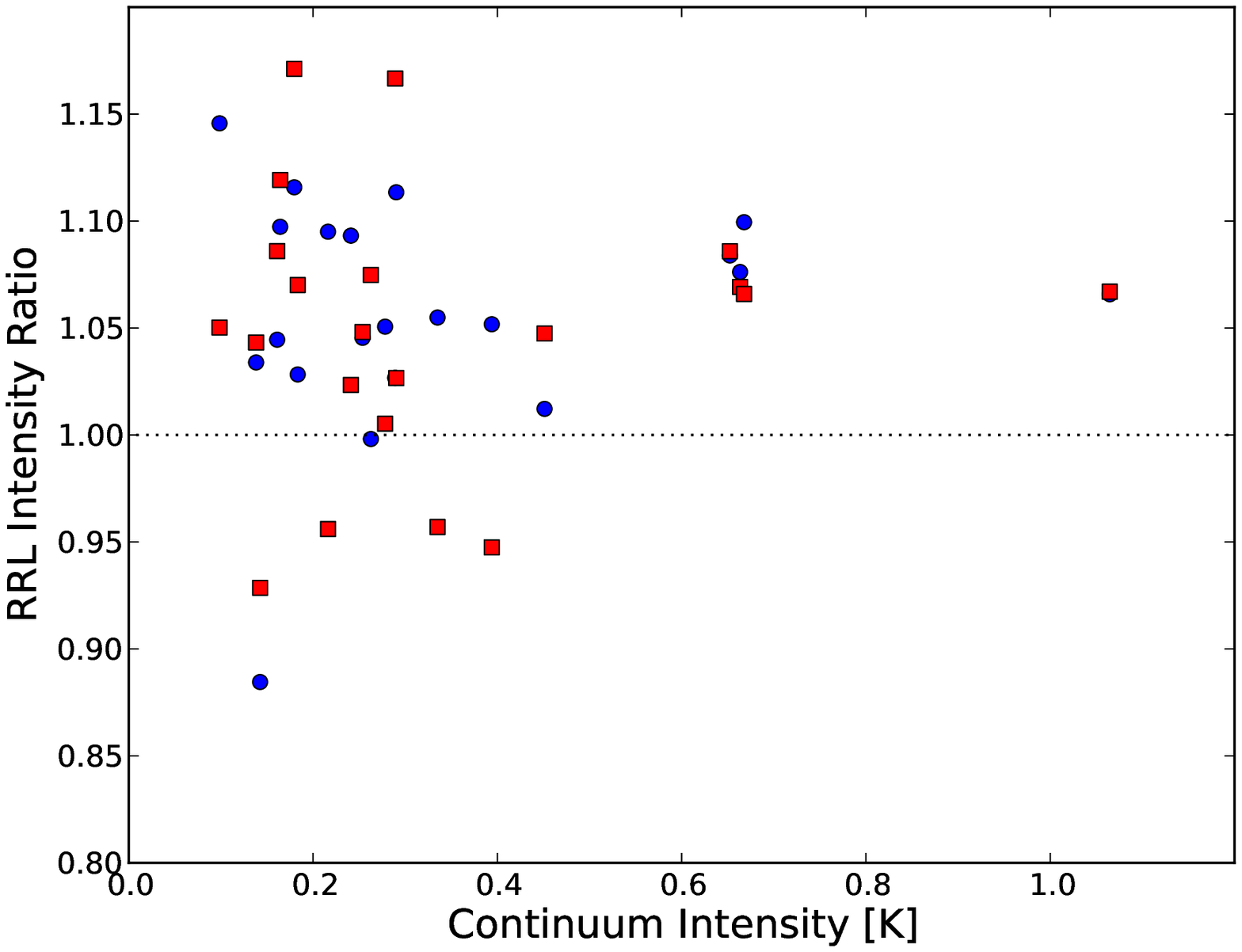}
\includegraphics[angle=0,scale=0.55]{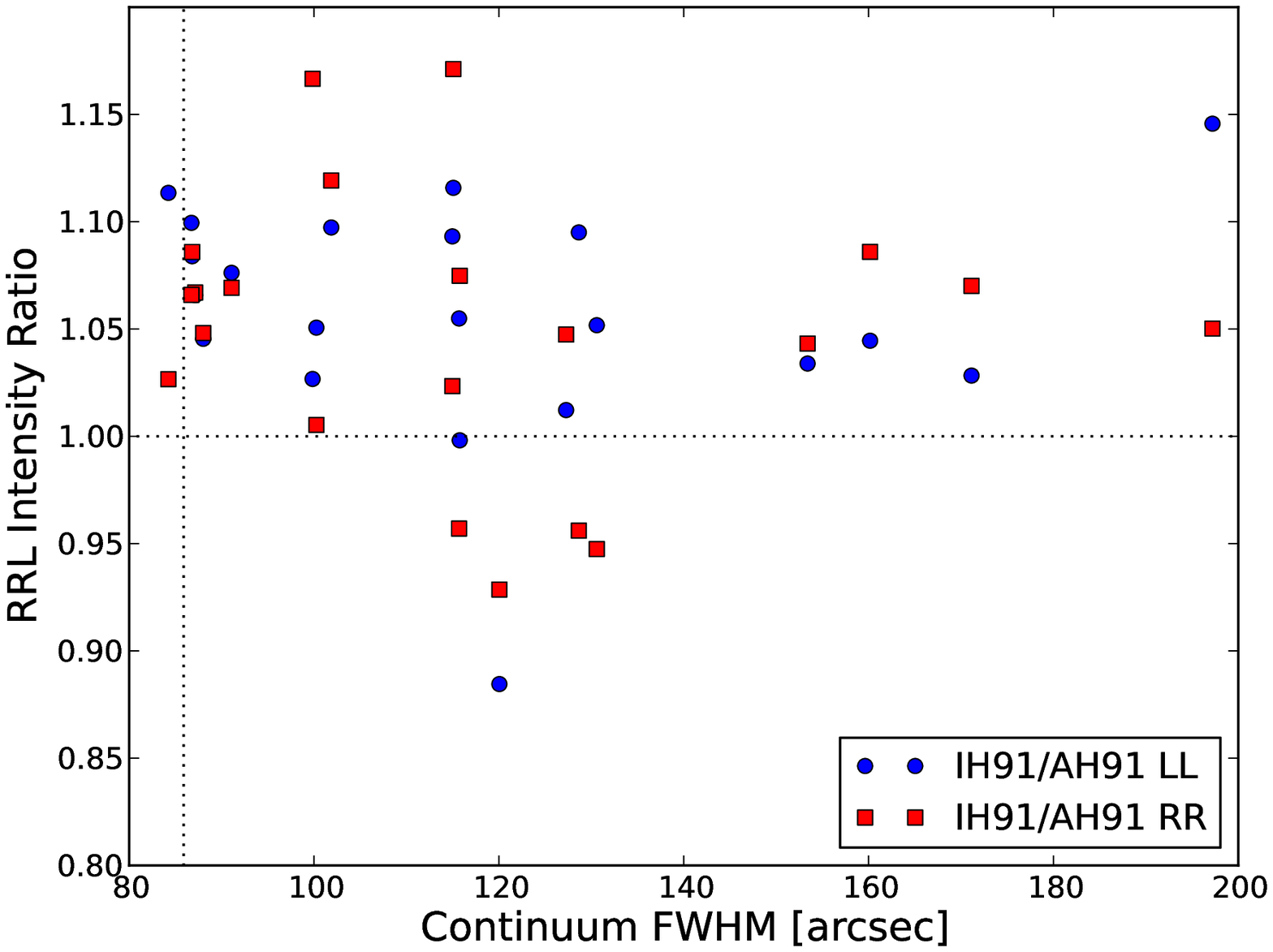}
\caption{RRL intensity ratios as a function of continuum intensity
  (top) and FWHM angular size (bottom).  Three different \hy91\
  intensities are calculated for all sources: H91 is the \hy91\
  intensity; IH91 is the \hy91\ intensity measured after the velocity
  scale has been interpolated; and AH91 is the \hy91\ intensity after
  the six adjacent RRLs have been interpolated and averaged.  Plotted
  are IH91/AH91 for each polarization.  The horizontal dashed line is
  a ratio of unity, whereas the vertical dashed line is the GBT's HPBW
  at 8.7\ghz.  We find that the IH91/AH91 ratios are typically less
  than 5-10\% from unity and are not correlated with either continuum
  intensity or angular size.}
\label{fig:ta}
\end{figure}

The electron temperature depends on the line-to-continuum ratio,
therefore, in principle, we only require a relative intensity
calibration.  This is not true here since we have only measured the
continuum at one frequency, near the \hy91\ RRL.  Following
\citet{balser11} we calculate an average RRL intensity relative to the
\hy91\ line.  The continuum data were taken at 8665\mhz\ over a
bandwidth that includes the \hy91\ transition, so we calibrated the
Hn$\alpha$ lines relative to the \hy91\ line.  In Figure~\ref{fig:ta},
we plot the ratio of the interpolated \hy91\ intensity (IH91) and the
average RRL intensity (AH91), scaled to the \hy91\ frequency as
discussed above, for all 21 HRDS sources.  The IH91/AH91 ratios are
$1.058 \pm\ 0.0533$ for LL and $1.048 \pm\ 0.0631$ for RR.  The ratios
are offset by less than 5-10\% from unity and are caused by errors in
the noise tube calibration ($T_{\rm cal}$) and any systematic errors
in our data reduction.  The dispersion in IH91/AH91 may be caused by
several factors that include systematic errors in the spectral
baseline fits, random noise, elevation gain fluctuations, weather
fluctuations, etc.  The dispersion here is three times larger than in
\citet{balser11} because the sources are weaker by more than a factor
of 5.  Nevertheless, we deem that the line-to-continuum calibration
scale is accurate to within $\sim 5$\%.  Using these procedures we
calculated and applied corrections factors of ${\rm CF} = 1.058$ for
LL and ${\rm CF} = 1.048$ for RR.

\begin{figure}
\includegraphics[angle=0,scale=0.55]{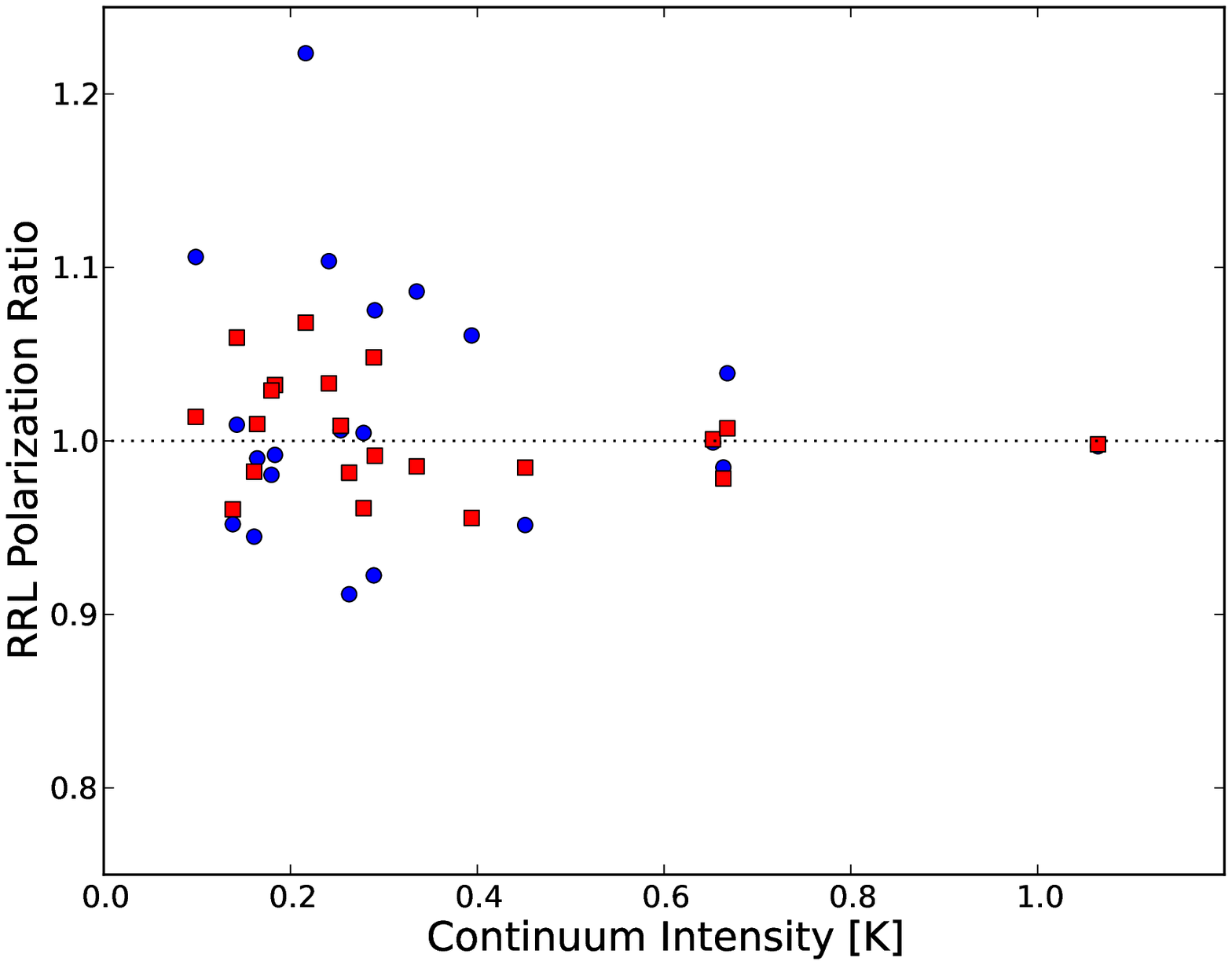}
\includegraphics[angle=0,scale=0.55]{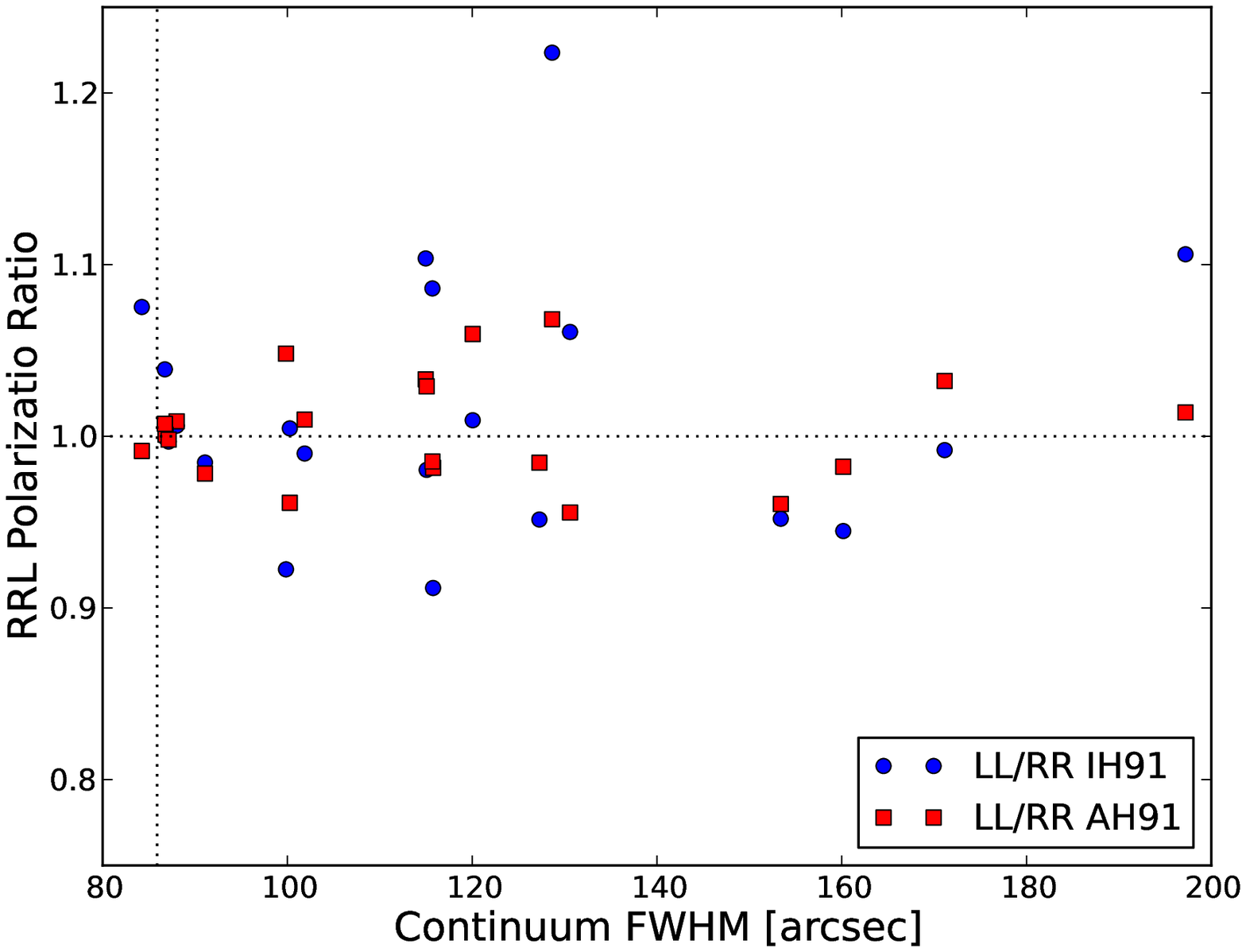}
\caption{RRL polarization ratios as a function of continuum intensity
  (top) and FWHM angular size (bottom).  Plotted are the polarization
  ratios (LL/RR) for the IH91 and AH91 RRL intensities (see
  Figure~\ref{fig:ta}).  The AH91 polarization ratios are consistent
  with random errors in $T_{\rm cal}$.}
\label{fig:pol}
\end{figure}

Comparing the RRL intensity for each polarization provides an
additional check on our calibration.  We derive polarization ratios of
$1.016 \pm\ 0.0714$ and $1.004 \pm\ 0.0310$ for IH91 and AH91,
respectively using all 21 HRDS sources (see Figure~\ref{fig:pol}).
These ratios are less than the expected $T_{\rm cal}$ uncertainties of
about 10\%.  Moreover, the dispersion in the distribution from IH91 to
AH91 decreases as $1/\sqrt{N}$ as expected for random Gaussian noise.

We used the IDL single-dish software package TMBIDL, written by our
team, to analyze these data\footnote{See
  http://www.bu.edu/iar/tmbidl/.}.  The spectral line parameters were
determined by first fitting the spectral baseline with a third-order
polynomial, which was subtracted from the data, and then fitting
Gaussian functions to the H, He, and C RRLs.  Three parameters were
derived for each RRL: the peak antenna temperature, the FWHM line
width, and the Local Standard of Rest (LSR)\footnote{We used the
  kinematic local standard of rest (LSR) frame with the radio
  definition of the Doppler shift.  The kinematic LSR is defined by a
  solar motion of 20.0\kms\ toward ($\alpha$, $\delta$) = ($18^{\rm
    h}$, $+30$\degree) [1900.0] \citep{gordon76}.} velocity center.
The continuum parameters were determined by first removing the
background zero-level with a third-order polynomial and then using
Gaussian functions to model the peak continuum intensity and the FWHM
angular source size.

\section{Results}\label{sec:results}

Here we use radio observations of \hii\ regions to determine the
metallicity structure of the Milky Way disk.  Accurate measurement of
the line-to-continuum ratio is necessary to derive the electron
(thermal) temperature.  Metals such as oxygen and nitrogen are the
primary coolants in \hii\ regions and are well correlated with
electron temperature \citep[e.g.,][]{shaver83}.  Therefore we use the
derived electron temperatures as a proxy for metallicity.  \hii\
region locations are determined from the LSR velocity and a model of the
rotation of the Galactic disk.  Together these provide the information
necessary to produce a map of the metallicity distribution in the
Galactic disk.

\begin{figure}
\includegraphics[angle=-90,scale=0.30]{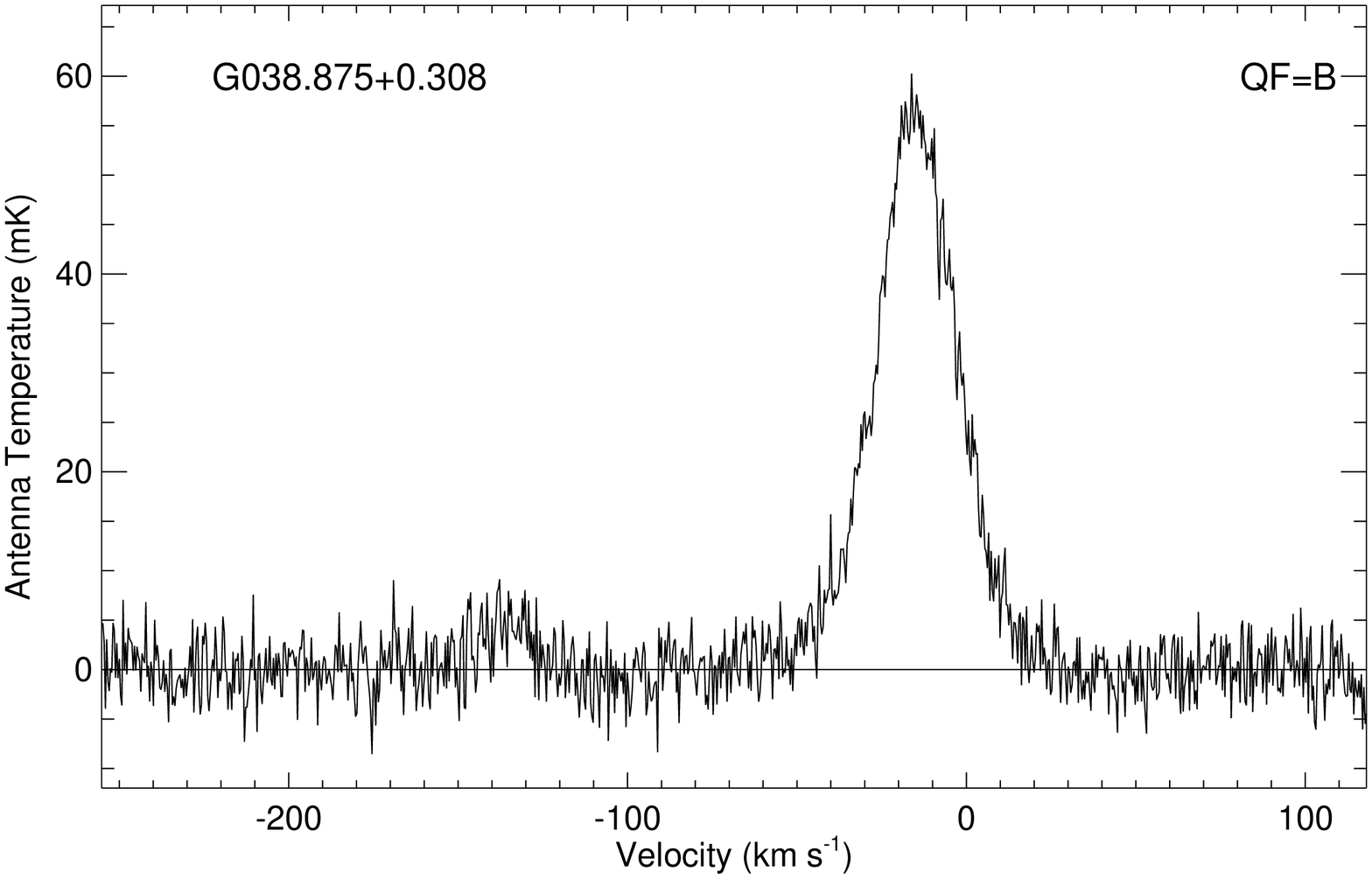} 
\includegraphics[angle=-90,scale=0.30]{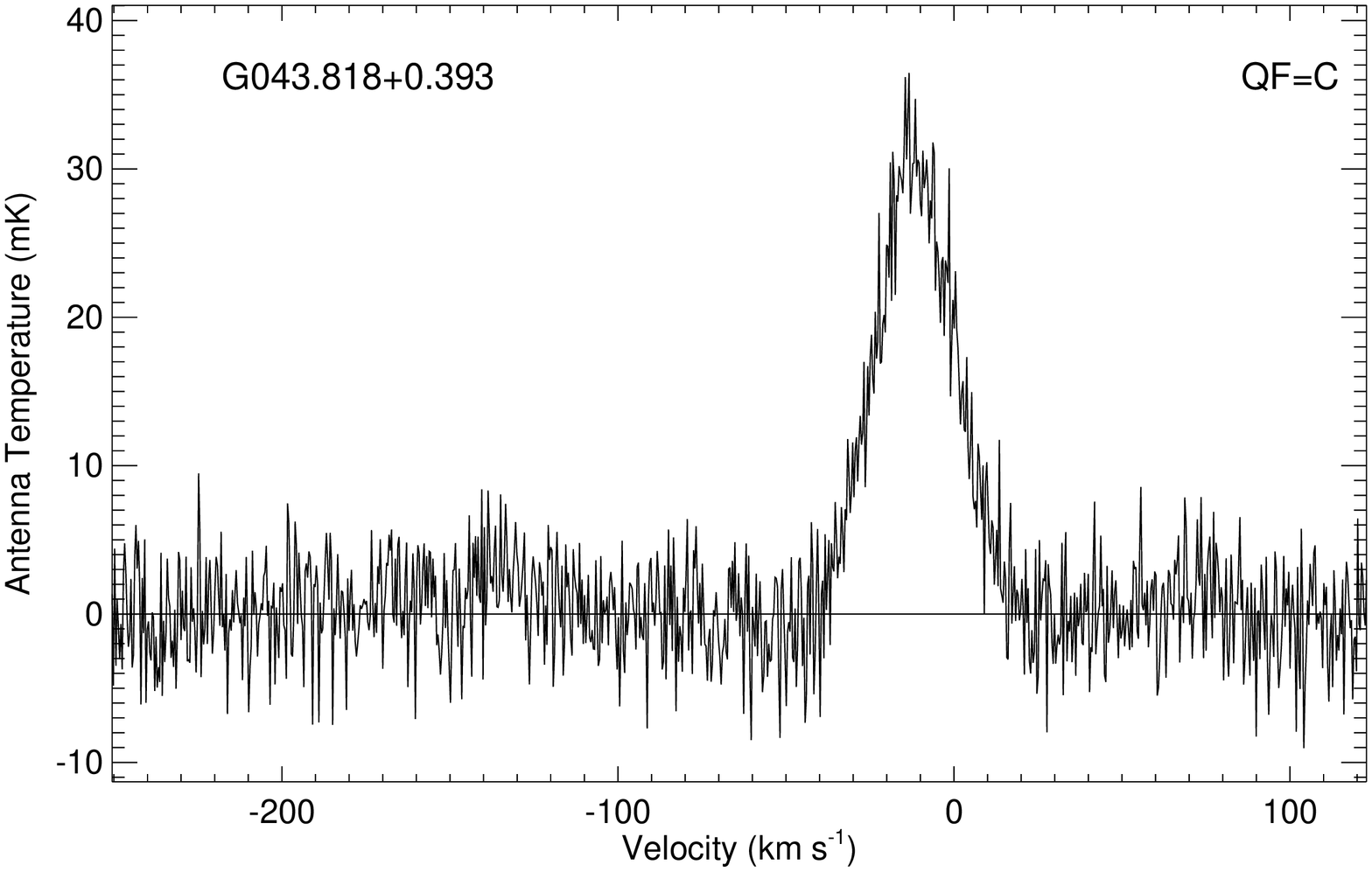} 
\includegraphics[angle=-90,scale=0.30]{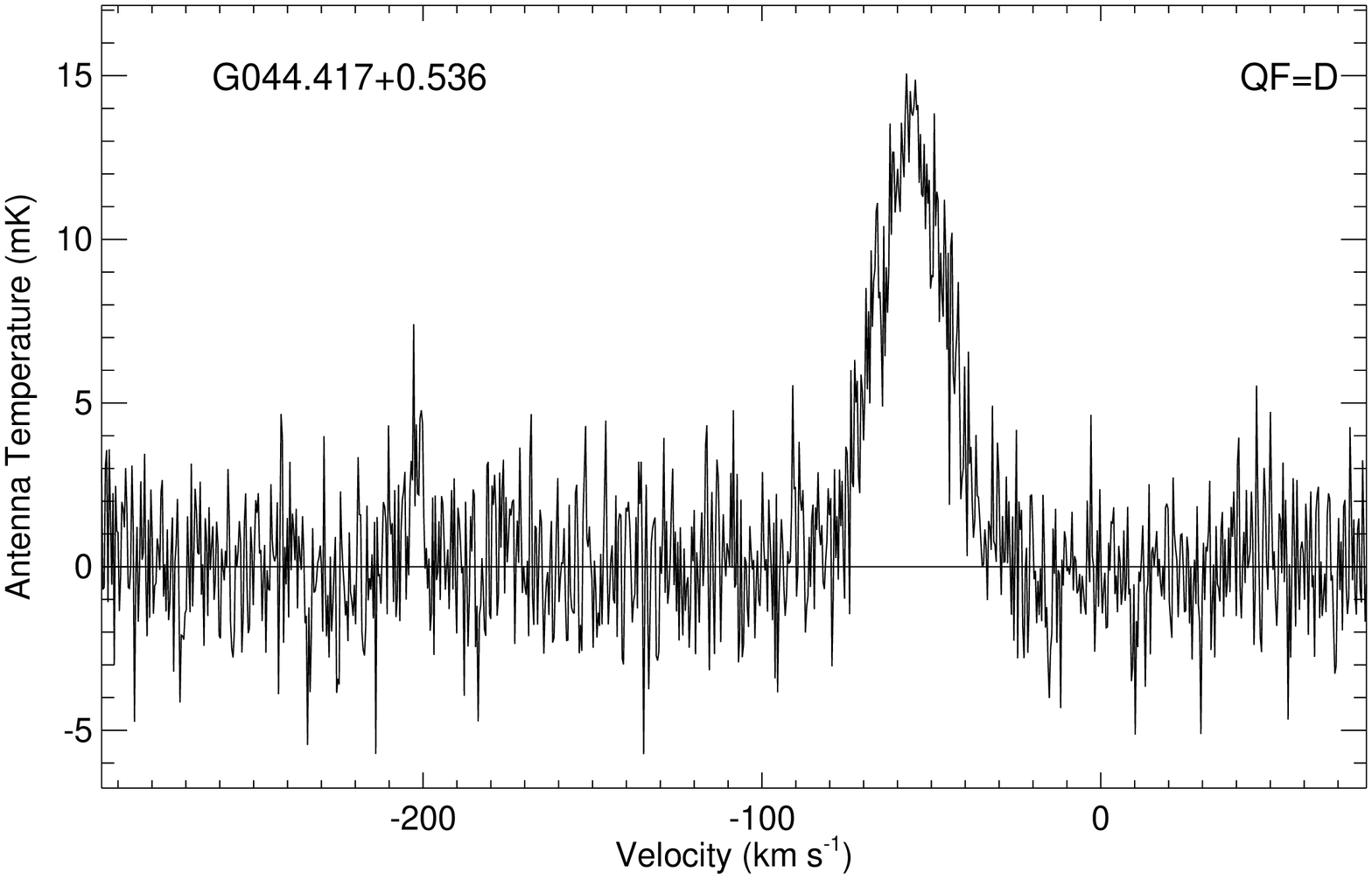} 
\includegraphics[angle=-90,scale=0.30]{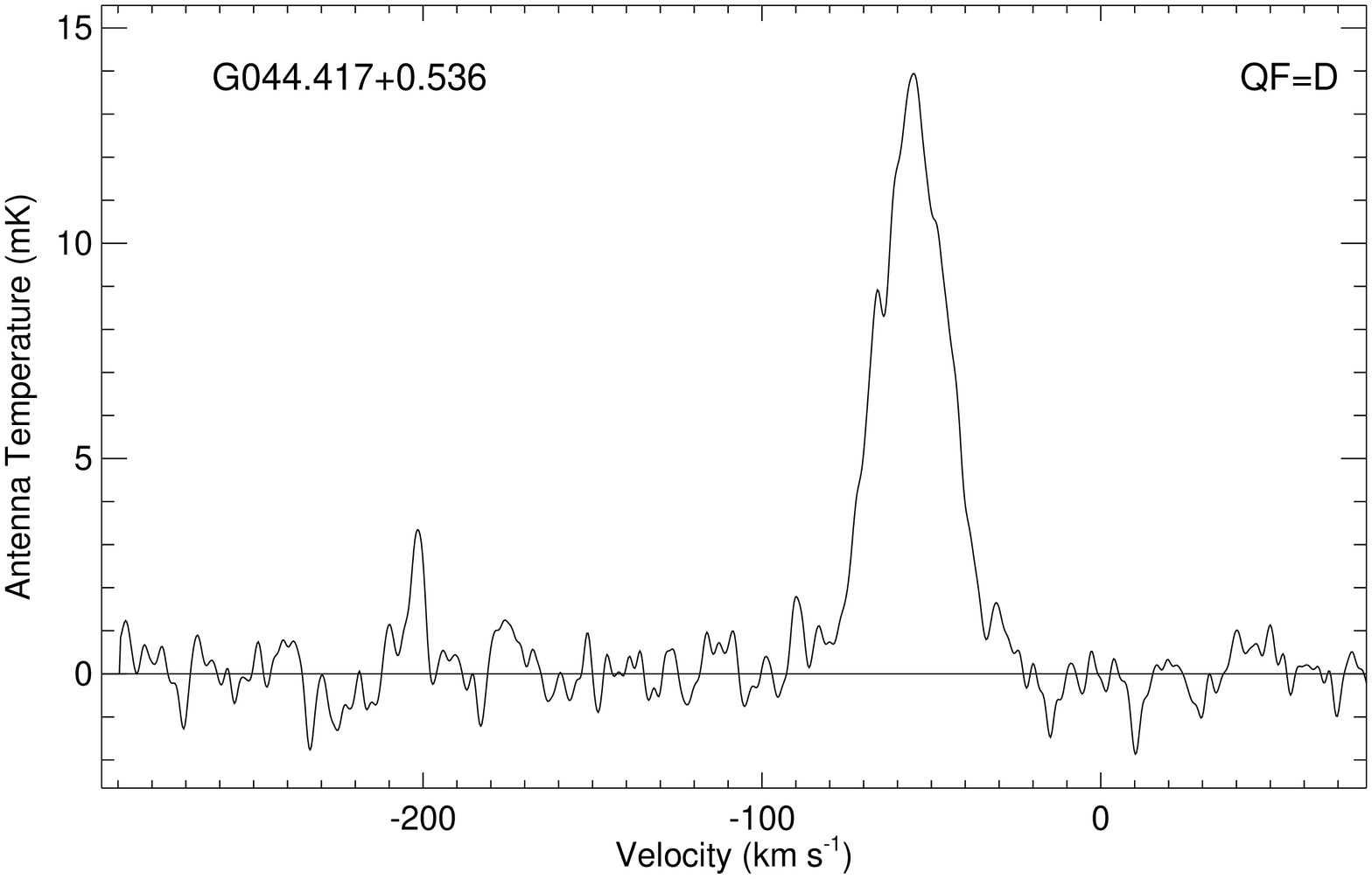} 
\caption{Sample \hii\ region RRL spectra.  The antenna temperature is
  plotted as a function of the LSR velocity.  The LSR velocity is
  referenced with respect to the H89$\alpha$ RRL.  For some sources
  the He and C RRLs, located about $-125$\kms\ from the H line, are
  detected.  The bottom right hand plot duplicates the spectrum for
  G044.417+0.536 where we smooth the spectrum to 2.61\kms\ to reveal the
  narrow carbon RRL.  A third order polynomial function is fit to the
  line free regions and removed from the data.  The quality factor is
  shown at the top right-hand corner of each plot, where QF=A is
  excellent and QF=D is poor.}
\label{fig:line}
\end{figure}

Table~\ref{tab:line} summarizes the RRL measurements for our 21 \hii\
region sample.  Listed are the source name, the element, Gaussian fit
parameters, the total integration time, the root-mean-square (rms)
spectral noise, and the quality factor (hereafter QF; see below).  The
Gaussian fits include the peak intensity, the full-width at
half-maximum (FWHM) line width, and the LSR velocity with their
associated errors.  The total integration time includes the average of
both polarizations for the \hy87, \hy88, \hy89, \hy91, \hy92, and
\hy93\ transitions. Figure~\ref{fig:line} shows sample spectra for
three sources where the antenna temperature is plotted as a function
of the LSR velocity.  Since the \hii\ regions in our sample are
distant, the peak intensities are weak, typically less than 50\mk.
Helium and carbon RRLs are detected in fewer than 20\% of the sample.

\begin{figure}
\includegraphics[angle=-90,scale=0.32]{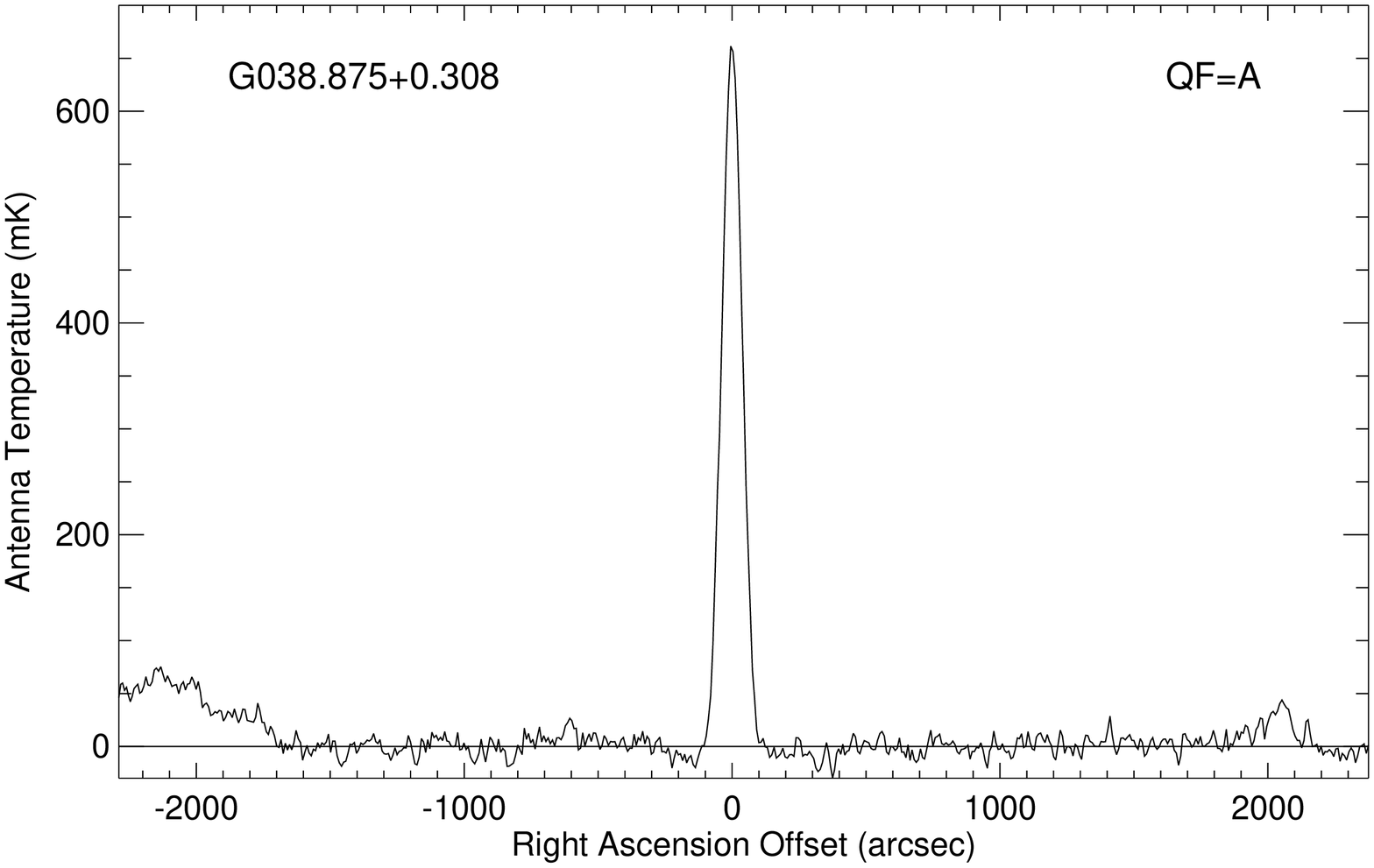} 
\includegraphics[angle=-90,scale=0.32]{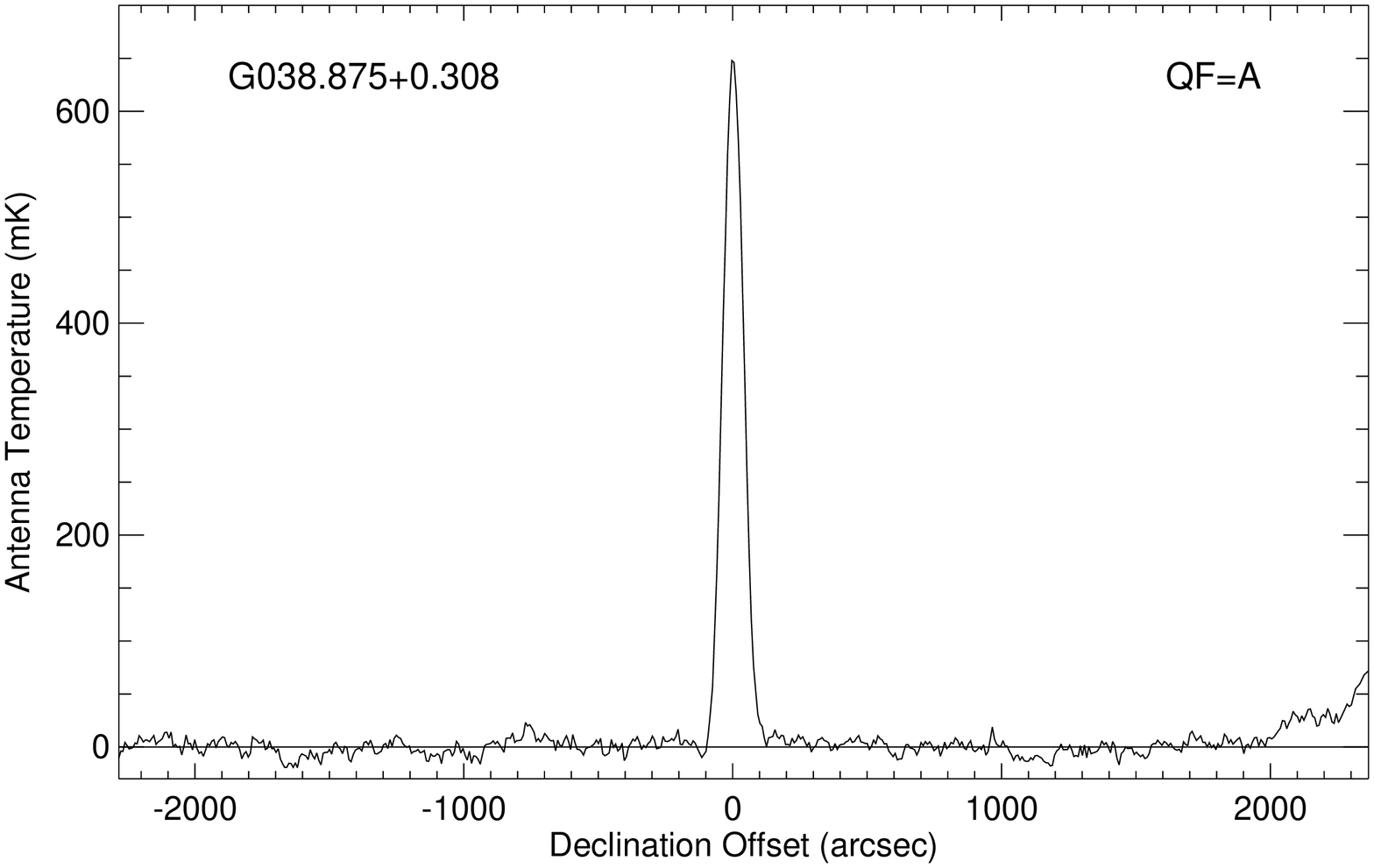} 
\includegraphics[angle=-90,scale=0.32]{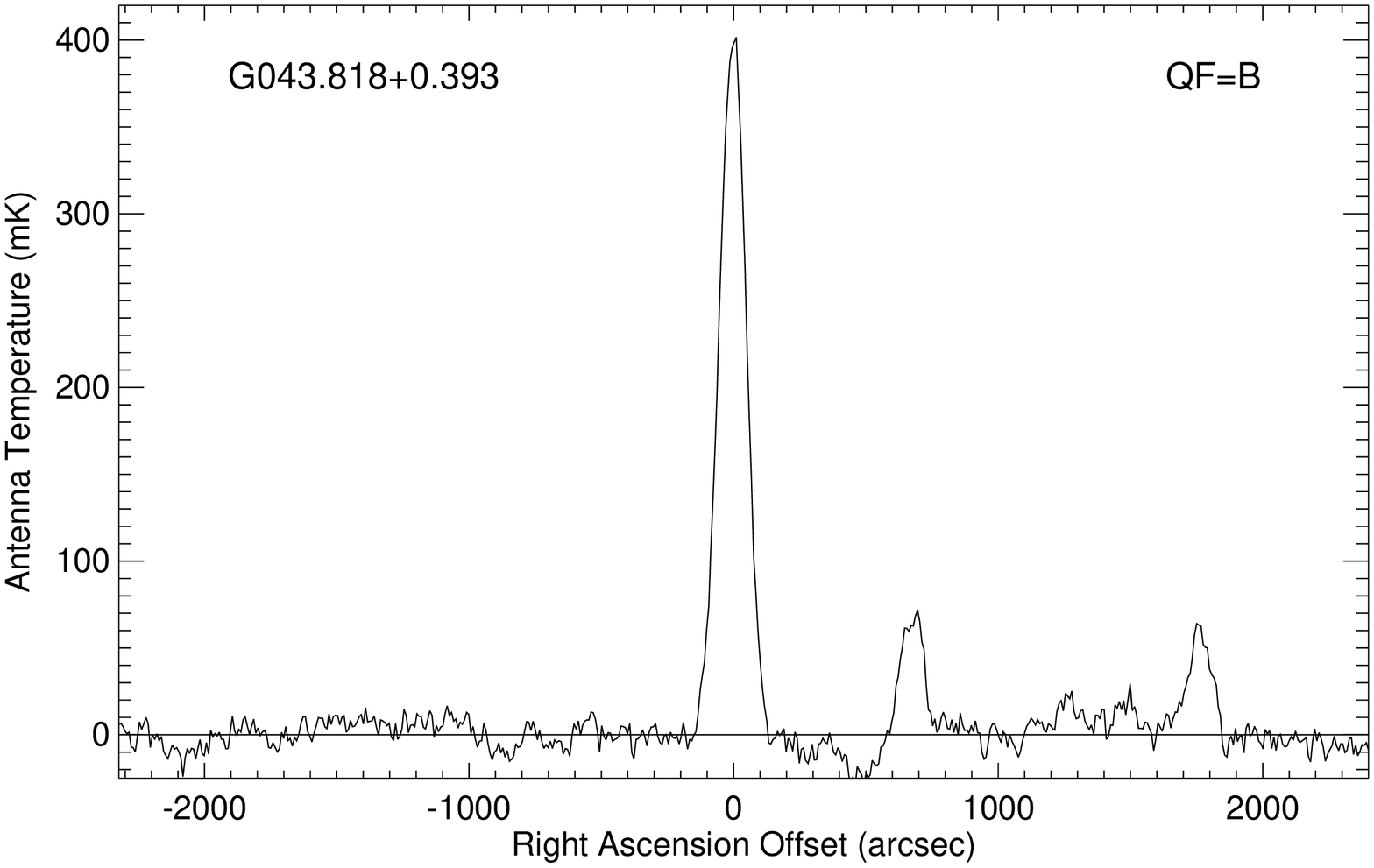} 
\includegraphics[angle=-90,scale=0.32]{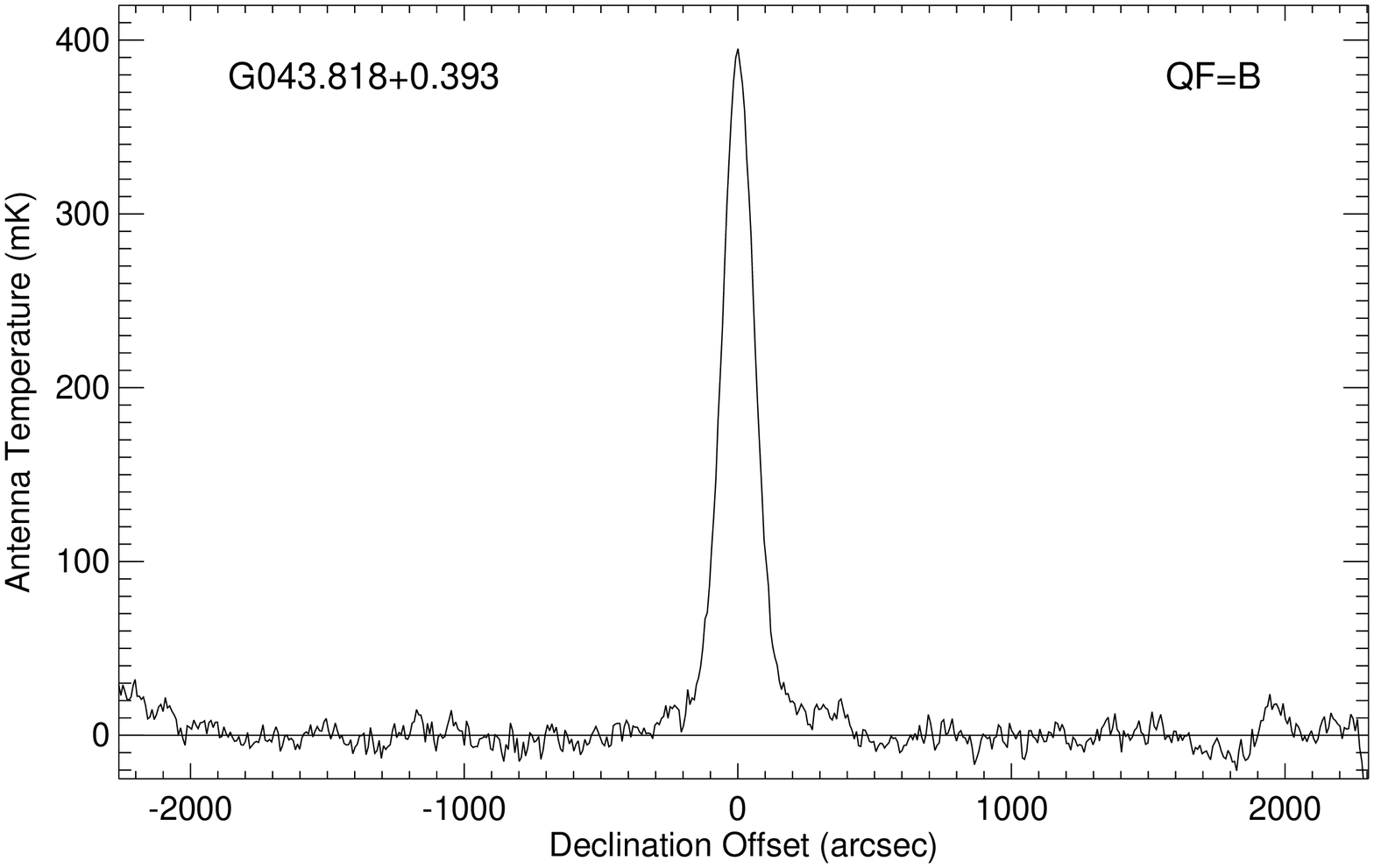} 
\includegraphics[angle=-90,scale=0.32]{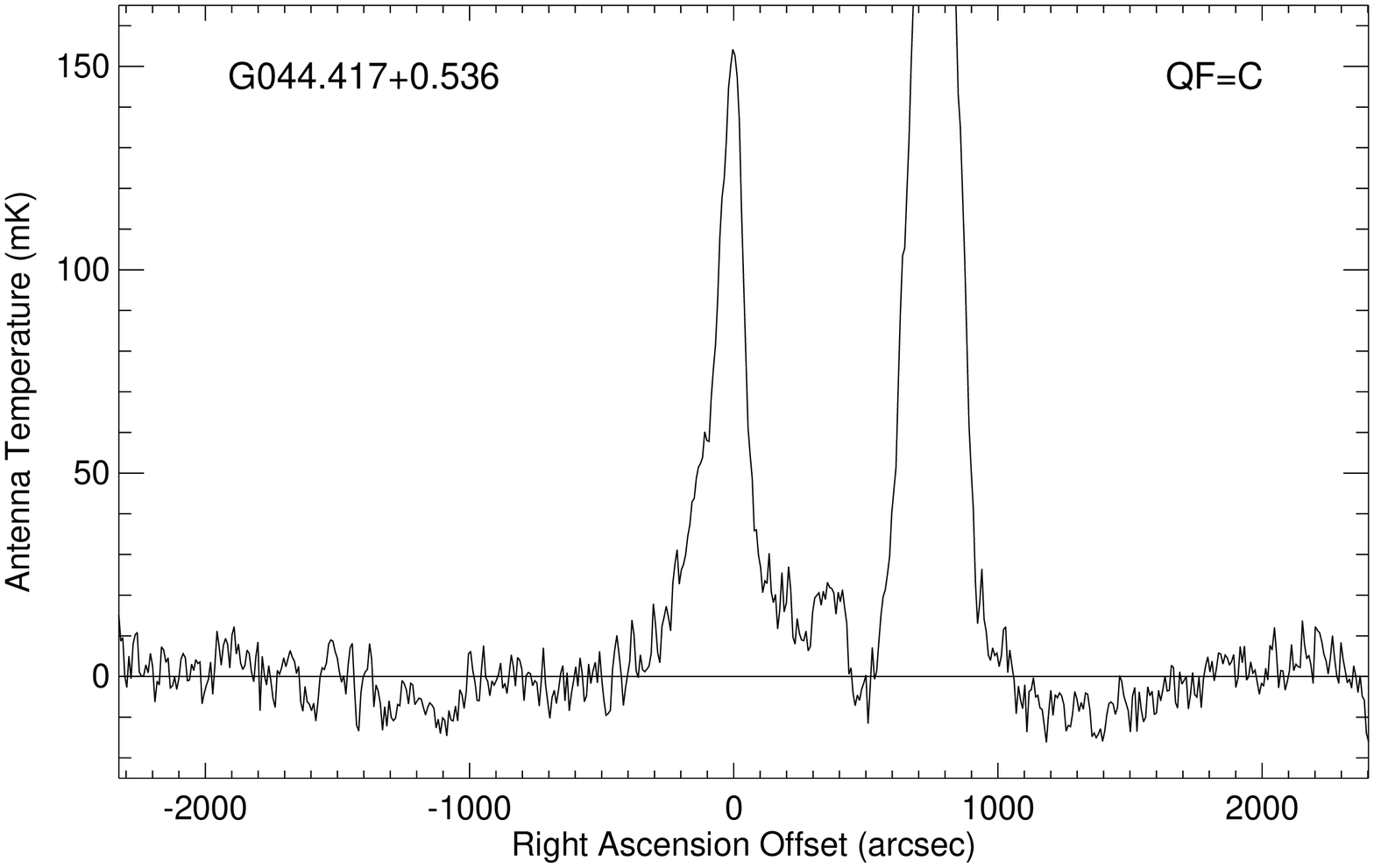} 
\includegraphics[angle=-90,scale=0.32]{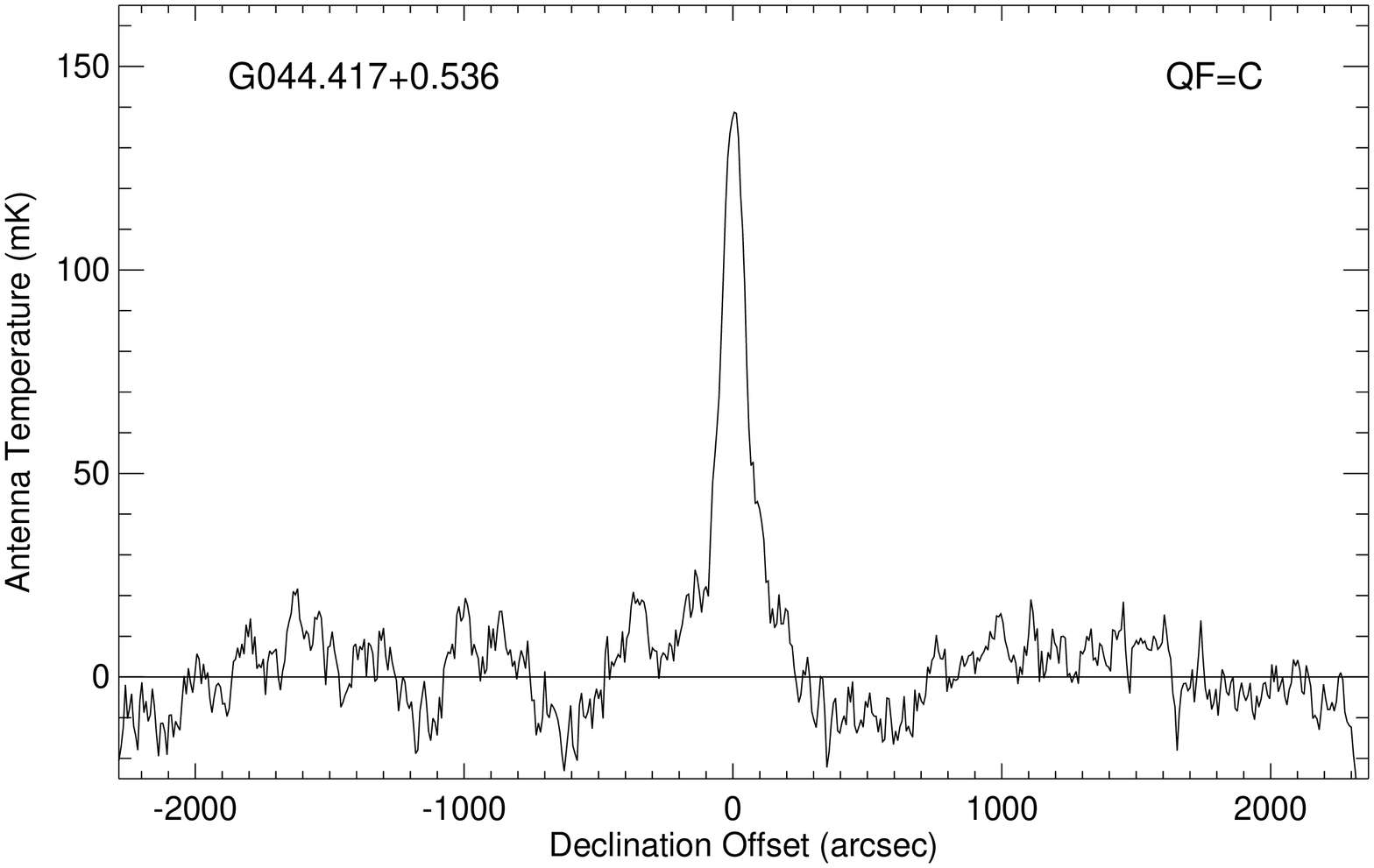} 
\caption{Sample \hii\ region continuum scans.  The antenna temperature
  is plotted as a function of the offset position relative to the
  target coordinates.  For each source the R.A. and Decl. scans are
  shown.  A third order polynomial function is fit to the baseline
  and removed from the data.  The quality factor is shown at the top
  right-hand corner of each plot, where QF=A is excellent and QF=D is
  poor.}
\label{fig:cont}
\end{figure}

Table~\ref{tab:cont} summarizes the continuum measurements for this
sample.  Listed are the source name, Gaussian fit parameters to the
continuum intensities in the R.A.  and Decl. scan directions, the
average Gaussian parameters, and the QF.  The peak intensity and FWHM
Gaussian fitted parameters are listed with their associated errors.
The average peak intensity and FWHM correspond to the arithmetic and
geometric mean, respectively.  Figure~\ref{fig:cont} shows the
continuum scans for the sources in Figure~\ref{fig:line}.  The antenna
temperature is plotted as a function of the offset R.A. and
Decl. directions relative to the nominal source position.

The QF is a qualitative assessment of the line and continuum
measurement where ``A'' is excellent and ``D'' is poor.  Baseline
structure and confusion are difficult to quantify and thus visual
inspection is used to assess the quality of line and continuum data.
Since the GBT RRL baselines are excellent with little line confusion
at X-band, we follow \citet{balser11} and calculate the QF for the
line data using the signal-to-noise ratio and the percent error of the
Gaussian fit of the line area.  For the continuum data the QF includes
a visual assessment of the baseline structure, source confusion, the
consistency of the peak intensity between the R.A. and Decl. scans,
and the repeatability between the forward and backward scans.  There
are no QF=A sources for the RRL data since the signal-to-noise ratio
is not sufficient for these more distant, weaker sources.  Since our
sample was selected based on the quality of the continuum data there
are no QF=D sources for the continuum data.

\begin{figure}
\includegraphics[angle=90,scale=0.30]{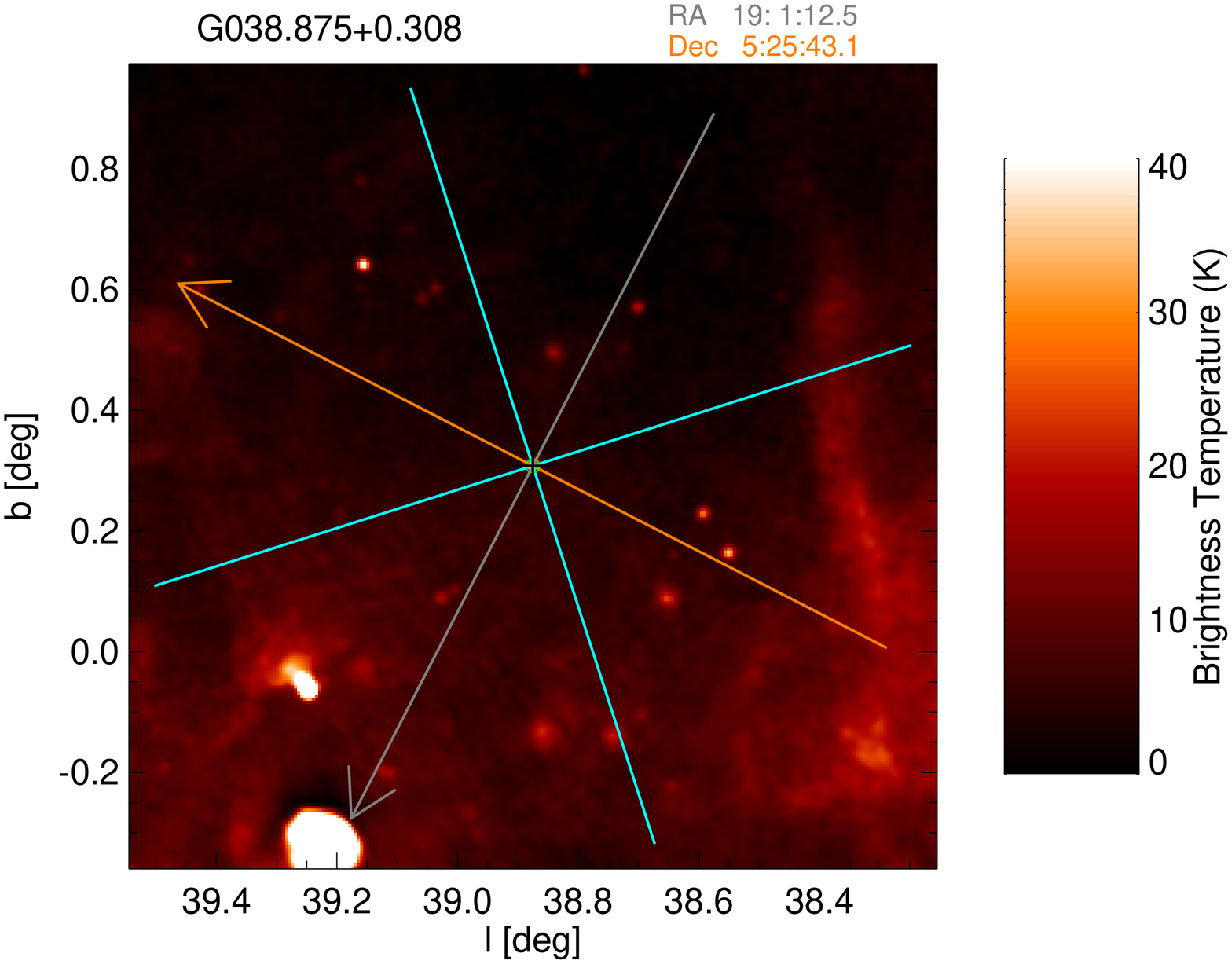} 
\includegraphics[angle=90,scale=0.30]{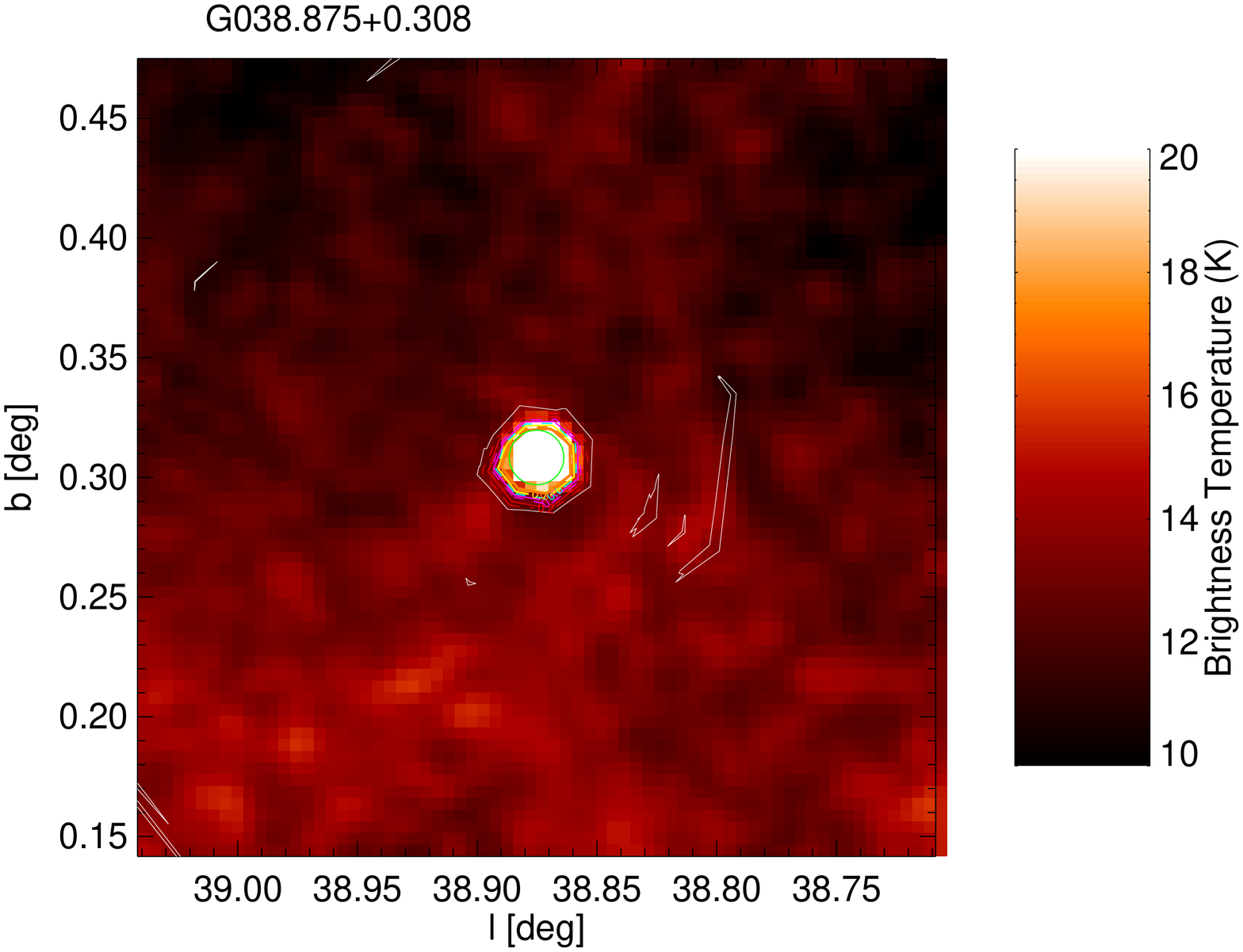} 
\includegraphics[angle=90,scale=0.30]{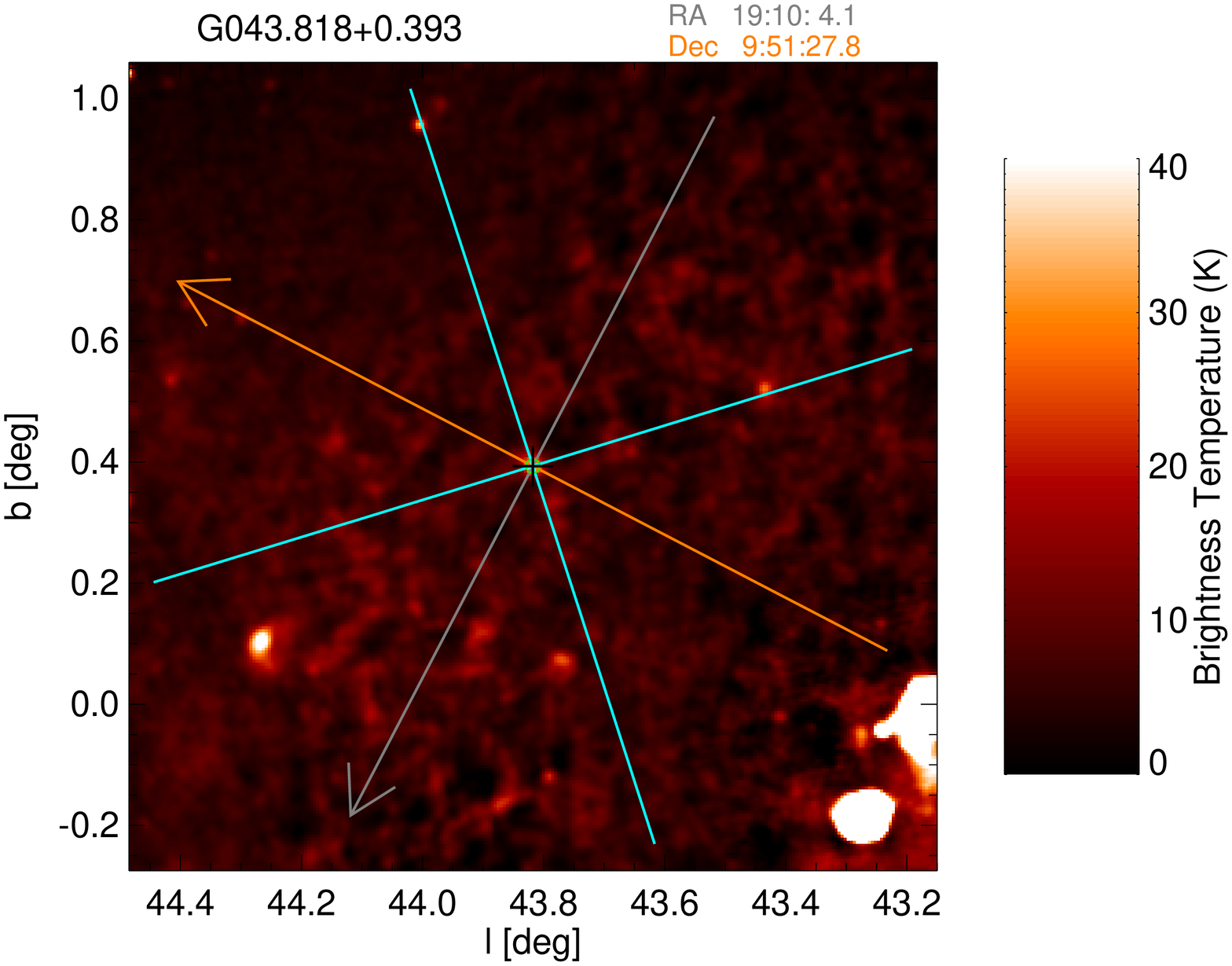} 
\includegraphics[angle=90,scale=0.30]{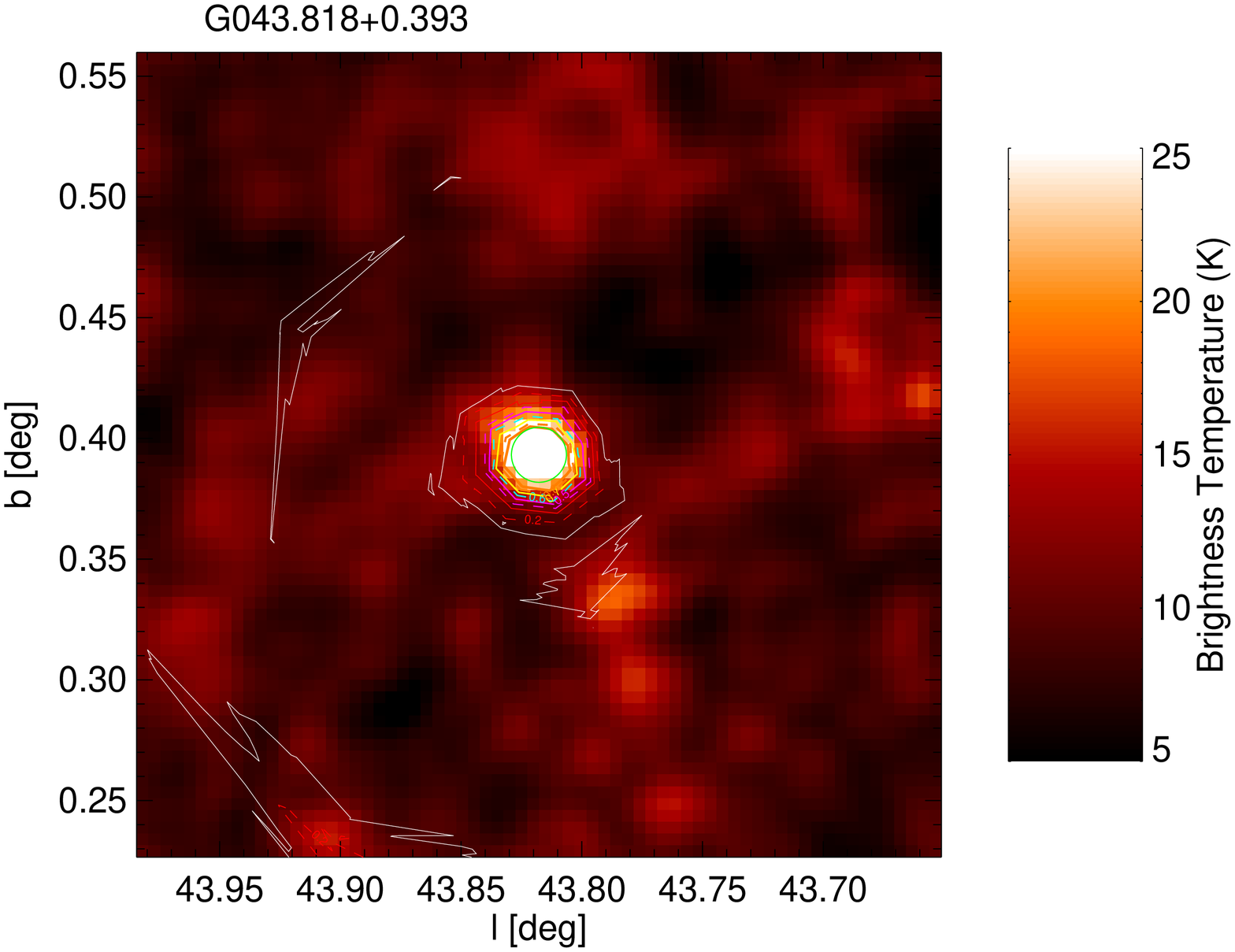} 
\includegraphics[angle=90,scale=0.30]{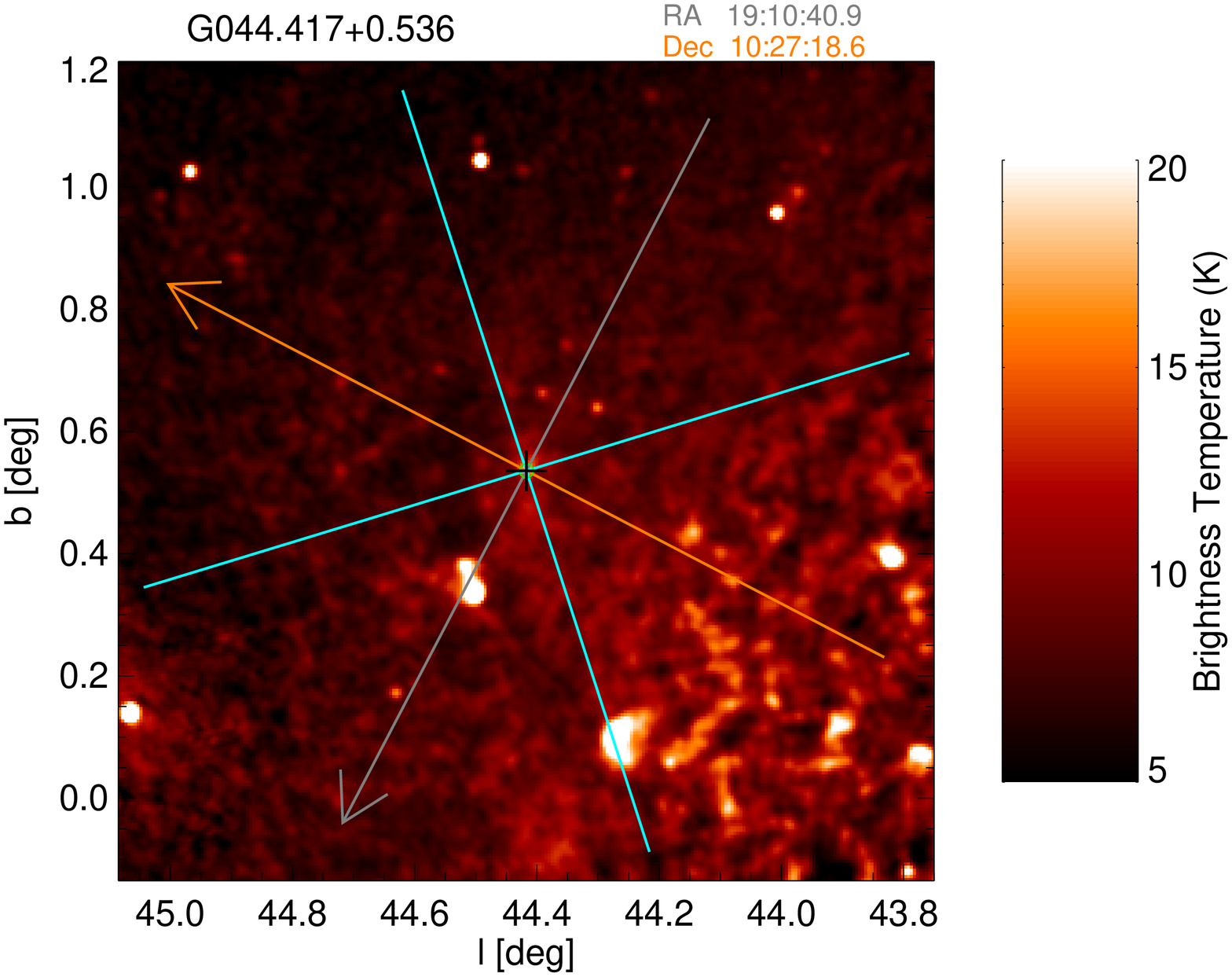} 
\includegraphics[angle=90,scale=0.30]{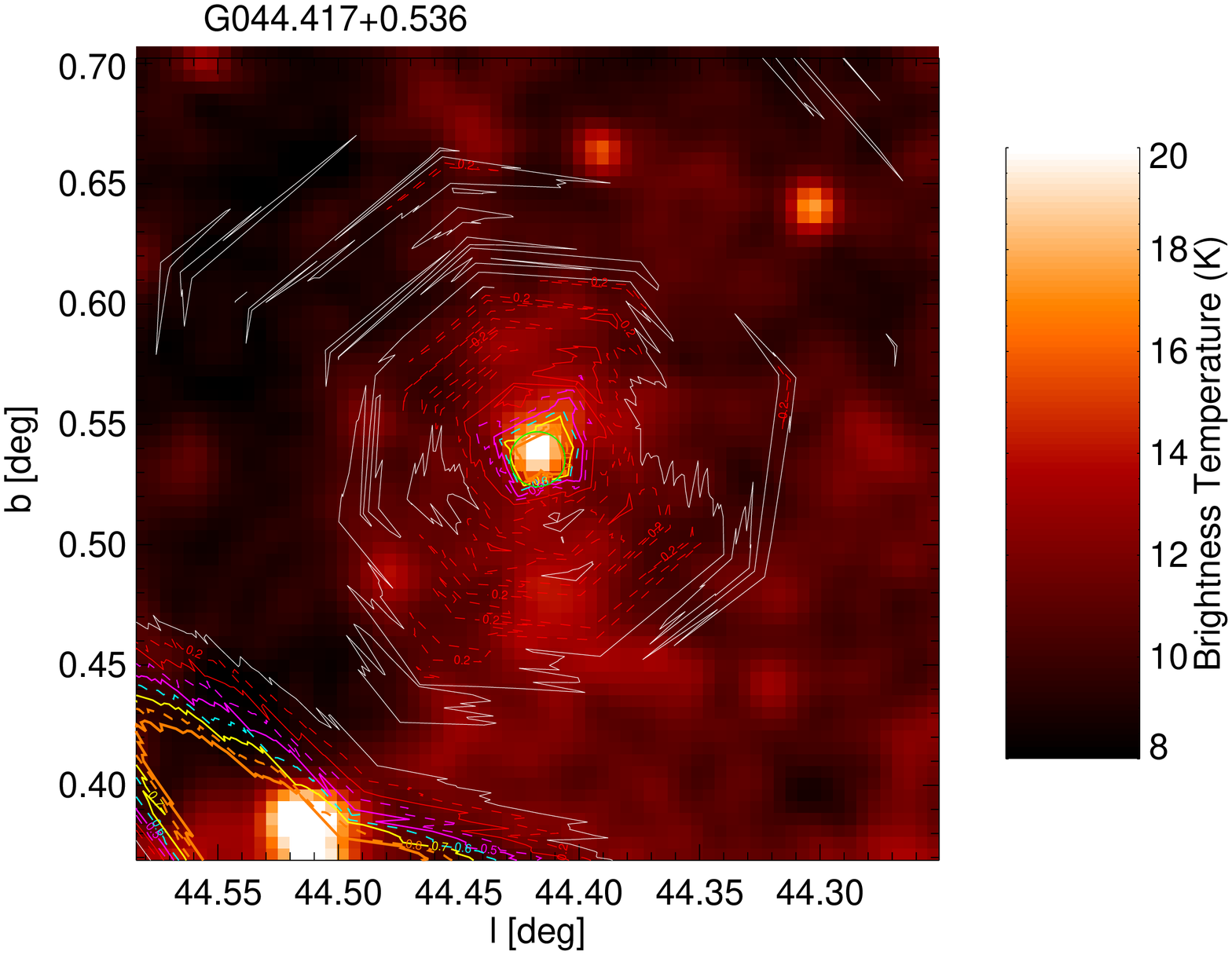} 
\caption{Sample \hii\ region continuum images in Galactic coordinates.
  Left panel: 1.4\ghz\ VGPS continuum image (80\degree\ $\times$
  80\degree) with the Spider scan orientation and extent shown as
  solid lines.  Increasing R.A. and Decl.  directions are indicated by
  the arrows.  The R.A. and Decl.  coordinates of the center of the
  map are shown at the top right corner of the plot.  Right panel:
  1.4\ghz\ VGPS continuum image (20\degree\ $\times$ 20\degree) with
  contours from the 8.7\ghz\ GBT continuum, interpolated from the
  Spider scan data.  The green circle at the center of the image
  indicates the size of the GBT HPBW. Contour levels range from 0.1 to
  0.9 of the peak in increments of 0.1.}
\label{fig:spider}
\end{figure}

To better assess the level of source confusion we made crude 8.7\ghz\
continuum images for each \hii\ region.  This was accomplished with
Spider continuum scans which consist of the R.A. and Decl. scans
described above plus two additional pairs of scans rotated by
45\degree\ \citep[see][]{robishaw09}.  There are 8 total scans for
each Spider measurement that probe 4 cardinal directions both in the
forward and backward direction.  Figure~\ref{fig:spider} shows images
produced from these Spider scans of the \hii\ regions in
Figure~\ref{fig:cont}.  The 1.4\ghz\ continuum emission from the VLA
Galactic Plane Survey \citep[VGPS][]{stil06} is displayed by the color
image.  The orientation and extent of the Spider scans are shown as
solid lines in the left panel.  The central region is well sampled but
there are large gaps in the outer regions where the data are
interpolated.  The right panel shows a magnified view of the central
region with contours representing the GBT 8.7\ghz\ continuum emission.
Based on these Spider scan observations all of the \hii\ regions in
our sample are clearly visible and reasonably well isolated from other
sources.

\subsection{Electron Temperature}\label{sec:te}

For an optically thin nebula that is in local thermodynamic
equilibrium (LTE), the hydrogenic RRL emission strength is
proportional to $n_{\rm e}^{2}\,T_{\rm e}^{-2.5}$, whereas the thermal
free-free continuum emission is proportional to $n_{\rm e}^{2}\,T_{\rm
  e}^{-1.35}$.  Here $n_{\rm e}$ and $T_{\rm e}$ are the electron
density and temperature, respectively.  The line-to-continuum ratio is
therefore proportional to $T_{\rm e}^{-1.15}$ and is an excellent
probe of the electron (thermal) temperature of the ionized gas
\citep{gordon09}.  Following \citet{quireza06}, for RRLs near 9\ghz\
the electron temperature is given by
\begin{equation}\label{eq:te}
{\left(T_{\rm e} \over K\right)} = \left[7103.3 {\left(\nu_{\rm L}
\over {\rm GHz}\right)}^{1.1} 
\left({T_{\rm C} \over\strut T_{\rm L} ({\rm H{^+}})} \right)
{\left(\Delta V ({\rm H{^+}}) \over \kms \right)}^{-1} 
\left( 1 + {y}\right)^{-1} \right]^{0.87}
\end{equation}
where
\begin{equation}\label{eq:y}
y \equiv {n({^4}{\rm He}{^+}) \over n({\rm H}{^+})} = {{T_L ({\rm
{^4}He{^+}})\, \Delta V ({\rm {^4}He{^+}})}\over{T_L ({\rm H{^+}})\,
\Delta V ({\rm H{^+}})}},
\end{equation}
and $\nu_{\rm L}$ is the H RRL rest frequency, $T_{\rm C}$ is the
continuum peak intensity, $T_{\rm L}$ is the H (or He) RRL intensity,
and $\Delta V$ is the H (or He) RRL FWHM line width.  We use $\nu_{\rm
  L} = 9$\ghz, the average H RRL frequency.  Following
\citet{balser11}, we assume $y = 0.08$ for sources with no detected
helium emission.  For typical \hii\ region electron densities and
temperatures in our sample non-LTE effects and pressure broadening by
electron impacts should be negligible at this frequency
\citep[c.f.,][]{balser99}.  Nevertheless, these effects on RRLs may
introduce additional uncertainty into our analysis discussed in
\S{\ref{sec:structure}}.

\begin{figure}
\includegraphics[angle=0,scale=0.8]{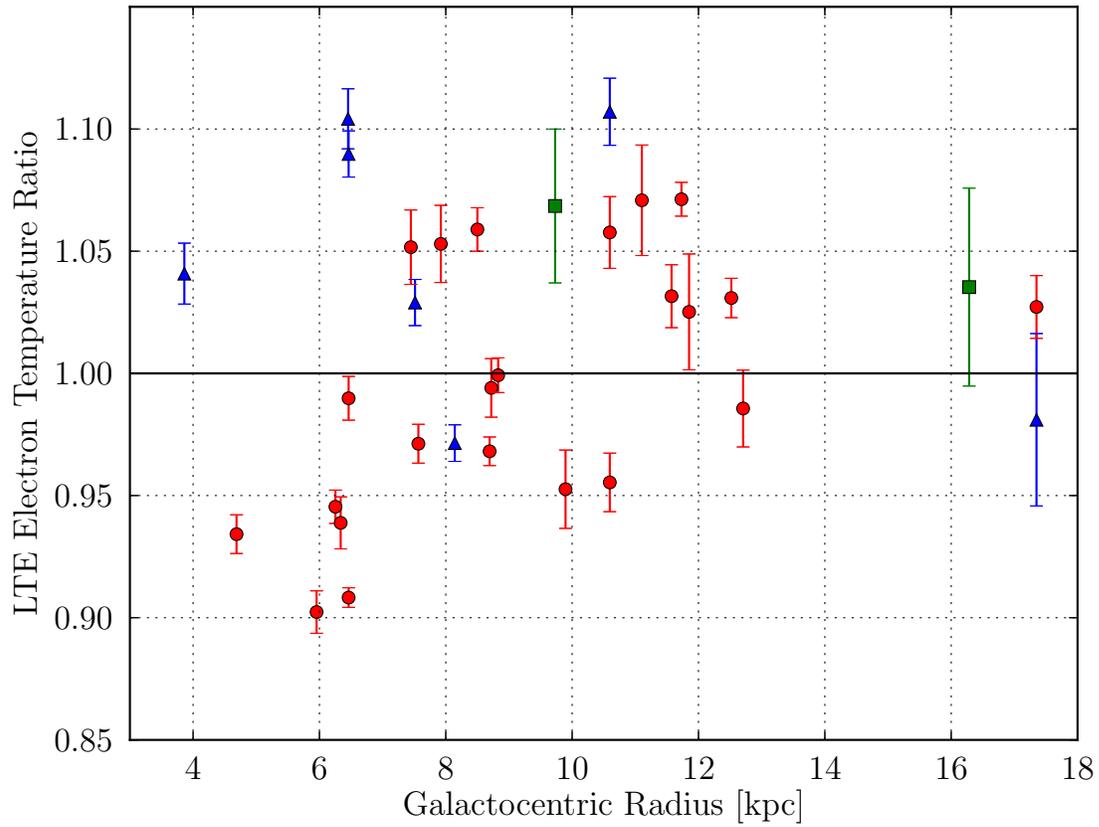} 
\caption{LTE electron temperature comparison.  Plotted is the electron
  temperature ratio as a function of Galactic radius for sources in
  common with the GBT (green squares), the 140 Foot (blue triangles),
  and with the GBT and 140 Foot samples (red circles).  Only sources
  with QF A, B, or C are included.}  
\label{fig:ratioTe}
\end{figure}

Table~\ref{tab:prop} lists values of $T_{\rm e}$ and $y$, together
with other properties (see below), for the Green Bank sample.  The RRL
and continuum parameters together with the electron temperatures are
derived using the same procedures and equations for {\bf all} nebulae
in the sample.  The uncertainties are calculated by propagating the
Gaussian fit errors through Equations~\ref{eq:te} and \ref{eq:y}.
These random errors are typically less than 5\%.  We estimate
systematic errors by comparing the results for sources observed twice
by both the same telescope and different telescopes.
Figure~\ref{fig:ratioTe} plots the LTE electron temperature ratio as a
function of the Galactic radius for these duplicate measurements.
Shown are ratios for common sources within the GBT sample, the 140
Foot sample, and between the GBT and 140 Foot samples.  We consider
two sources to be the same if their positions are within the HPBW/2.
When searching for GBT-140 Foot duplicates we use the larger 140 Foot
HPBW/2 value.  We only plot QFs between A--C. The rms of the electron
temperature ratio for these duplicate measurements is 5\%, larger than
the random errors indicated by the error bars, but comparable to our
calibration uncertainty.

\subsection{\hii\ Region Distances}\label{sec:dist}

There are three main methods used to determine \hii\ region distances:
geometrically (parallax), kinematically, and spectrophotometrically.
Geometric distances are derived by measuring the parallax of molecular
masers and associating them with \hii\ regions using position and
radial velocity.  They provide the most accurate distance measurement
and are obviously preferred.  There are 42 unique sources with
parallax distances for the Green Bank sample.  Distances determined
using spectrophotometry can have lower formal errors than kinematic
distances, but there are often large discrepancies between different
studies.  Moreover, many of our sources are optically obscured.  Here,
we only consider parallax and kinematic distances.

For consistency, we have redone the kinematic distance determination
for all sources in the Green Bank sample following the procedures of
\citet{anderson14}.  We use the \citet{brand86} rotation curve model
with two different sets of Galactic parameters: (1) a distance to the
Galactic Center of $R_{\rm o} = 8.5$\kpc\ and a Solar circular
rotation speed of $\Theta_{\rm o} = 220$\kms\ (IAU parameters;
hereafter, Kinematic-IAU); and (2) the parameters from
\citet{reid14} where $R_{\rm o} = 8.3$\kpc\ and $\Theta_{\rm o} =
240$\kms\ (hereafter, Kinematic-Reid).  \hii\ regions in the inner
Galaxy have a kinematic distance ambiguity (KDA).  A KDA resolution
(KDAR) was made using three methods: \hi\ emission absorption
(\hiea), \htwoco\ absorption, and \hi\ self-absorption (\hisa).
Following \citet{anderson14} we employ these KDAR methods with the
following priority: far distance \htwoco\ KDARs, \hiea\ KDARs,
\hisa\ KDARs, and near distance \htwoco\ KDARs.

Table~\ref{tab:prop} summarizes the \hii\ region properties of the
Green Bank sample.  Listed are the source name, the Galactic
coordinates ($\ell, b$), the distance (\rsun) and Galactocentric
coordinates ($Az$, \rgal), the helium abundance by number ($y$), the
LTE electron temperature ($T_{\rm e}$), and the telescope used for the
RRL and continuum observations.  Three sets of (\rsun, $Az$, \rgal)
are shown: Kinematic-IAU, Kinematic-Reid, and Best.  The Best column
gives values determined using parallax measurements when available,
otherwise Kinematic-IAU parameters are used.  Distances derived from
parallax measurements are taken from the compilation in
\citet{reid14}.

\begin{figure}
\includegraphics[angle=0,scale=0.8]{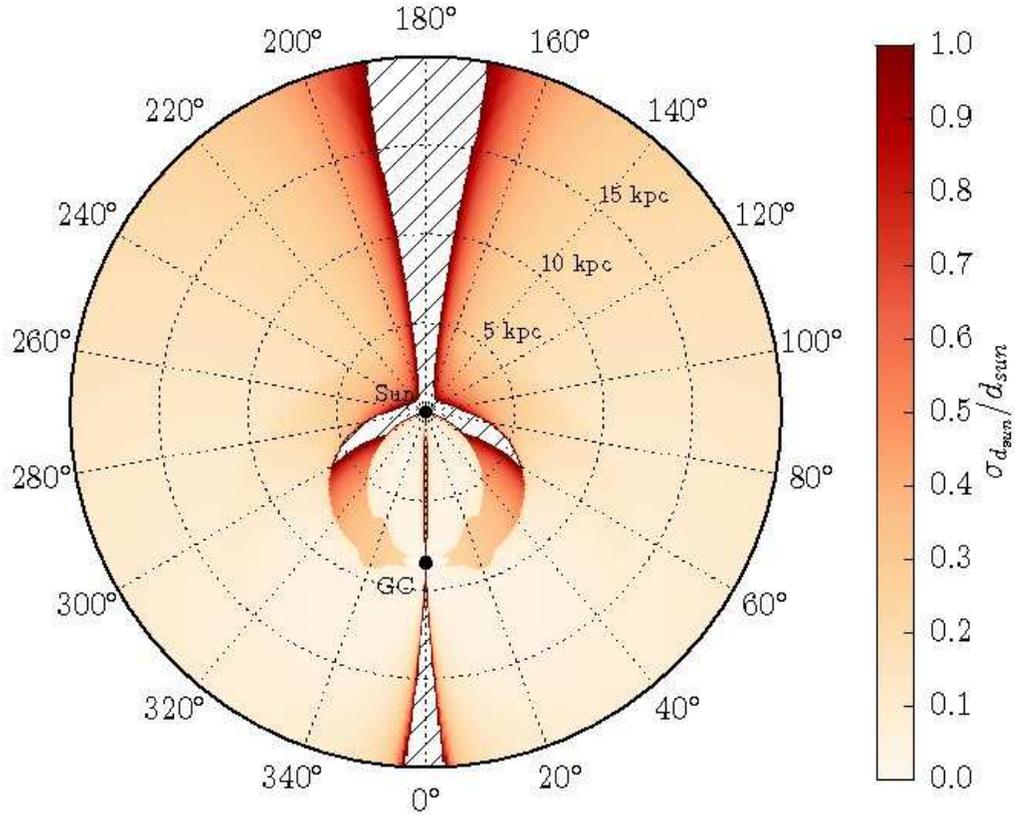} 
\caption{Face-on Galactic map of distance uncertainties from Wenger et
  al. (2015, in preparation) based on the analysis of
  \citet{anderson14}.  The distance uncertainties were calculated by
  exploring different rotation curves, streaming motions, and a change
  to the Solar circular rotation speed. Shown are the fractional
  distance uncertainties.  The hatched regions have fractional
  uncertainties larger than unity.}
\label{fig:distError}
\end{figure}

Kinematic distance uncertainties are estimated by exploring different
rotation curves, streaming motions, and a change to the Solar circular
rotation speed \citep{anderson12}.  Figure~\ref{fig:distError} is a
face-on Galactic plot of the fractional distance error from Wenger et
al. (2015, in preparation).  The largest errors occur when the radial
velocity is small: toward the Galactic Center or anti-center and near
the Sun.  Distances are also not very well determined near the tangent
point where the radial velocity is a maximum.  This zone occupies a
relatively small area.

\begin{figure}
\includegraphics[angle=0,scale=0.40]{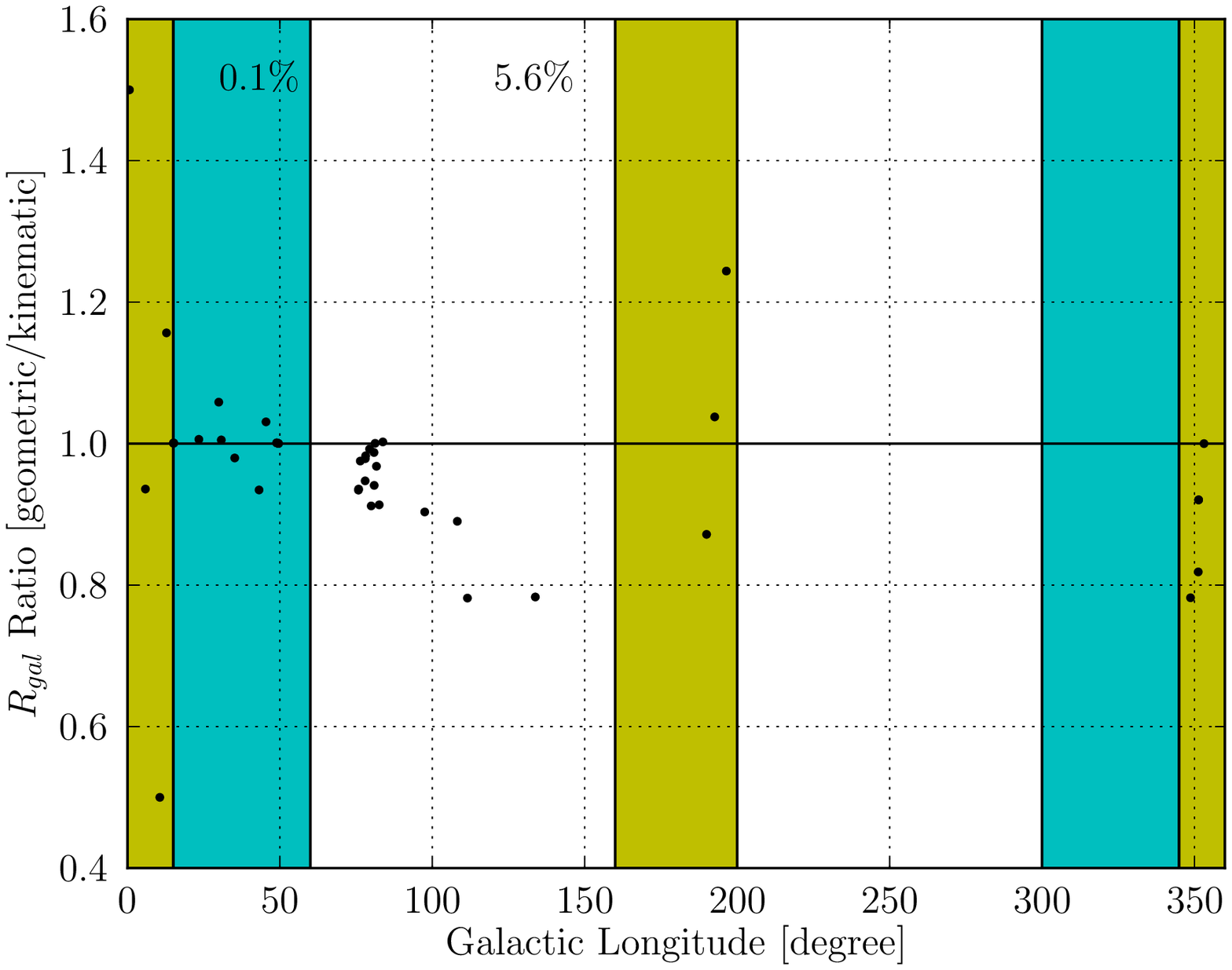} 
\includegraphics[angle=0,scale=0.40]{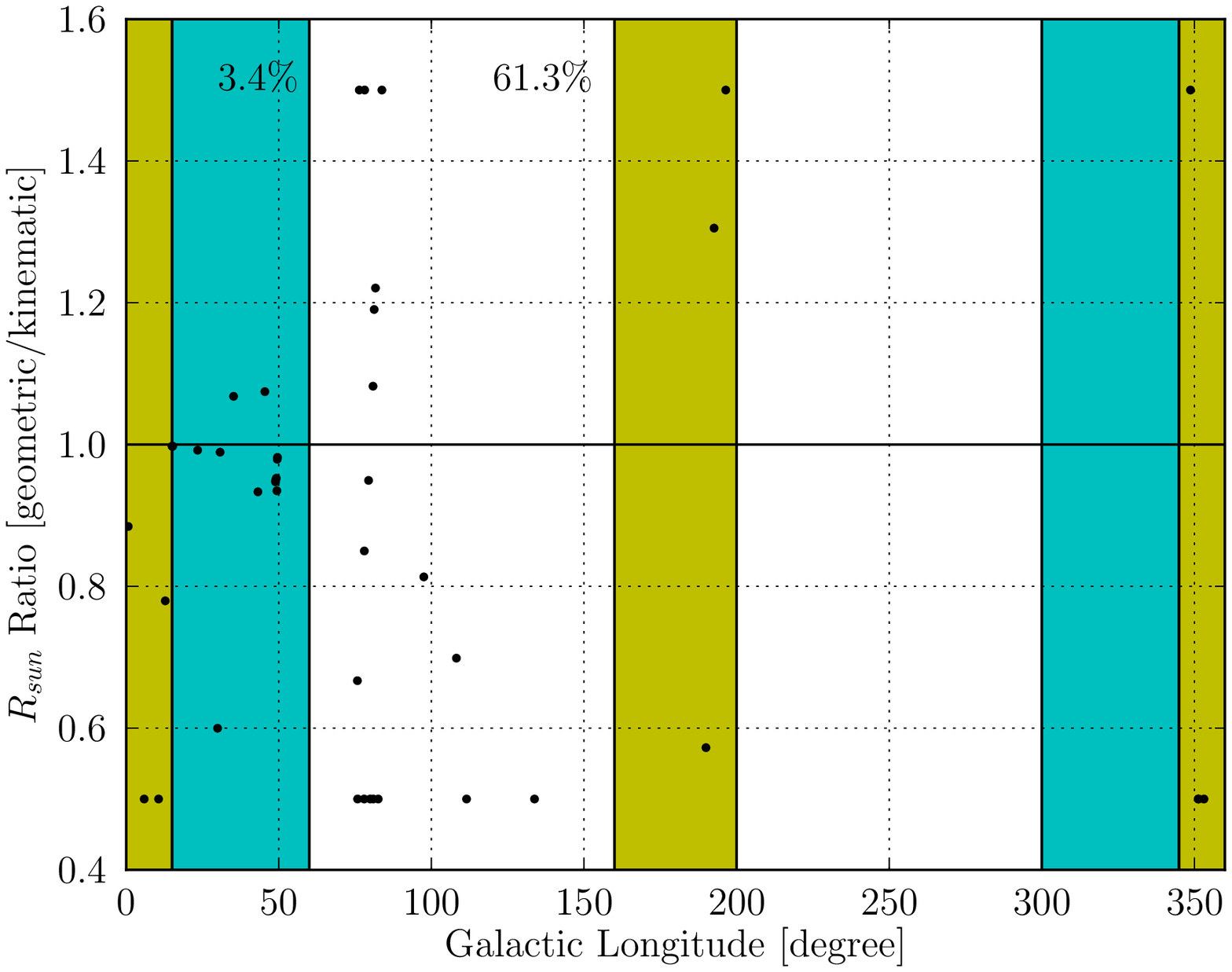} 
\caption{Comparison of parallax distances with kinematic distances as
  a function of Galactic longitude.  The Kinematic-IAU values are used
  for consistency.  Left: Galactic distance ratio (\rgal).  Right:
  Solar distance ratio (\rsun).  The green regions are within
  15\degree\ of the Galactic Center or 20\degree\ of the Galactic
  anti-center.  Kinematic distances are not very reliable over these
  longitudes since the radial velocity from Galactic rotation is near
  zero for all distances.  The cyan regions are between
  $15-60$\degree\ in the first Galactic quadrant or between
  $300-345$\degree\ in the fourth Galactic quadrant.  In this
  longitude zone we derive a median percent difference between the
  kinematic and geometric methods of 0.1\% and 3.4\% for \rgal\ and
  \rsun, respectively.  The remaining longitude range, shown in white,
  has larger kinematic distance uncertainties with a median percent
  difference of 5.6\% and 61.3\% for \rgal\ and \rsun. respectively.
  These median percent difference numbers are shown at the top of each
  plot.  Ratio values are truncated to be between 0.5 and 1.5 for
  clarity.}
\label{fig:distCompare}
\end{figure}

There are regions in the first and fourth quadrants, a zone that
encompasses a large fraction of the Galactic disk, that have accurate
(within $\sim\ 15$\%) kinematic distances.
Figure~\ref{fig:distCompare} plots \rgal\ and \rsun\ ratios as a
function of Galactic longitude for the Green Bank sample using
geometric and kinematic methods.  The Kinematic-IAU distances are used
here for consistency of these comparisons.  Values of \rgal\ derived
using a kinematic model are typically within 5\% of geometrically
determined values, whereas there are significant deviations in \rsun\
distances \citep[Figure~\ref{fig:distCompare}; also see][]{reid14}.
The trends are similar when using the Kinematic-Reid distances.  From
Figures~\ref{fig:distError} and \ref{fig:distCompare} it is clear that
the largest kinematic distance uncertainties exist toward the Galactic
Center and anti-center.  We therefore remove any sources from our
analysis that have longitudes with 15\degree\ of the Galactic Center
or within 20\degree\ of the Galactic anti-center.  These Galactic
zones are indicated in Figure~\ref{fig:distCompare} by the green
shaded region.  The longitude zone with the lowest distance
uncertainties, shown by the cyan shaded region in
Figure~\ref{fig:distCompare}, are between 15\degree$-$60\degree\ in
the first quadrant and 300\degree$-$345\degree\ in the fourth
quadrant.  We measure an rms of 3.4\% between the kinematic and
parallax distances within this zone in the first quadrant.  For the
longitude zone that remains, encompassing most of the second and third
quadrants, we measure an rms of 61.3\%.

\subsection{Electron Temperature Structure}\label{sec:structure}

\hii\ region electron temperatures are primarily set by the effective
temperature of the ionizing star, the electron density of the
surrounding medium, dust-particle interactions, and heavy elements
(e.g., oxygen and nitrogen).  \hii\ region models indicate that the
metallicity is the dominant factor, where heavy elements within the
ionized gas increase the cooling by the emission of collisionally
excited lines \citep{rubin85}.  \citet{churchwell75} were the first to
discover a positive radial electron temperature gradient in the Milky
Way disk using RRLs.  They interpreted this gradient as due to a
negative radial metallicity gradient, where higher metal abundances at
smaller radii produce lower electron temperatures.  More recent RRL
observations with the GBT revealed azimuthal electron temperature
structure that may indicate the Galaxy is not well mixed at a given
radius \citep{balser11}.  Here we expand the \hii\ region sample using
the HRDS to explore electron temperature structure over a larger area
of the Galaxy.

\begin{figure}
\includegraphics[angle=0,scale=0.43]{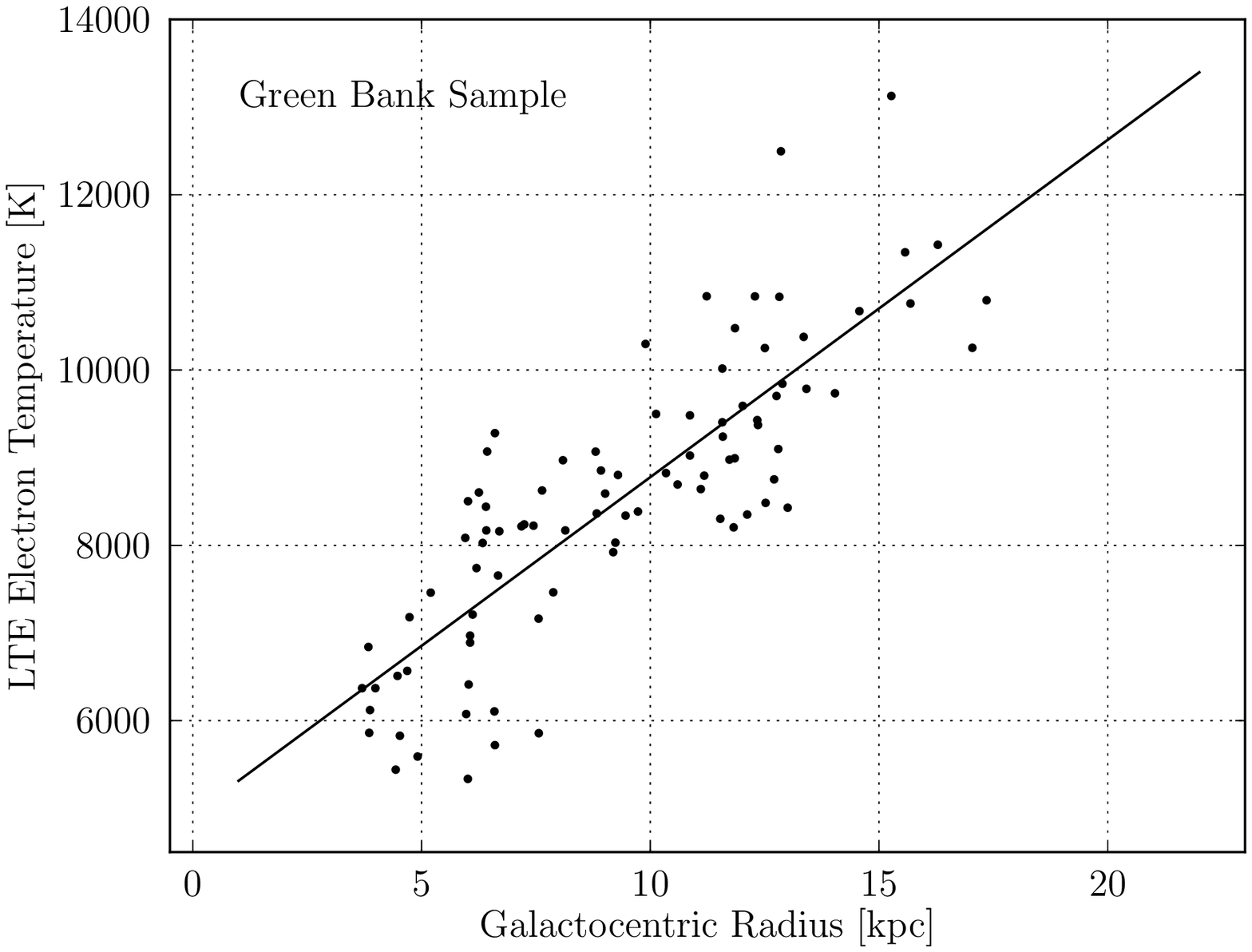} 
\includegraphics[angle=0,scale=0.43]{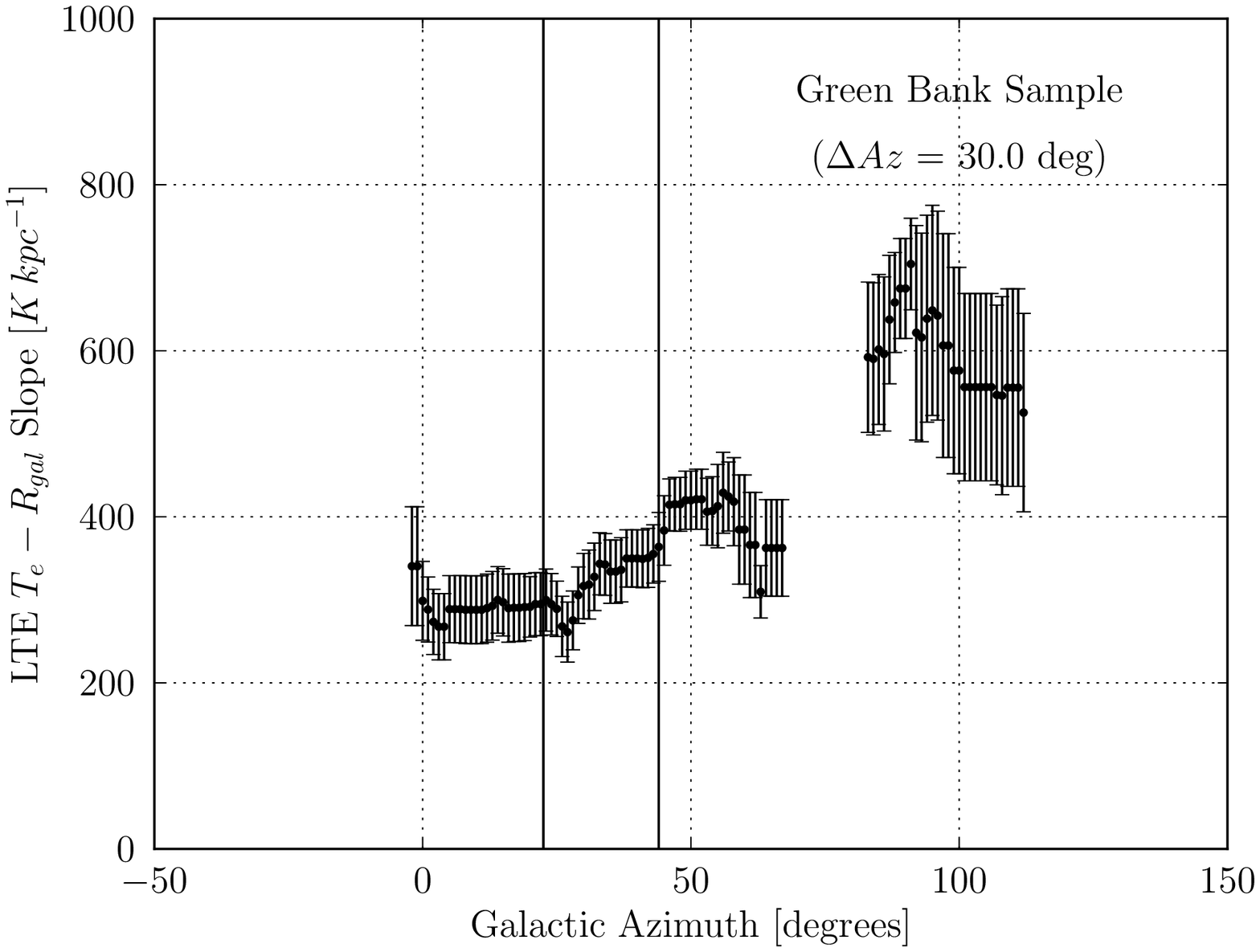} 
\includegraphics[angle=0,scale=0.43]{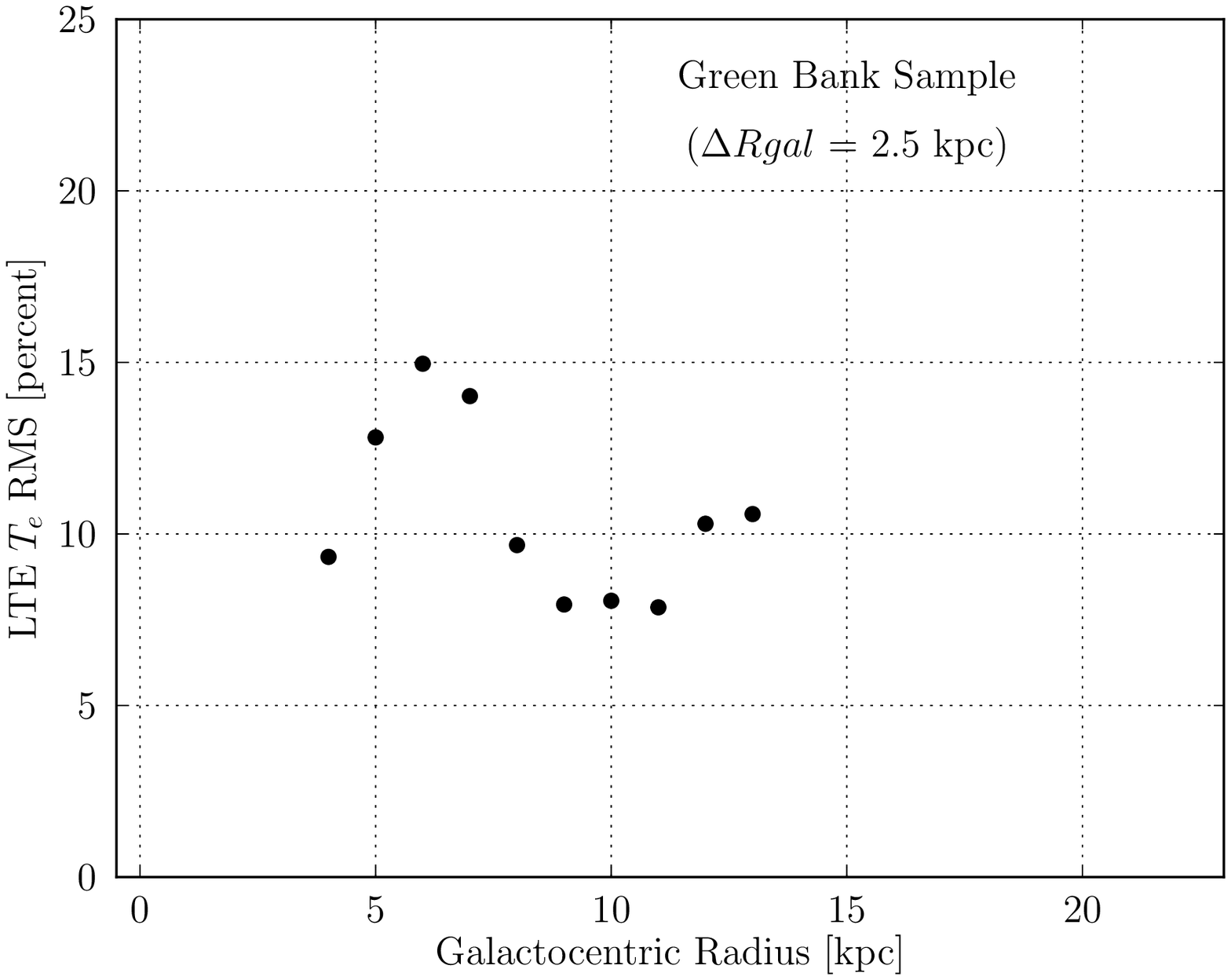} 
\includegraphics[angle=0,scale=0.50]{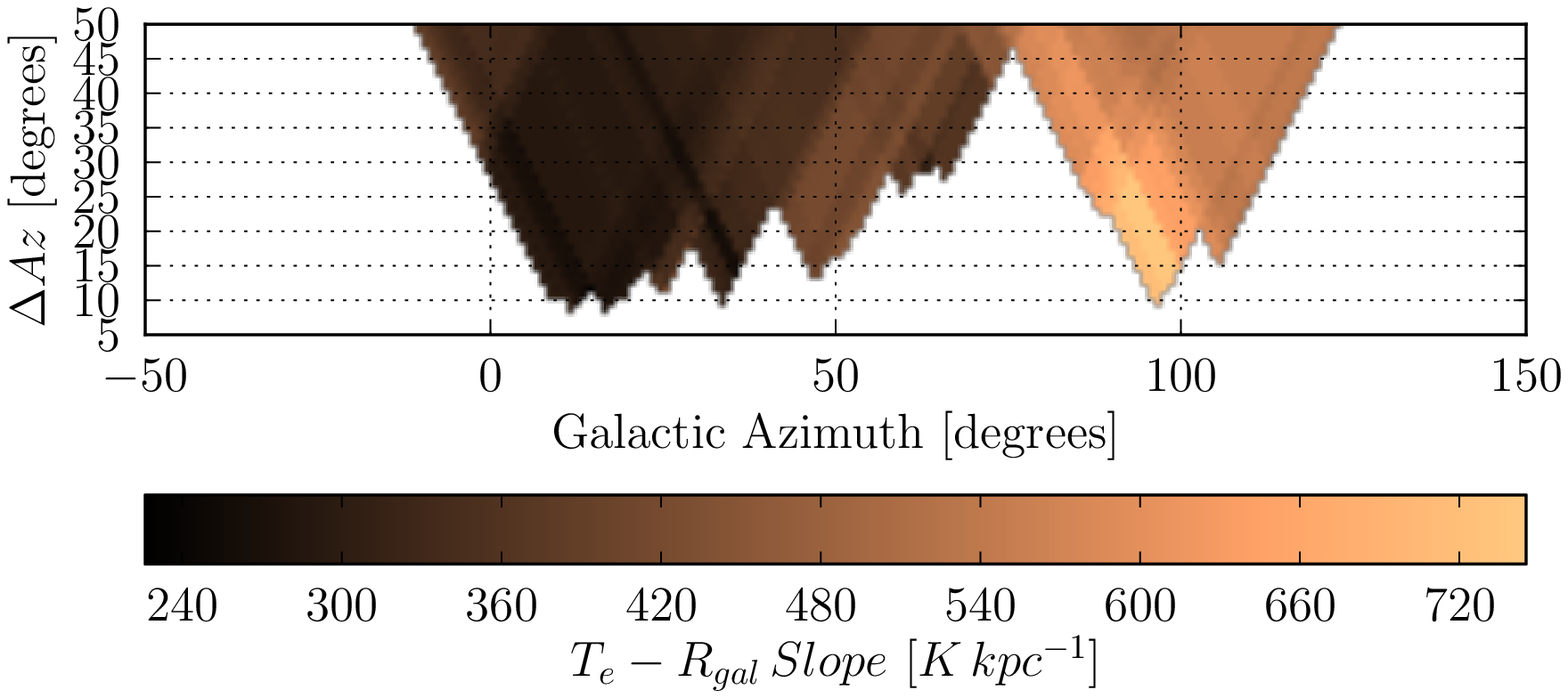} 
\caption{Electron temperature structure using Kinematic-IAU distances
  for the Green Bank sample. Top left: LTE electron temperature as a
  function of Galactic radius.  The points are data and the solid line
  is a fit to these data with $T_{\rm e} = (4928 \pm\ 277) + (385 \pm\
  29)\,R_{\rm gal}$.  Errors bars are not included for clarity.
  Bottom left: LTE electron temperature rms within Galactic radius
  bins of $\Delta R_{\rm gal} = 2.5$\kpc.  The electron temperature
  uncertainty is $\sim 5$\%.  Only bins that include 10 or more
  sources spanning 40\degree\ or more in azimuth are plotted.  This
  requirement is not met for $R_{\rm gal} > 14$\kpc\ and therefore no
  data points are plotted for these radii.  Top right: LTE \te--\rgal\
  slope as a function of azimuth.  Azimuth bin sizes of $\Delta{Az} =
  30$\degree\ are centered at the indicated azimuth.  The error bars
  are derived using the SLOPES algorithm with jackknife resampling.
  Only bins that include 10 or more sources spanning an \rgal\ range
  larger than 10\kpc\ are shown.  These requirements are not met for
  $Az \sim\ 75$\degree\ which explains the gap in data points near
  this azimuth.  The vertical solid lines mark the orientation of the
  bar ($Az \sim 25$\degree) and long bar ($Az \sim 45$\degree)
  \citep{benjamin08}.  Bottom right: LTE \te--\rgal\ slope color map
  as a function of bin size ($\Delta{Az}$) and azimuth ($Az$).  Here
  we explore the effects of using different bin sizes.}
\label{fig:KinBrand_4panel}
\end{figure}

\begin{figure}
\includegraphics[angle=0,scale=0.43]{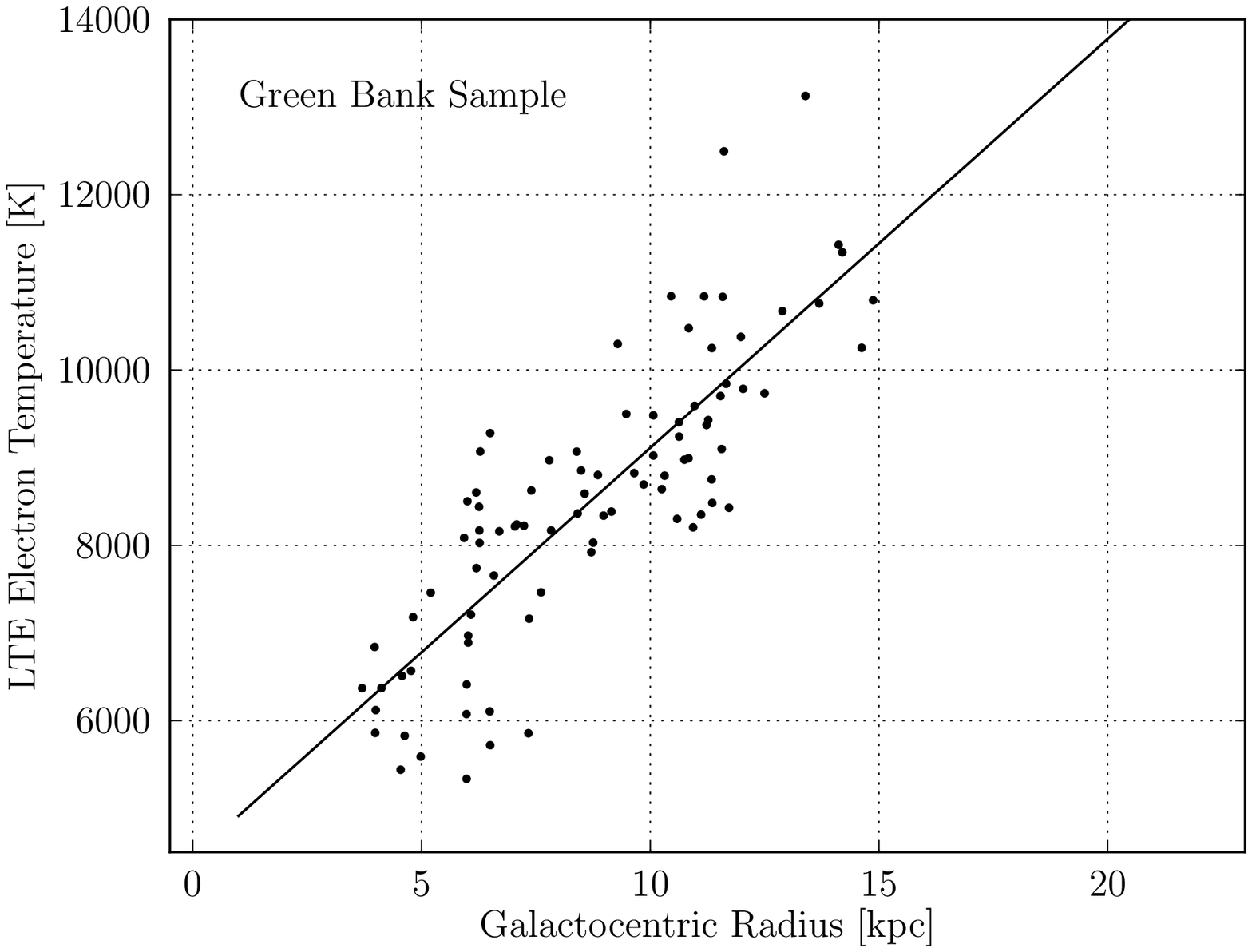} 
\includegraphics[angle=0,scale=0.43]{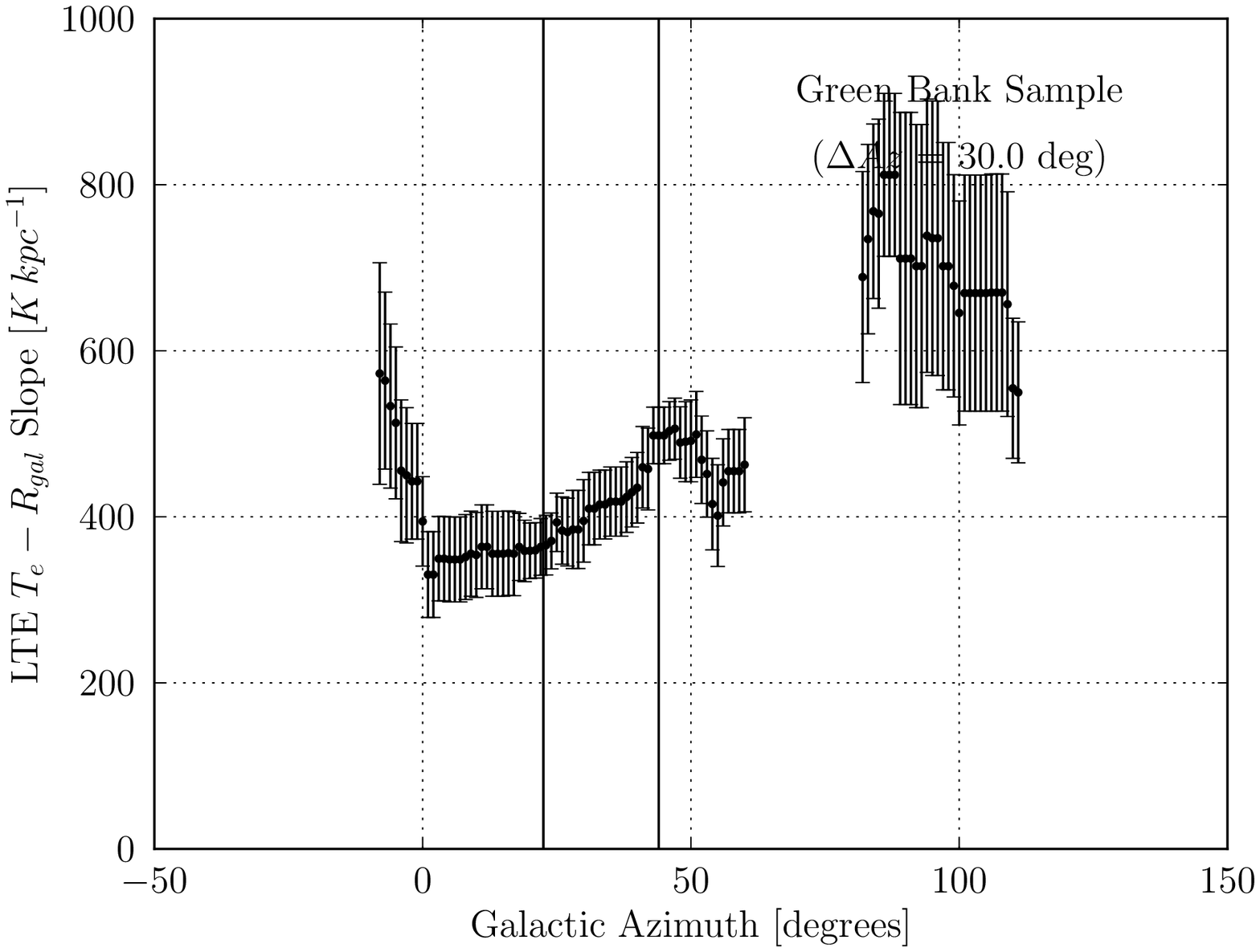} 
\includegraphics[angle=0,scale=0.43]{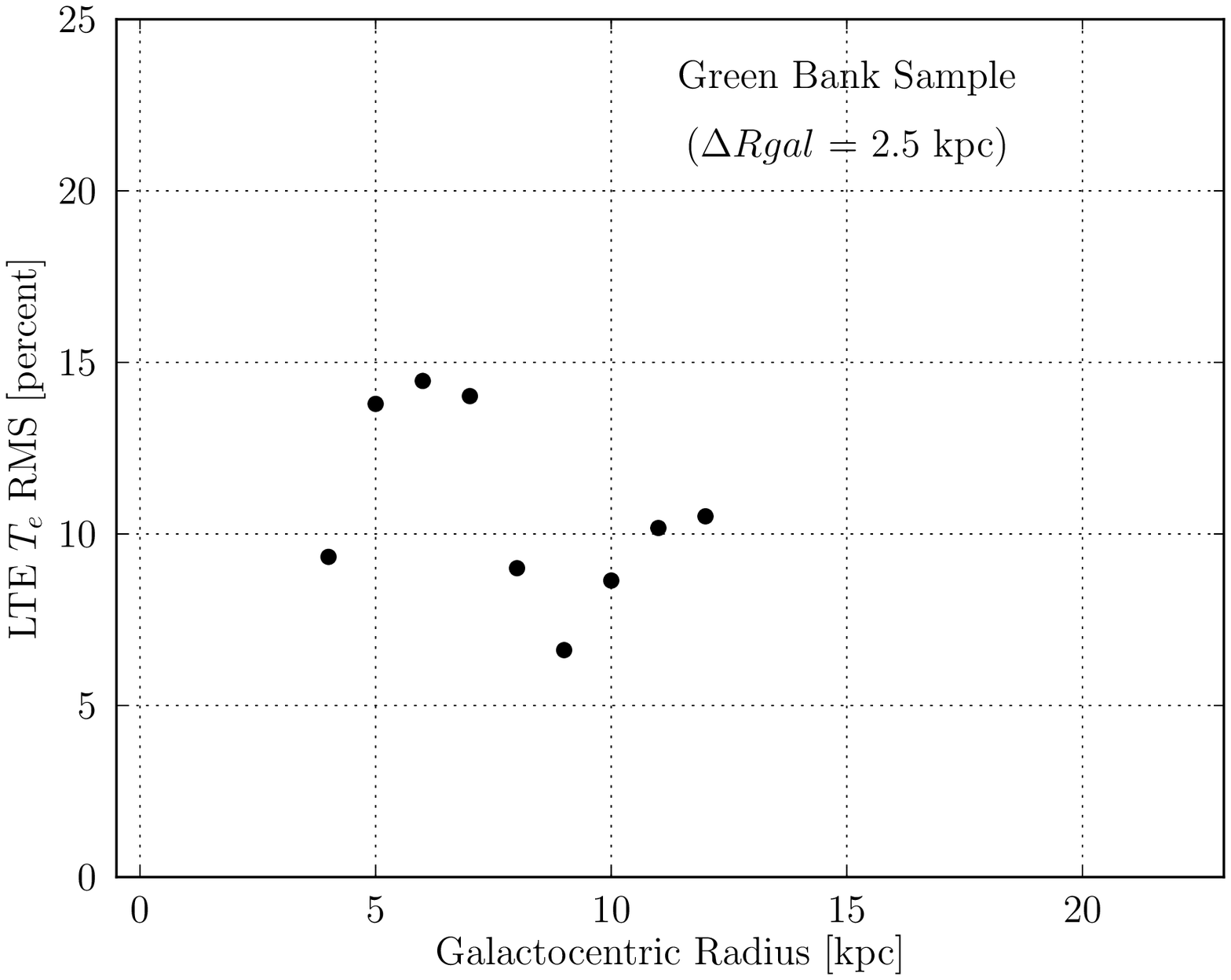} 
\includegraphics[angle=0,scale=0.50]{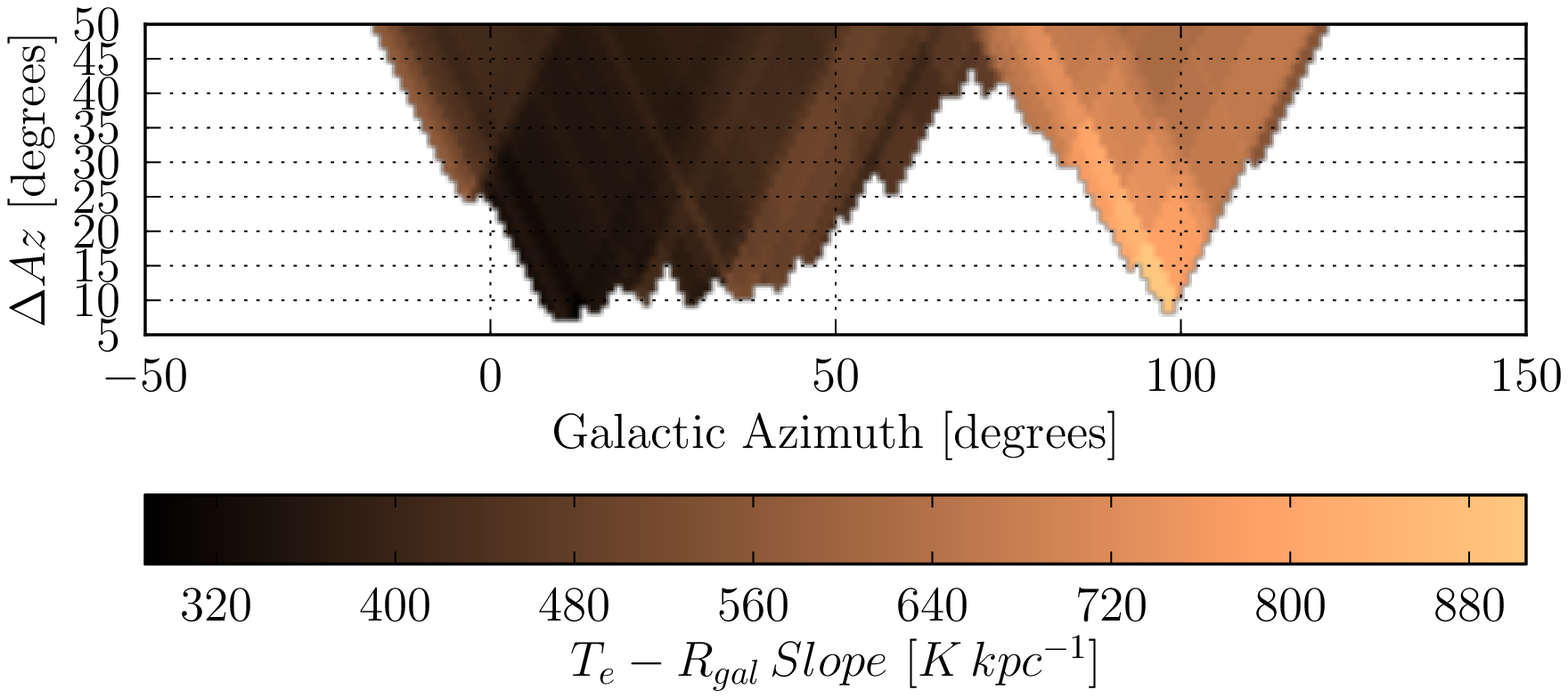} 
\caption{Electron temperature structure using Kinematic-Reid distances
  for the Green Bank sample.  See Figure~\ref{fig:KinBrand_4panel} for
  details.  The fit to the data in the top left plot is $T_{\rm e} =
  (4446 \pm\ 301) + (467 \pm\ 34)\,R_{\rm gal}$.}
\label{fig:KinReid_4panel}
\end{figure}

\begin{figure}
\includegraphics[angle=0,scale=0.43]{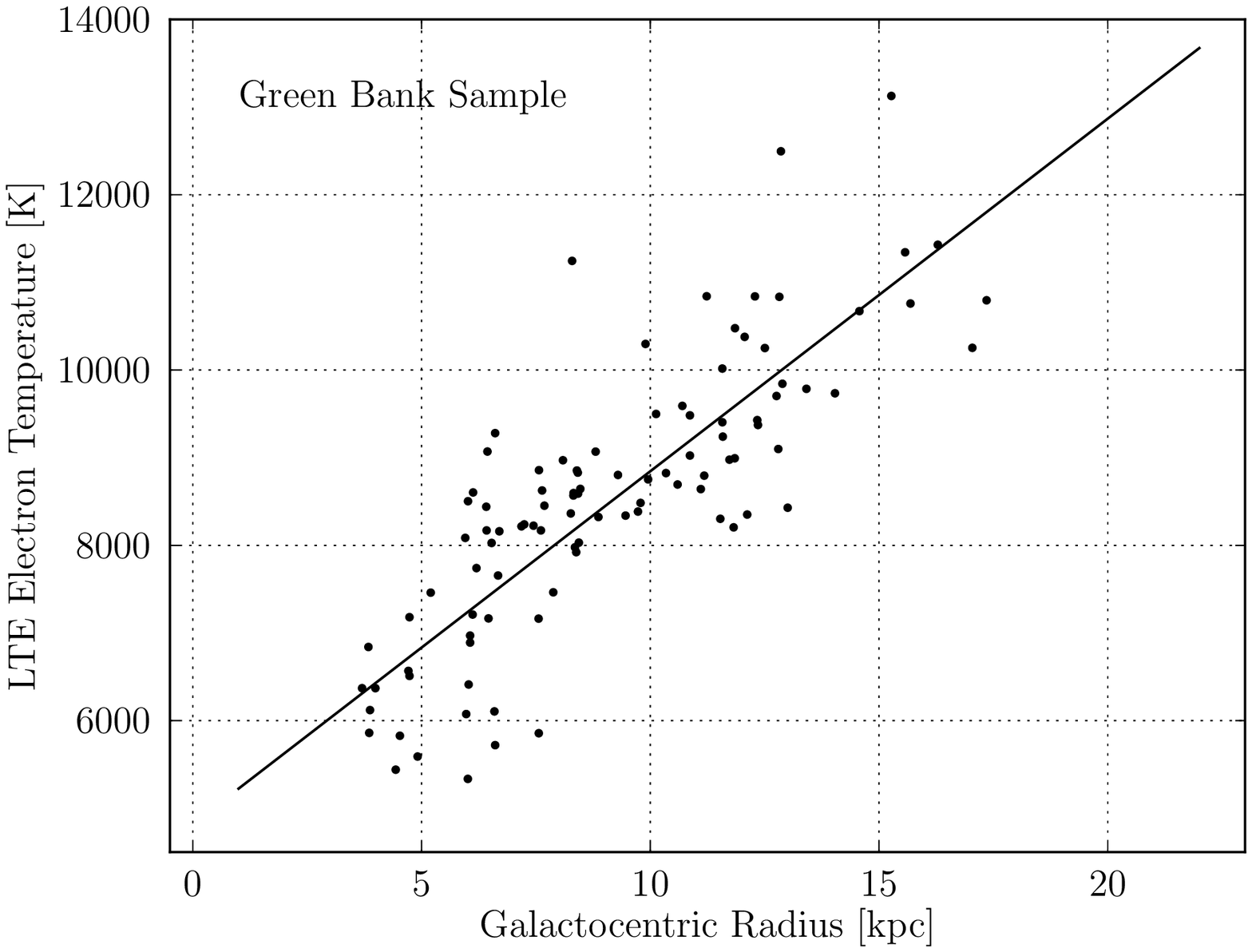} 
\includegraphics[angle=0,scale=0.43]{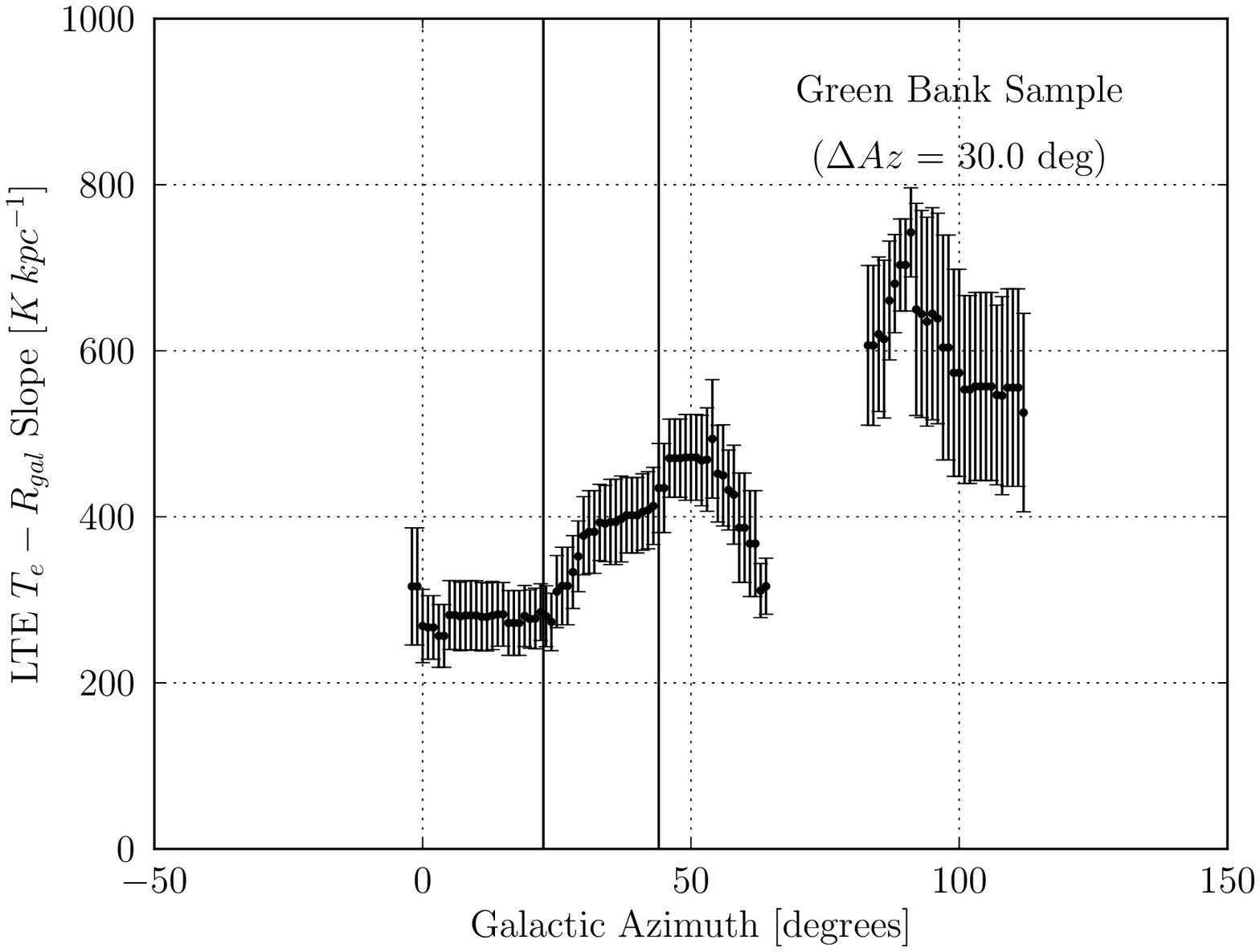} 
\includegraphics[angle=0,scale=0.43]{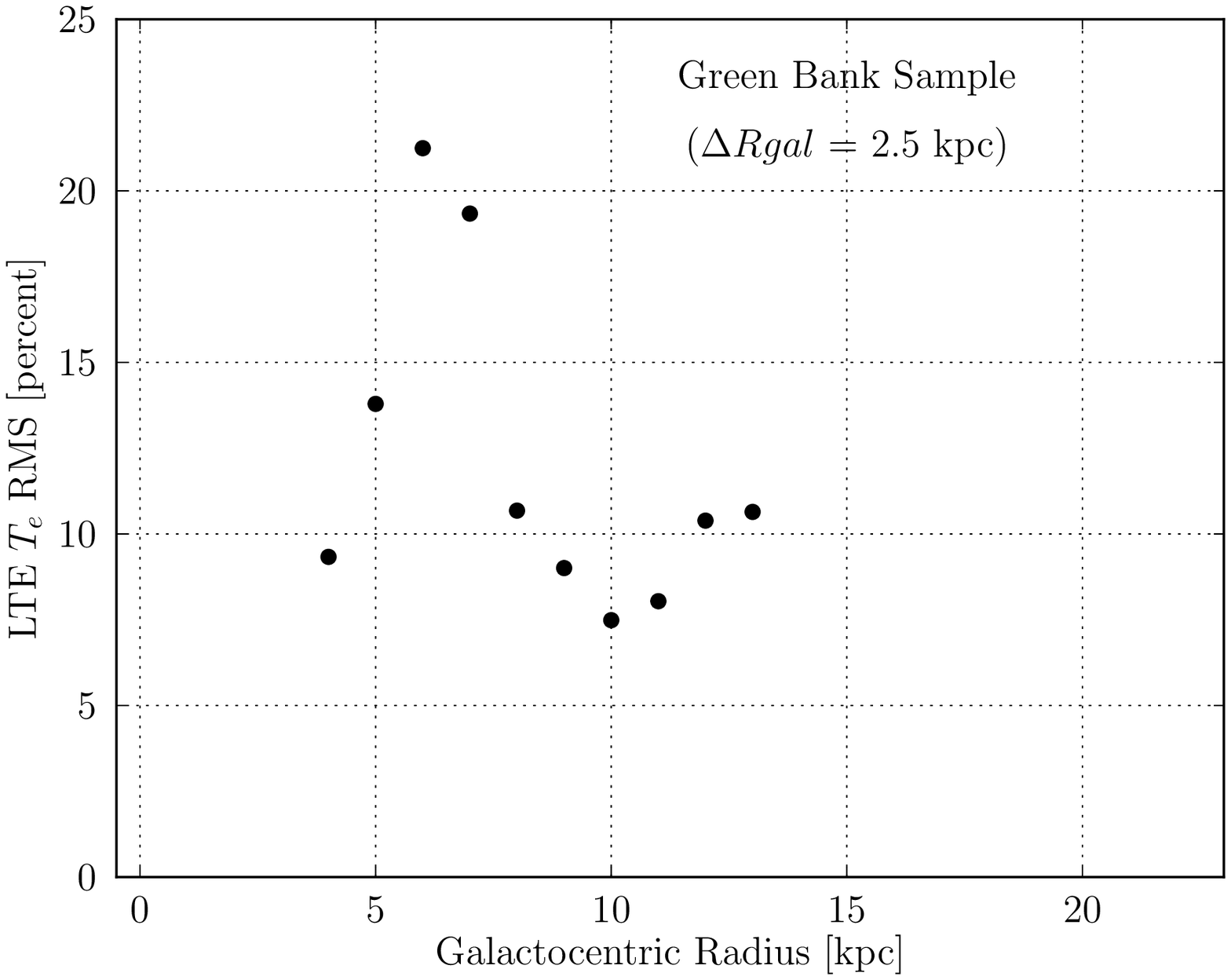} 
\includegraphics[angle=0,scale=0.50]{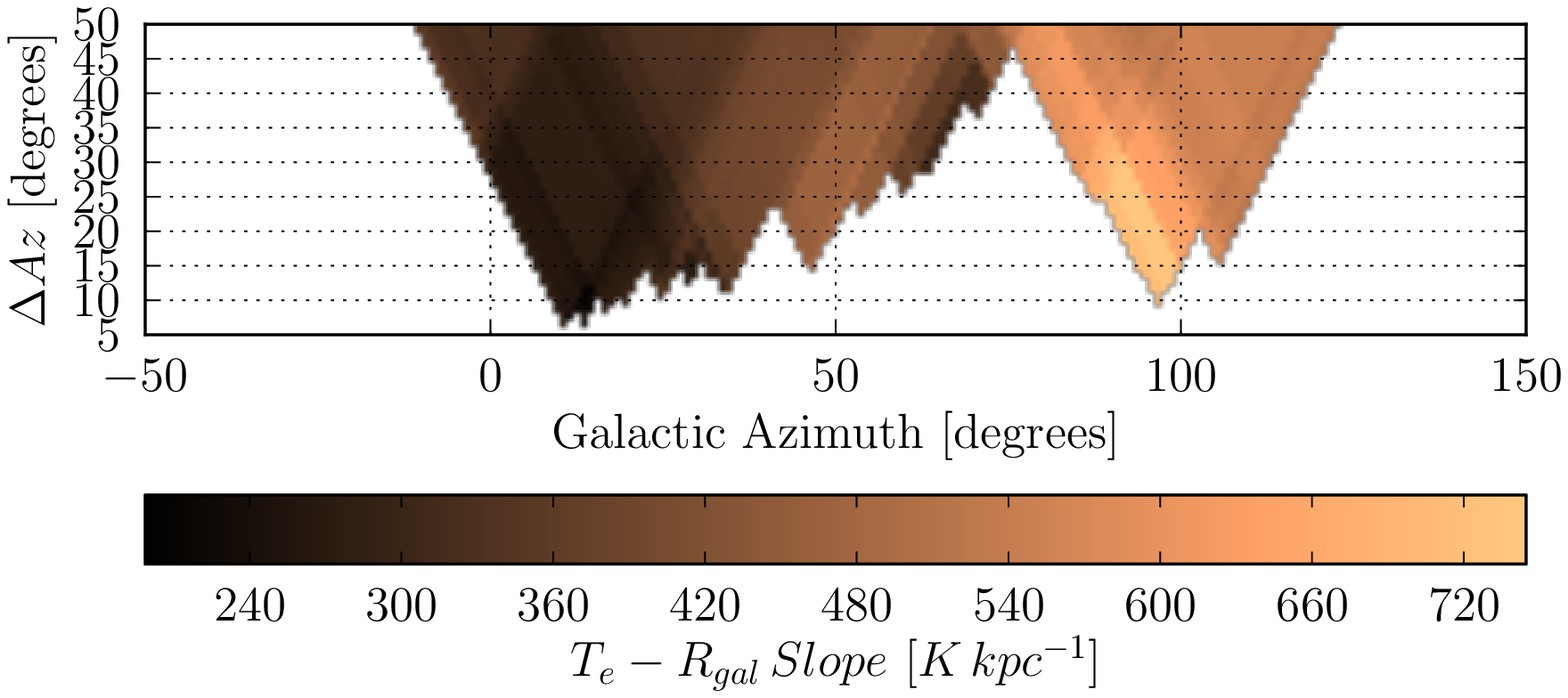} 
\caption{Electron temperature structure using Best distances for the
  Green Bank sample.  See Figure~\ref{fig:KinBrand_4panel} for
  details.  The fit to the data in the top left plot is $T_{\rm e} =
  (4821 \pm\ 328) + (402 \pm\ 33)\,R_{\rm gal}$.}
\label{fig:Best_4panel}
\end{figure}

The electron temperature structure in the Galactic disk is derived for
the three rotation curves: Kinematic-IAU
(Figure~\ref{fig:KinBrand_4panel}), Kinematic-Reid
(Figure~\ref{fig:KinReid_4panel}), and Best distances
(Figure~\ref{fig:Best_4panel}).  For this analysis we only include
sources with distance errors less than 50\%, QFs between A--C, and
directions that exclude zones within 15\degree\ of the Galactic Center
or within 20\degree\ of the Galactic anti-center.  The gradient in
electron temperature as a function of Galactic radius is defined by
$T_{\rm e} = a + b\,R_{\rm gal}$, where $b$ is the ``LTE \te--\rgal\
slope''.  Following \citet{balser11} we use the algorithm
SLOPES\footnote{See
  http://www.astro.psu.edu/users/edf/research/stat.html.} to make
ordinary least-squares fits \citep{isobe90, feigelson92}.  In all
cases we use jackknife resampling to derive more accurate
uncertainties.

Each figure is comprised of four plots.  The electron temperature as a
function of Galactic radius is shown in the top left.  The solid line
is a fit to the data.  The typical uncertainty for both \te\ and
\rgal\ is less than 5\% (see Table~\ref{tab:prop}).  In the bottom
left panel the electron temperature rms is plotted as a function of
Galactic radius.  The rms is calculated within $\Delta R_{\rm gal} =
2.5$\kpc\ bins that include 10 or more sources spanning 40\degree\ or
more in azimuth.  N.B., this requirement is not met for $R_{\rm gal} >
14$\kpc\ and therefore no data points are plotted for these radii.
The rms scatter at specific \rgal\ values is larger than the typical
\te\ uncertainty of 5\%, suggesting electron temperature azimuthal
structure is present.  The LTE \te--\rgal\ slope as a function of
azimuth is shown in the top right.  Here we derive the slope within
$\Delta{Az} = 30$\degree\ bins that include 10 or more sources
spanning an \rgal\ range larger than 10\kpc.  The error bars are
derived from SLOPES using jackknife resampling.  The vertical solid
lines mark the direction of the bar ($Az \sim 25$\degree) and long bar
($Az \sim 45$\degree) \citep{benjamin08}.  The bottom right panel is a
color map of the LTE \te--\rgal\ slope as a function of both the
azimuth bin size and azimuth.  Here we explore different azimuth bin
sizes to check if the bin size $\Delta{Az} = 30$\degree\ is special in
any way.  Large bin sizes act to smooth variations in the LTE
\te--\rgal\ slope but provide more points in each bin and therefore
better statistics.  For the Kinematic-Reid analysis
(Figure~\ref{fig:KinReid_4panel}) we reduced the \rgal\ range
constraint from 10\kpc\ to 7\kpc\ since the Galactic parameters
adopted shrink the size of the Galactic disk.

Several trends are clear from these figures regardless of the
distances used.  There is a positive electron temperature radial
gradient.  This is well known and has been correlated with a radial
metallicity gradient in the Galactic disk \citep[see][and
below]{shaver83}.  There is evidence for electron temperature
azimuthal structure.  Variations in \te\ measured by the rms at a
given Galactic radius are larger than the uncertainty.  This is most
significant near \rgal\ = 6\kpc.  Azimuthal electron temperature
structure is also revealed by variations in the LTE \te--\rgal\ slope
as a function of azimuth.  The slope gradually increases from Galactic
azimuth 0\degree\ to 60\degree, but is almost a factor of two higher
near $Az \sim 100$\degree.

\citet{balser11} derived LTE \te--\rgal\ slopes of $392 \pm\ 84$\Kkpc,
$228 \pm\ 26$\Kkpc, and $404 \pm\ 42$\Kkpc\ between azimuth zones of
$330$\degree$-360$\degree, $0$\degree$-30$\degree, and
$30$\degree$-60$\degree, respectively.  Here we have discarded many
\hii\ regions between $Az = 330$\degree$-360$\degree\ because of large
distance uncertainties and therefore do not plot these.  For
Figure~\ref{fig:KinBrand_4panel}, we derive slopes of $297 \pm\
41$\Kkpc and $384 \pm\ 42$\Kkpc\ for azimuth zones
$0$\degree$-30$\degree\ and $30$\degree$-60$\degree, respectively.
For the analysis here, the trend is the same but the amplitude of
these \te\ variations is less.  The new HRDS sources populate
$90$\degree$-130$\degree\ where we calculate a slope of $551 \pm\
92$\Kkpc, significantly larger than those at lower azimuths.

There are some differences between the three analyses explored here.
The Kinematic-Reid distances shrink the overall size of the Galactic
disk and this increases the LTE \te--\rgal\ slopes by about 20\%.  The
LTE \te\ rms peaks near $R_{\rm gal} = 6$\kpc\ for all analyses, but
the value is 30\% higher for the Best distances compared with
Kinematic-IAU or Kinematic-Reid.  We do not know whether this is real
or due to some yet to be identified source of systematic uncertainty.

\begin{figure}
\includegraphics[angle=0,scale=0.40]{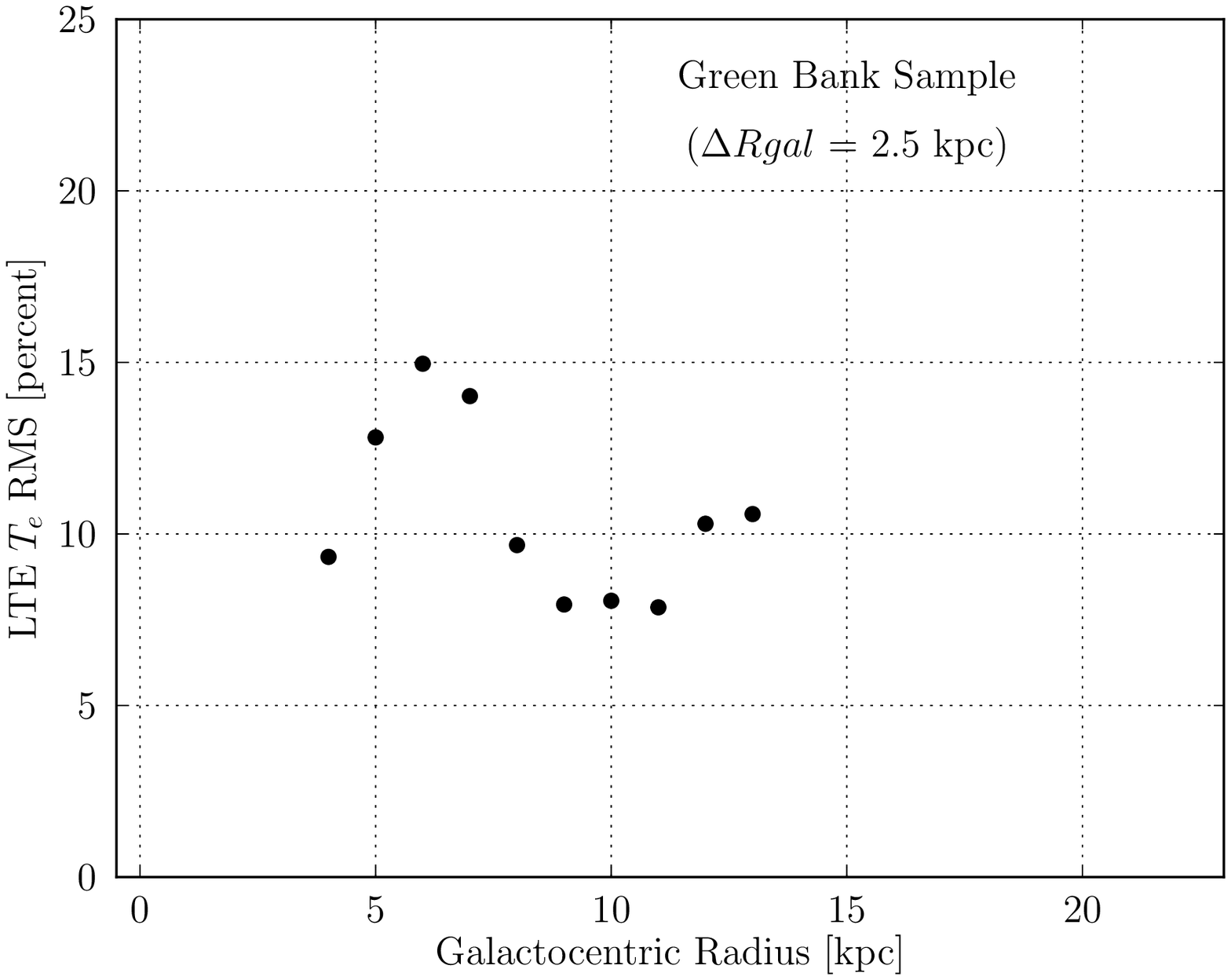} 
\includegraphics[angle=0,scale=0.40]{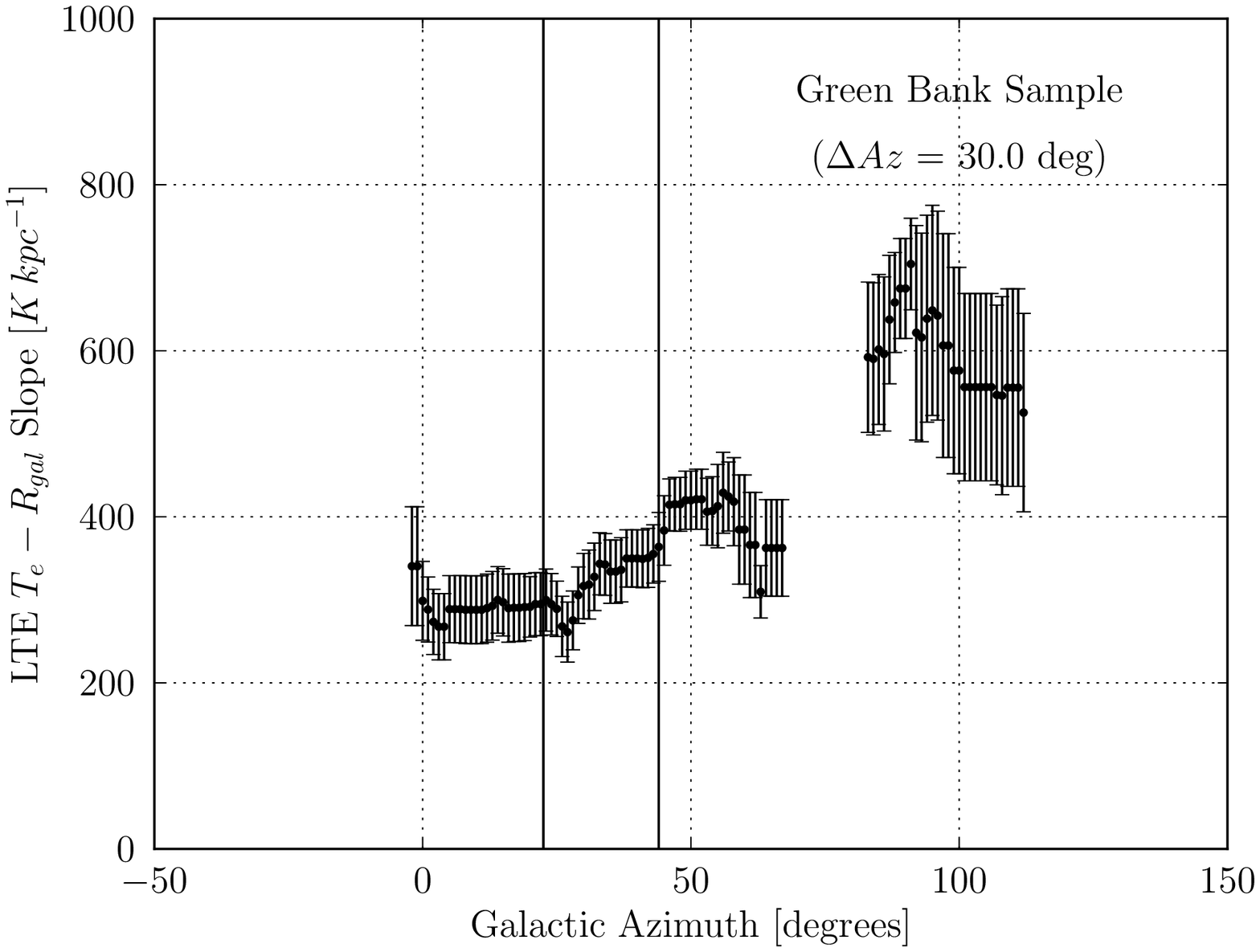} 
\includegraphics[angle=0,scale=0.40]{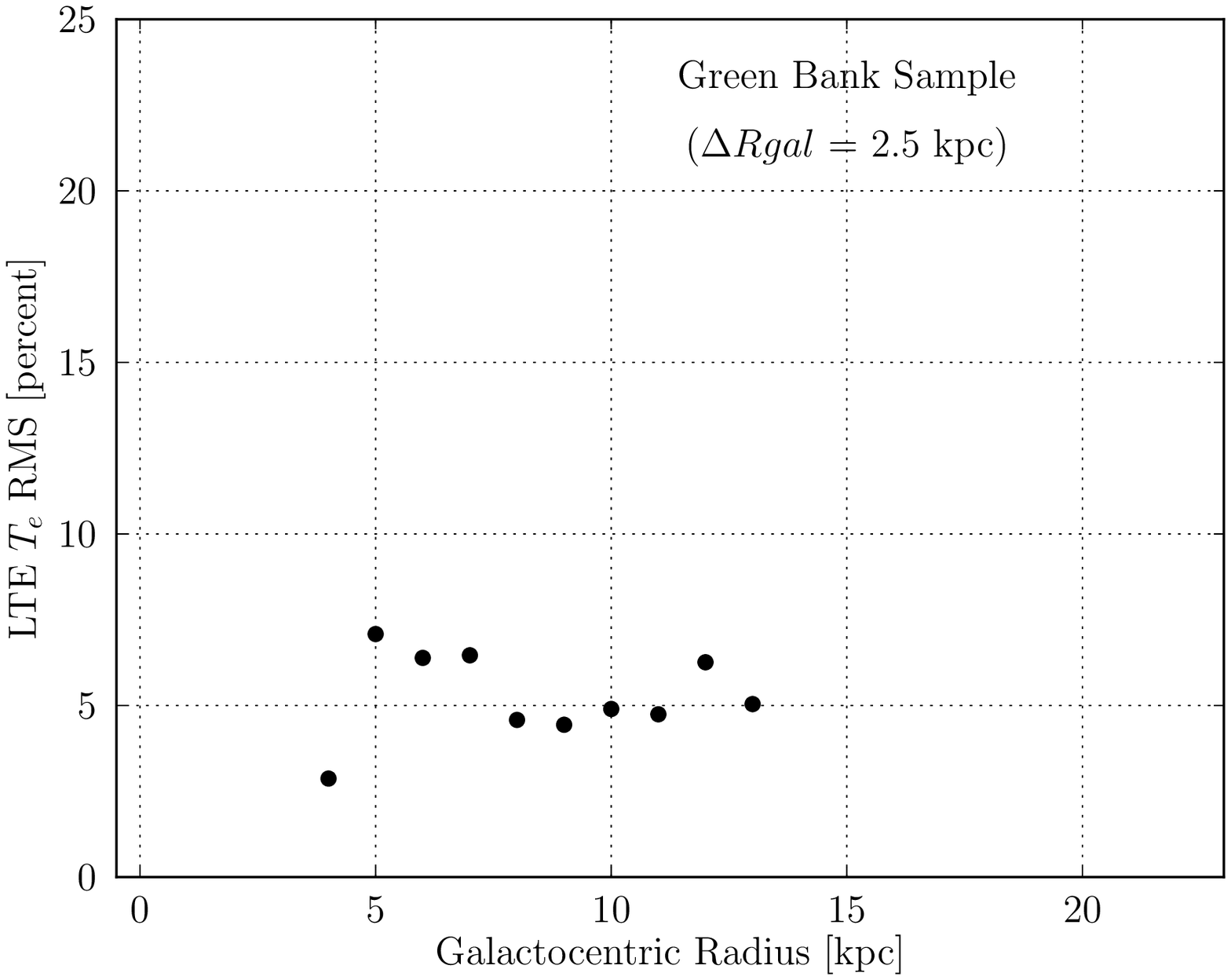} 
\includegraphics[angle=0,scale=0.40]{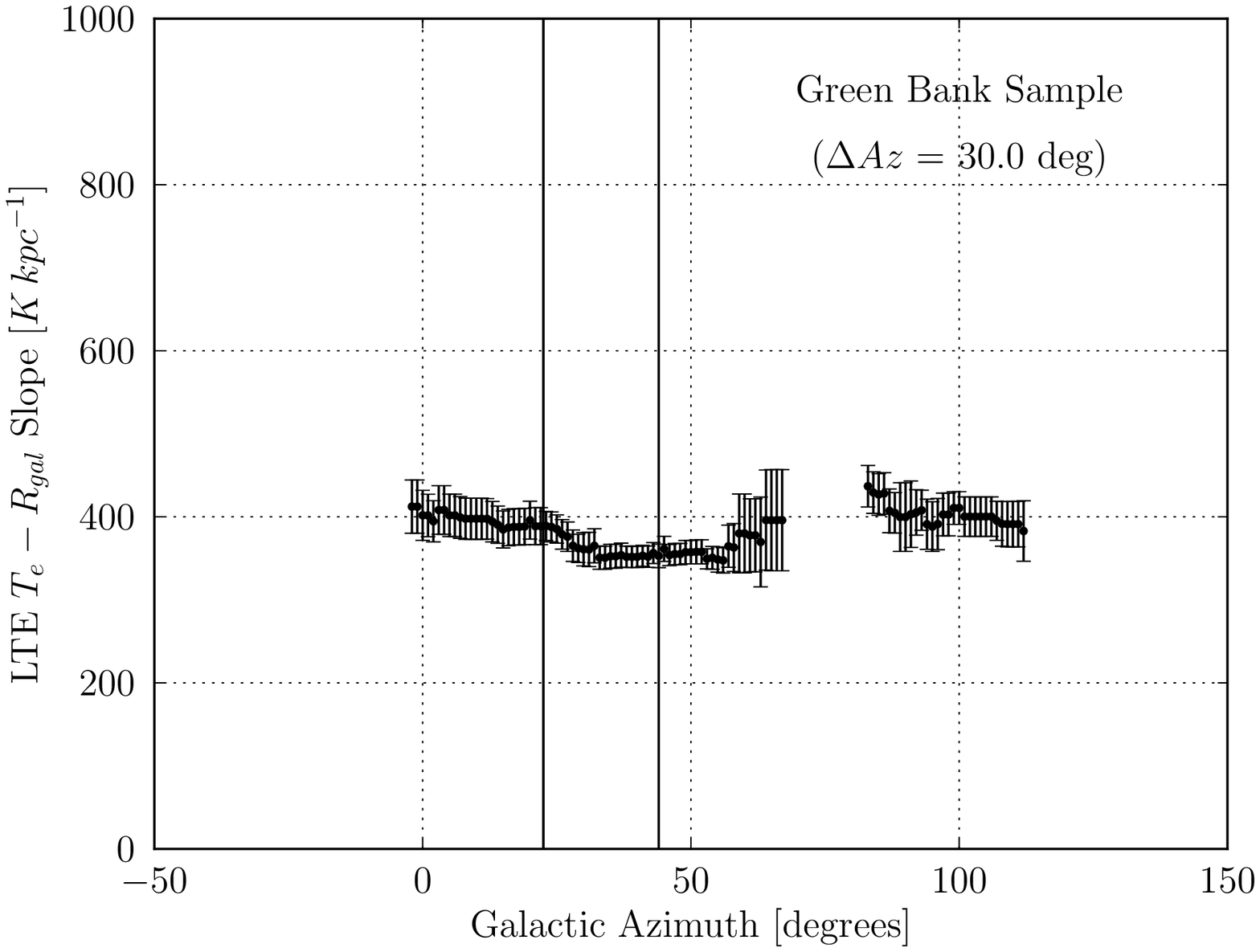} 
\includegraphics[angle=0,scale=0.40]{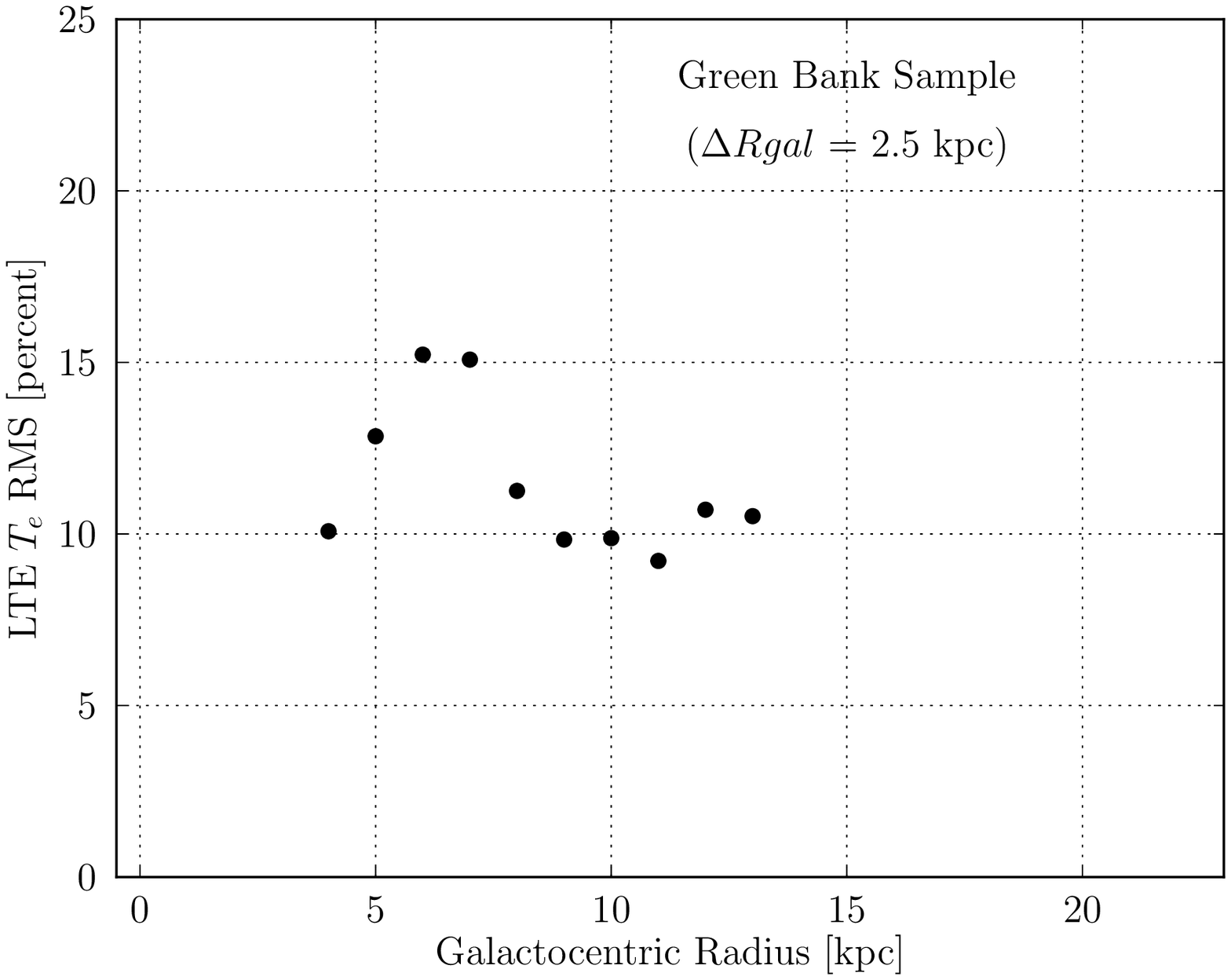} 
\includegraphics[angle=0,scale=0.40]{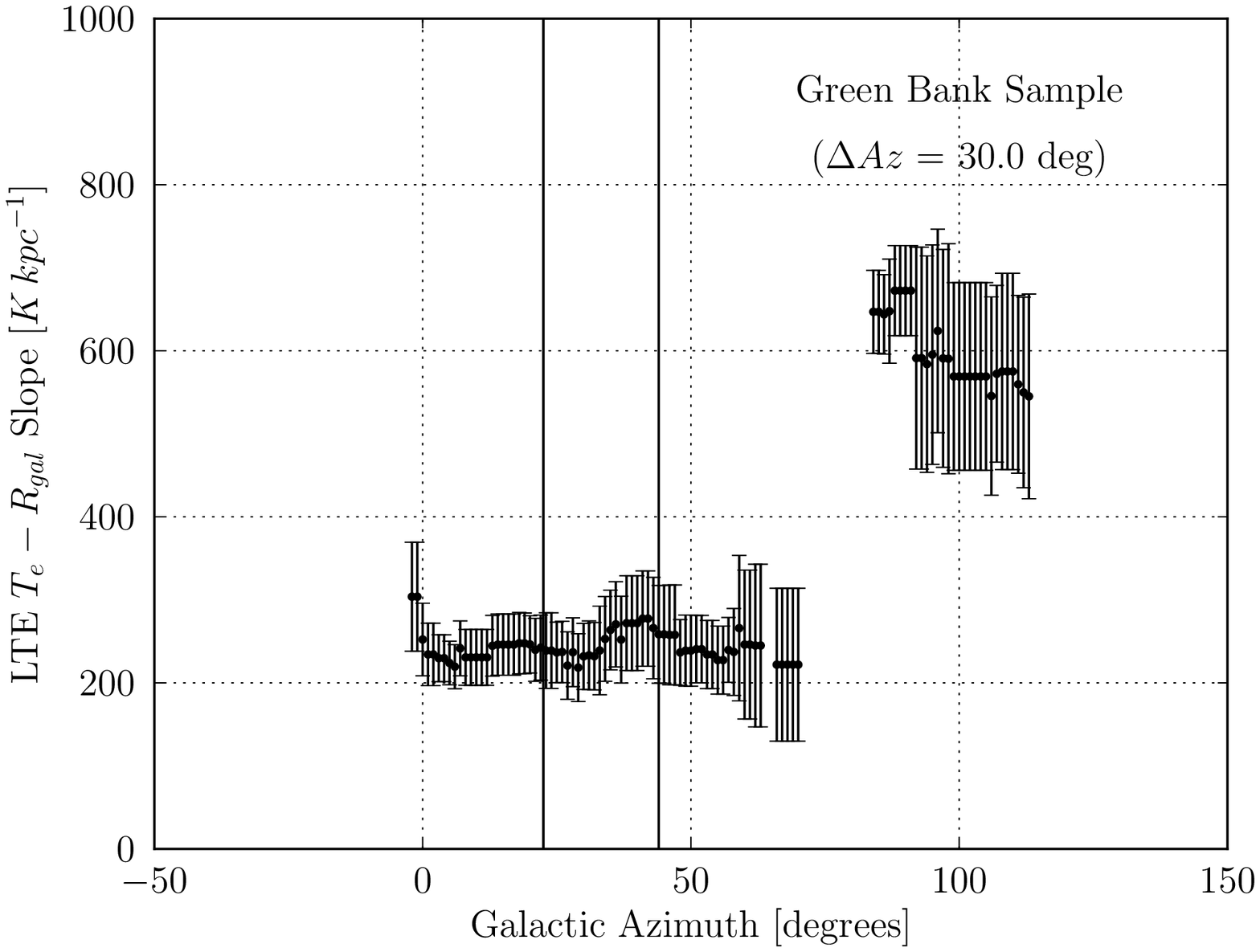} 
\caption{Monte Carlo simulations of the electron temperature and
  kinematic distance uncertainties.  The Kinematic-IAU distances are
  used for the Green Bank sample.  Left: LTE electron temperature rms
  as a function of Galactic radius.  Right: LTE \te--\rgal\ slope as a
  function of Galactic azimuth. Top: same plots as shown in
  Figure~\ref{fig:KinBrand_4panel} for comparison.  Middle: results
  from randomly generating the electron temperatures assuming $T_{\rm
    e} = a + b\,R_{\rm gal}$.  The values for $a$ and $b$ are taken
  from the fits to the Green Bank sample using the Kinematic-IAU
  distances.  Bottom: results from a Monte Carlo simulation that
  randomly adds Gaussian uncertainties to the \te\ and the Galactic
  position (see text).}
\label{fig:Random_4panel}
\end{figure}

Our \hii\ regions are not uniformly distributed in the Galactic plane
(see Figure~\ref{fig:dist}).  Could this non-uniform distribution
produce azimuthal electron temperature structure where none exists?
We explore such a possibility by randomly generating the electron
temperatures for each \hii\ region location in our sample and assuming
that only a radial \te\ gradient exits.  We assume $T_{\rm e} = a +
b\,R_{\rm gal}$, where $a = 4928 \pm\ 277$\K\ and $b = 385 \pm\
29$\Kkpc.  The values for $a$ and $b$ are taken from the fits to the
Green Bank sample using the Kinematic-IAU distances.  We perform this
experiment 100 times.  The middle panels in
Figure~\ref{fig:Random_4panel} show a typical result.  We plot the LTE
\te\ rms as a function of Galactic radius (left) and the LTE
\te--\rgal\ slope as a function of azimuth (right).  For comparison we
show the results from Figure~\ref{fig:KinBrand_4panel} in the top two
panels.  The LTE \te\ rms values are around 5\% for our experiment,
consistent with uncertainties in the electron temperature.  The LTE
\te--\rgal\ slope is approximately constant around 400\Kkpc\ with no
significant variations with azimuth.  We calculate a mean LTE
\te--\rgal\ slope over the 100 random experiments of $386 \pm\
15$\Kkpc\ and $384 \pm\ 34$\Kkpc\ for $Az = 0$\degree$-60$\degree\ and
$Az = 90$\degree$-130$\degree, respectively.  We therefore conclude
that the non-uniform \hii\ region distribution of our sample cannot
artificially produce azimuthal metallicity structure.

For most of the \hii\ regions in our sample we must rely on kinematic
distances (\rsun).  For some Galactic locations, kinematic distances
uncertainties can be larger than 50\% (see \S{\ref{sec:dist}}).  How
do these uncertainties affect electron temperature structure within
the Galactic disk?  Could these uncertainties be artificially
producing the observed electron temperature structure?  The
uncertainties for \rgal\ are typically less than 5\% and when combined
in quadrature with the 5\% error in \te\ should produce radial
electron temperature gradients with an accuracy of better than 10\%.
But determining Galactic azimuth requires \rsun, and therefore
azimuthal electron temperature structure has larger uncertainties.

To explore how kinematic distance uncertainties affect our results we
perform Monte Carlo simulations using the Kinematic-IAU distances for
the Green Bank sample.  We assume the errors in \te\ and \rsun\ have a
Gaussian distribution.  For each simulation we modify the values of
\te\ and \rsun\ by adding randomly generated, Gaussian errors to each
source.  For \te\ we use a Gaussian FWHM of 5\%, the rms value derived
from the Figure~\ref{fig:ratioTe} data.  For \rsun\ we use a Gaussian
FWHM of 3.4\% for $15 < \ell < 60$\degree\ and 61\% for $61 < \ell <
160$\degree\ based on the results in Figure~\ref{fig:distCompare}.  We
then propagate these new \rsun\ values to $Az$ and \rgal.  We ran 100
Monte Carlo simulations.  The bottom panel in
Figure~\ref{fig:Random_4panel} shows typical results for one
simulation.  The variations in the LTE \te--\rgal\ slope between $Az =
0$\degree$-60$\degree\ have been significantly reduced, producing a
constant slope to within the rms of the fits.  The much larger
gradients near $Az \sim\ 100$\degree, however, remain for most
simulations.  We calculate a mean LTE \te--\rgal\ slope over the 100
Monte Carlo simulations of $246 \pm\ 28$\Kkpc\ and $507 \pm\ 79$\Kkpc\
for $Az = 0$\degree$-60$\degree\ and $Az = 90$\degree$-130$\degree,
respectively.

We conclude that the azimuthal electron temperature structure over $Az
= 0$\degree$-60$\degree\ detected by \citet{balser11}, and confirmed
here, is not statistically significant.  This is due to the large
kinematic distance uncertainties in this azimuth zone (see
Figure~\ref{fig:distError}).  The much higher LTE \te--\rgal\ slopes
between $Az = 90$\degree$-130$\degree, however, appear to be
significant and indicate real azimuthal electron temperature
structure.  Table~\ref{tab:fit} summarizes these results where the
average LTE \te-\rgal\ slope is given, over these two azimuth ranges,
for the Kinematic-IAU data and the two simulations described above.

\subsection{O/H Abundance Ratio}\label{sec:o2h}

\begin{figure}
\includegraphics[angle=0,scale=0.7]{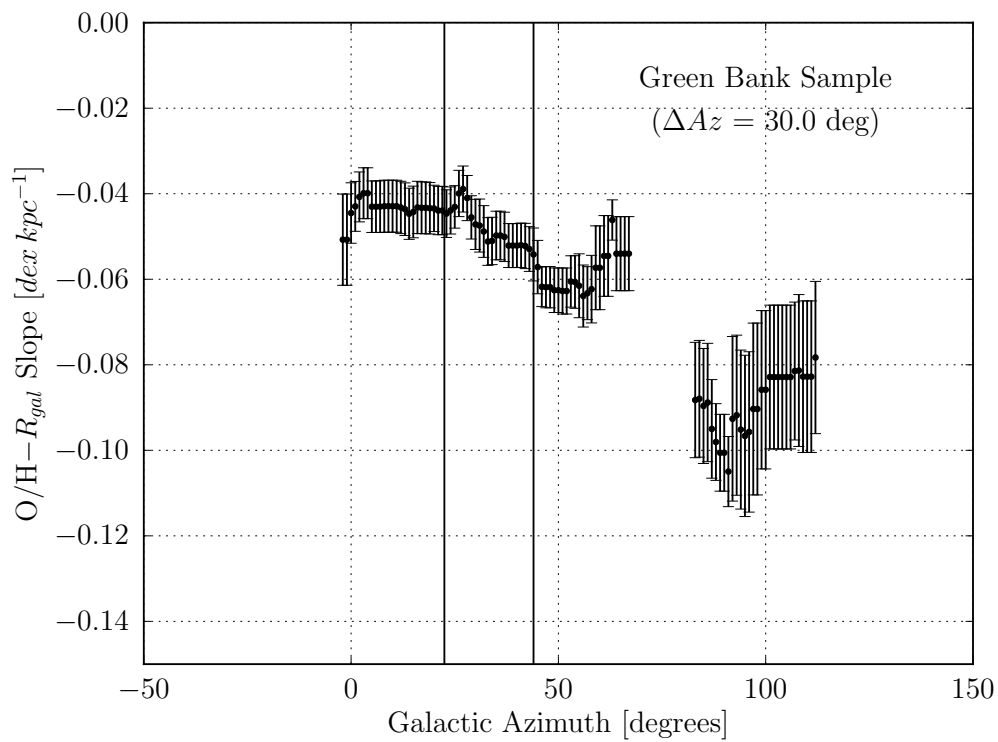} 
\caption{[O/H]--\rgal\ slope as a function of Galactic azimuth using
  the Kinematic-IAU distances.  Azimuth bin sizes of $\Delta{Az} =
  30$\degree\ are centered at the indicated azimuth.  The error bars
  are the uncertainties calculated by the SLOPES algorithm using
  jackknife resampling.  The vertical solid lines mark the orientation
  of the bar ($Az \sim 25$\degree) and long bar ($Az \sim 45$\degree)
  \citep{benjamin08}.}
\label{fig:o2h_gradient}
\end{figure}

\citet{churchwell75} were the first to suggest that the radial
electron temperature gradient observed in the Milky Way is a result
of an abundance gradient of heavy elements.  Using CELs from oxygen at
optical wavelengths along with electron temperatures from RRLs,
\citet{shaver83} derived a relationship between Log(O/H) and \te:
\begin{equation}\label{eq:cel}
[{\rm O/H}] \equiv 12 + {\rm Log(O/H)} = (9.82 \pm 0.02) - (1.49 \pm 0.11)\,T_{\rm e}/10^{4}\,{\rm K}. 
\end{equation}
More recently, \citet{afflerbach96, afflerbach97} observed higher
frequency RRLs and far IR CELs toward a sample of ultracompact (UC)
\hii\ regions to derive radial abundance gradients.  These data
have several advantages compared to \citet{shaver83}: the RRLs have
higher signal-to-noise ratios; the higher frequency RRLs are less
likely to suffer from non-LTE effects; and the far IR lines are less
susceptible to extinction by dust.  In principle we could use the
Afflerbach et al. results to derive O/H abundances from our \te\
values.  The higher electron densities in UC \hii\ regions, however,
increase \te\ by $\sim 1000$\K\ compared to the classical \hii\
regions observed by \citet{shaver83} and that are typical of our
sample \citep{rubin85, afflerbach96}.  Therefore, following
\citet{balser11} we use Equation~\ref{eq:cel} to calculate [O/H]
abundances across the Galactic disk.  Since this is a linear
relationship, the electron temperature structure discussed in
\S{\ref{sec:structure}} will pertain to the [O/H] abundance ratio,
except that the radial gradient will be negative with [O/H] decreasing
with Galactic radius.  We define the gradient in [O/H] as a function
of Galactic radius by $[{\rm O/H}] = c + d\,R_{\rm gal}$, where $d$ is
the ``[O/H]--\rgal\ slope''.  Figure~\ref{fig:o2h_gradient} plots the
[O/H]--\rgal\ slope as a function of Galactic azimuth.  The
[O/H]--\rgal\ slope increases from around $[{\rm O/H}] \sim -0.04$
between $Az = 0$\degree$-60$\degree\ to $[{\rm O/H}] \sim -0.08$
between $Az = 90$\degree$-130$\degree.

\begin{figure}
\includegraphics[angle=0,scale=0.45]{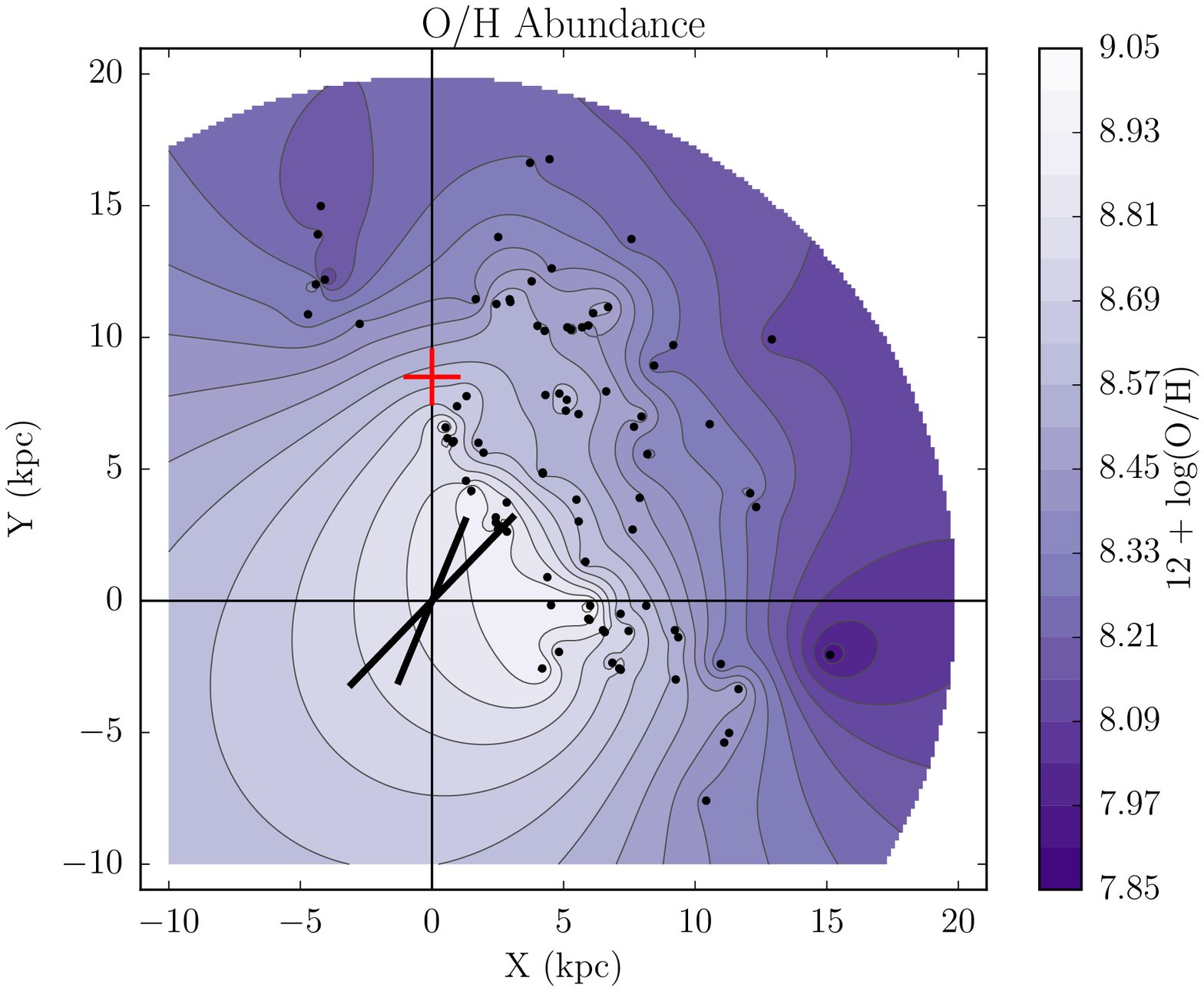} 
\includegraphics[angle=0,scale=0.45]{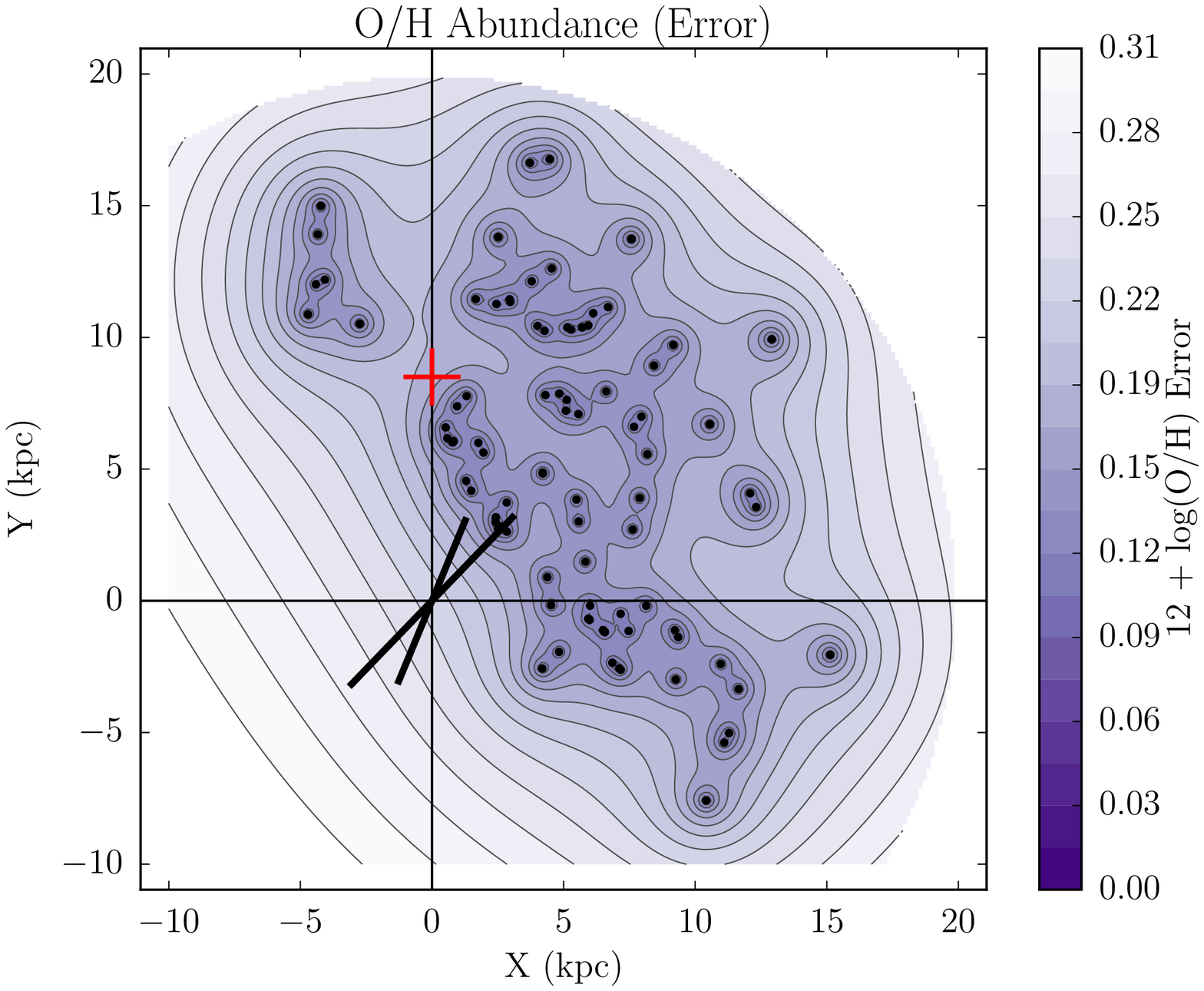} 
\caption{Face-on Galactic [O/H] abundance ratio image using Kriging to
  interpolate between the discrete \hii\ region values (left), and the
  corresponding $1\sigma$ errors (right).  A power law variogram model
  produced the best fit.  The points indicate the location of the
  discrete \hii\ regions.  The solid lines intersect at the Galactic
  Center.  The red lines mark the location of the Sun.  The thick
  lines correspond to the central locii of the putative ``short'' and
  ``long'' bars \citep{benjamin08}.  The very low abundance near ($Az
  = 97.7$\degree, \rgal\ = 15.3\kpc) is dominated by G55.11+2.4 (S83)
  \citep{balser11}.  This \hii\ region has a continuum QF = A and a
  spectral line QF = B, and therefore we have no reason to exclude it
  in our analysis.}
\label{fig:o2h_dist}
\end{figure}

\begin{figure}
\includegraphics[angle=0,scale=0.45]{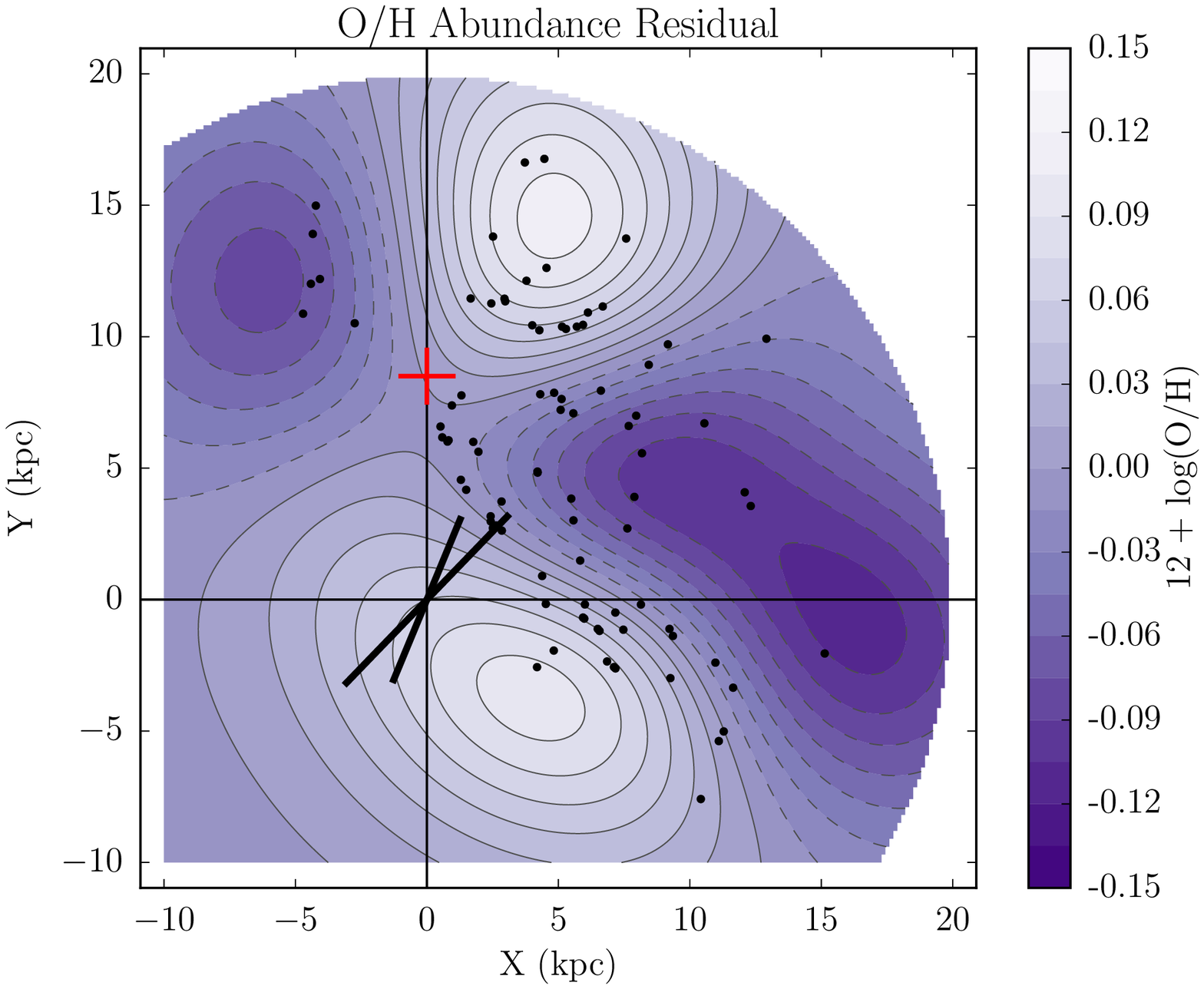} 
\includegraphics[angle=0,scale=0.45]{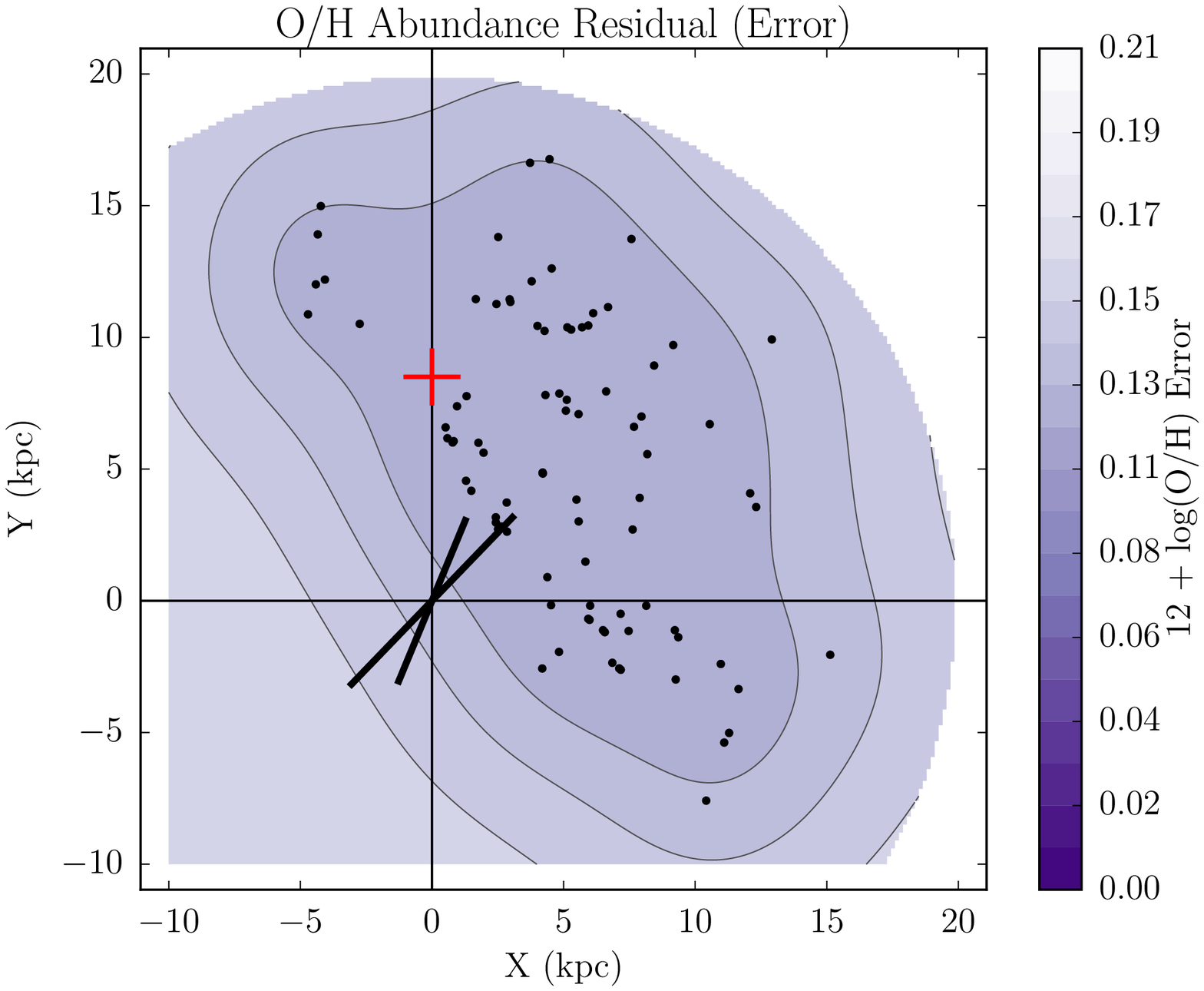} 
\caption{Face-on Galactic [O/H] abundance residual image using Kriging
  to interpolate between the discrete \hii\ region values (left), and
  the corresponding $1\sigma$ errors (right).  A Gaussian variogram
  model produced the best fit.  (See Figure~\ref{fig:o2h_dist}.)}
\label{fig:o2h_fit}
\end{figure}

We use the software package {\it pyKrige}\footnote{See
  https://github.com/bsmurphy/PyKrige.}, which employs Kriging
\citep[see][]{feigelson12}, to interpolate the distribution of [O/H]
abundances from the Green Bank sample of \hii\ regions to produce an
image (Figure~\ref{fig:o2h_dist}).  The [O/H] values near the Sun are
similar to the solar value of $8.69 \pm\ 0.05$ \citep{asplund09}.
Therefore, both the \te--[O/H] calibration from Shaver et al. and our
derived electron temperatures are consistent with Solar measurements.
The negative [O/H] radial gradient is evident with larger abundances
near the Galactic center.  We find that $[{\rm O/H}] = (9.086 \pm\
0.041) + (-0.0573 \pm\ 0.0043)\,R_{\rm gal}$ using the Green Bank
sample.  This radial gradient slope is consistent, to within the
uncertainties, with \citet{afflerbach97}.  We produce an [O/H]
residual image by subtracting this fit from the data
(Figure~\ref{fig:o2h_fit}). The [O/H] residuals span about 0.3 dex.
The residuals are smaller along the direction of the long bar,
oriented toward $Az = 44 \pm\ 10$\degree\ \citep{benjamin08}.

\section{Discussion}\label{sec:dis}

The most prominent metallicity structure in the Milky Way disk is the
decrease of metallicity with increasing Galactic radius---the radial
metallicity gradient.  All of the major tracers reveal radial
gradients, typically with slopes between $-0.03$ to $-0.09$\dexkpc.
The radial metallicity gradient can be explained by an inside-out
galaxy formation where the disk grows via a radially dependent gas
infall rate and star formation rate \citep[e.g.,][]{matteucci89}.  But
why is there such a wide range of radial gradient slopes measured?
There are several possibilities.  (1) {\it Measurement uncertainty.}
Measurement errors will cause some variations, but it is unlikely to
produce a factor of three difference in slope.  Homogeneity in
observing procedures and data analysis may be more important given the
variations in abundance that can be derived for the same source
\citep[for further discussion see][]{rudolph06, henry10, balser11}.
(2) {\it Dynamical evolution.} Radial gradients calculated with
stellar tracers may be affected by radial migration, where stars are
scattered into different orbits \citep{sellwood02}.  Radial migration
should flatten the radial metallicity gradient, but there is some
evidence that this may not be a large factor for stars in the Milky
Way \citep{dimatteo13, kubryk13, bovy14}.  (3) {\it Temporal
  evolution.}  Many tracers have a wide range of age and therefore are
probing the Milky Way disk at different times.  For example, the
radial gradient has been observed to flatten with time when using Open
clusters \citep[e.g.,][]{friel02}.  But accurate ages are crucial to
separate out these temporal effects.  Planetary nebulae (PNe) studies
indicate a flattening of the radial gradient with time
\citep{maciel03}, a steepening with time \citep{stanghellini10}, or no
temporal variation \citep{henry10}.  These vastly different
conclusions reveal the difficulty in deriving accurate PNe ages and
distances.  (4) {\it Sample Evolution.}  \citet{hayden13} measure a
flattening of the radial gradient away from the Galactic mid-plane.
Therefore, including sources out of the Galactic plane, which may be
from an older population, may alter the derived radial gradient slope
\citep[see][]{minchev14}.  Azimuthal abundance variations have been
reported \citep[e.g.,][\S\ref{sec:structure}]{pedicelli09,
  balser11}. If real, these will complicate any analysis of radial
gradients.  Here we detect metallicity gradient slopes over different
azimuth ranges that span the values observed in the literature.  The
usual assumption is that the disk is well mixed at a given radius but
this may not be true.

\citet{twarog97} were the first to measure a break in the radial
gradient around 10\kpc, and a flattening at larger radii.  We detect
no such flattening here.  \citet{balser11} suggested that many of the
studies showing breaks in the radial gradient use heterogeneous data
sets, or that the flattening disappears when restricting sources to be
near the mid-plane.  Abundance studies in external disk galaxies,
however, detect a flattening of the radial gradient beyond twice the
effective radius, the radius that contains half the total light
\citep[e.g.,][]{bresolin09, bresolin12, marino12}.
\citet{pilyugin03} suggested that a flattening could be artificially
produced by using the strong line method (e.g., $R_{23}$) to derive
the [O/H] abundance ratio.  But there are examples where the more
accurate direct method, involving the [O III]$\lambda$4363 auroral
line, was used to derive the [O/H] abundance ratio and the flattening
persists \citep[e.g.,][]{bresolin09}.

Explanations for the flattening include variations in gas density, the
presence of a bar (e.g., radial migration), the corotation resonance
(the radius where the rotation speed of the spiral arm equals that of
the material in the disk), and mergers.  \citet{sanchez14} measured
the oxygen abundance radial gradient in the disks of 306 galaxies
using \hii\ regions.  They find a slope that is independent of
morphology, the presence of bars, absolute magnitude, or mass; but the
gradient is flatter for interacting galaxies.  Most galaxies with
sufficient data indicate a flattening of the gradient beyond two
effective radii.  \citet{sanchez14} attribute this flattening to
secular evolution (e.g., radial migration or mergers).
\citet{scarano13} analyzed the metallicity gradient and the corotation
radius in 16 galaxies.  They find a correlation between the corotation
radius and the radius where there exists a metallicity break in the
radial gradient for most galaxies.  Nevertheless, the empirical
evidence for any radial break in the metallicity is not clear in the
Milky Way.

Azimuthal metallicity structure has been claimed by some authors.
Observations of Cepheids indicate metallicity variations across
Galactic quadrants II and III near \rgal\ $\approx 10$\kpc\
\citep{luck06, lemasle08, pedicelli09}.  More recently, however,
\citet{genovali14} find that these azimuthal residuals are correlated
with candidate Cepheid groups.  These are spatial overdensities of
Cepheids related to star clusters, and they may be responsible for the
observed variations \citep[also see][]{luck11}.  \citet{balser11}
detected similar metallicity structure using \hii\ regions, along with
variations in the radial gradient with Galactic azimuth between $Az =
330 - 60$\degree.  Here we determine kinematic distances for all
sources in a consistent way and use Monte Carlo simulations to assess
how distance uncertainties will effect our results.  Although we
detect the same azimuthal metallicity structure as \citet{balser11},
we conclude that this structure is not statistically significant
because of the large kinematic distance uncertainties between $Az =
330 - 60$\degree.  Nevertheless, the doubling of the [O/H]--\rgal\ slope
from $Az = 0$\degree$-60$\degree\ to $Az = 90$\degree$-130$\degree\ is
statistically significant (see Figure~\ref{fig:o2h_gradient}).  The
[O/H]--\rgal\ slope increases from around $[{\rm O/H}] \sim -0.04$
between $Az = 0$\degree$-60$\degree\ to $[{\rm O/H}] \sim -0.08$
between $Az = 90$\degree$-130$\degree.

Measuring radial gradients requires accurate determination of the
Galactic radius (\rgal), whereas probing azimuthal structure depends
on the distance from the Sun, \rsun.  Figure~\ref{fig:distCompare}
shows that the kinematic method is able to determine accurate \rgal\
values for most of the Galaxy.  Determining accurate \rsun\ distances,
however, is much more difficult.  This problem is avoided for
extragalactic objects where relative locations within the disk are
easier to determine.  Some authors have claimed azimuthal metallicity
variations in extragalactic sources; for example, in the nearby spiral
galaxies M31 \citep{sanders12}, M33 \citep{rosolowsky08}, and M101
\citep{kennicutt96}.  But when the more accurate direct method is used
to derive the [O/H] abundance ratio, non-radial metallicity structure
is not detected: M31 \citep{zurita12}, M33 \citep{bresolin11}, and
M101 \citep{li13}.  Unlike the Milky Way these galaxies do not contain
a bar.

Galactic chemical evolution (GCE) models have become more complex over
the last decade.  Early GCE models divided the disk into annuli with
no radial exchange of material, and typically employed a
Schmidt-Kennicutt star formation law that included infalling gas
\citep[e.g.,][]{chiappini97}.  These models can generally reproduce
the observed radial metallicity gradient but they include free
parameters that can be tuned to match the observations.  It has been
known for many years that non-axisymmetric forces, such as spiral
arms, can change the angular momentum of stars in the Galaxy
\citep{barbanis67}.  \citet{sellwood02} showed that the dominant
effect of spiral arms is the resonant scattering of stars located at
corotation; stars can therefore migrate radially while maintaining a
circular orbit.

Chemodynamical models consider both the chemical and dynamical
evolution of the Galaxy \citep[e.g.,][]{samland97, schonrich09,
  kubryk13, minchev14}.  They produce radial metallicity gradients
that are within the range observed, but, as discussed above, the
derived gradient slopes vary by a factor of three between the
different studies and is therefore not well constrained.
\citet{kubryk13} considered the chemical evolution of both the stars
and gas in a bar dominated disk galaxy.  There are no azimuthal
variations in either the stars or gas at early times, but at late
times (10\gyr) the corotation of the bar moves outward allowing metal
poor gas to flow inward creating azimuthal variations up to a factor
of two in the gas in the outer regions (\rgal = 12\kpc).  This may
explain the lower radial metallicity gradients we observe in the
direction of the bar.  Since most of the stars have already been
formed at all radii before radial gas flows are induced by the bar,
stars show little azimuthal metallicity structure in these simulations
\citep[also see][]{dimatteo13}.  This is consistent with metallicity
measurements of red clump stars in the Galaxy \citep{bovy14}.  One
caveat is that these simulations do not include infall, which could
also create azimuthal variations.

\section{Summary}\label{sec:summary}

We measure the RRL and continuum emission with the GBT toward 21
Galactic \hii\ regions between Galactic azimuth $Az =
90-130$\degree\ at X-band ($8-10\,$\ghz).  We derive electron
temperatures from the line-to-continuum ratio assuming LTE, and use
the \te--[O/H] calibration from \citet{shaver83} to determine the
metallicity.  We combine these results with our previous efforts using
the NRAO 140 Foot \citep{quireza06} and GBT \citep{balser11} to
produce a high quality sample of 90 sources over $Az = 0-130$\degree,
spanning Galactic radii \rgal\ = $4-17$\kpc.  

We find radial metallicity gradients, [O/H] vs \rgal, between $-0.04$
and $-0.08$\dexkpc\ depending on the azimuth range sampled.  We see no
evidence for a break or flattening of the radial gradient near \rgal\
= 10\kpc.  We confirm the azimuthal metallicity structure detected by
\citet{balser11} between $Az = 0-60$\degree, but the magnitude of the
metallicity variations is reduced by 25\%.  Monte Carlo simulations,
however, indicate that the azimuthal metallicity structure in this
azimuth zone is not statistically significant.  Nevertheless, the
large variations in the [O/H]--\rgal\ slope, by almost a factor of two,
between azimuth zone $Az = 0$\degree$-60$\degree and $Az =
90$\degree$-130$\degree\ is significant.  We suggest that the lower
radial metallicity gradients over $Az = 0-60$\degree\ may be the
result of radial mixing from the bar.

Determining metallicity structure using RRLs in \hii\ regions has many
advantages.  \hii\ regions are very young and probe metallicity at the
current epoch---there is no need to correct for age.  They are not
influenced by dynamical effects such as radial migration.  RRLs can
probe \hii\ regions across the Galactic disk and therefore provide a
more comprehensive view of chemical evolution.  There are, however,
two main drawbacks.  First, the method of deriving metallicities is
indirect; we calculate electron temperature and determine [O/H]
abundances derived from optical data.  We plan to update this
calibration using better optical data and IR data from {\it Herschel},
along with \hii\ region models using the program {\it Cloudy} to
better understand the calibration.  Second, kinematic distances are
used to determine the \hii\ region locations and can have large
errors.  But many sources now have very accurate distances determined
from maser parallax measurements. As reported here we have a much
better understanding of the quality of our kinematic distances.
Future parallax measurements will improve the situation.

\acknowledgements

We thank an anonymous referee for constructive comments that improved
this paper.  The editor provided useful feedback from an expert on
statistics on the limitations of using Shepard's method for
interpolation.  We thank Fred Schwab for discussions about various
interpolation algorithms.

{\it Facility:} \facility{GBT}

\clearpage

%
%

%
%

\clearpage

\begin{deluxetable}{lcrrrrrrrrc}
\tabletypesize{\scriptsize}
\tablecaption{Radio Recombination Line Parameters of Galactic \hii\ Regions \label{tab:line}}
\tablewidth{0pt}
\tablehead{
\colhead{} & \colhead{} & \colhead{$T_{\rm L}$} & \colhead{$\sigma\,T_{\rm L}$} & 
\colhead{$\Delta{V}$} & \colhead{$\sigma\,\Delta{V}$} & 
\colhead{$V_{\rm LSR}$} & \colhead{$\sigma\,V_{\rm LSR}$} & 
\colhead{$t_{\rm intg}$} & \colhead{rms} & \colhead{} \\ 
\colhead{Name} & \colhead{Element} & \colhead{(mK)} & \colhead{(mK)} & 
\colhead{(\kms)} & \colhead{(\kms)} & \colhead{(\kms)} & \colhead{(\kms)} & 
\colhead{(hr)} & \colhead{(mK)} & \colhead{QF} 
} 
\startdata 
 G027.562+0.084 & H &   40.63 &   0.47 &  21.63 &  0.29 & $  87.16$ &   0.12 & 11.1 &  2.40 & C \\ 
 G031.274+0.485 & H &    9.30 &   0.33 &  23.20 &  0.99 & $  -6.49$ &   0.40 & 23.8 &  1.86 & D \\ 
 G032.272-0.226 & H &   59.00 &   0.40 &  25.83 &  0.20 & $  21.14$ &   0.09 & 11.9 &  2.26 & B \\ 
                & He &    8.35 &   0.49 &  17.15 &  1.17 & $  19.70$ &   0.50 & \nodata & \nodata & \nodata \\ 
                & C &    6.21 &   0.88 &   5.32 &  0.88 & $  20.70$ &   0.37 & \nodata & \nodata & \nodata \\ 
 G032.733+0.209 & H &   22.63 &   0.40 &  21.04 &  0.44 & $  16.16$ &   0.18 & 19.0 &  2.14 & C \\ 
 G032.876-0.423 & H &   29.27 &   0.60 &  18.17 &  0.43 & $  49.56$ &   0.18 &  9.5 &  2.75 & C \\ 
 G032.928+0.607 & H &   44.37 &   0.37 &  27.40 &  0.27 & $ -38.49$ &   0.11 & 14.3 &  2.23 & C \\
\enddata 
\end{deluxetable}

\begin{deluxetable}{lrrrrrrrrrrrrc}
\tabletypesize{\scriptsize}
\tablecaption{Radio Continuum Parameters of Galactic \hii\ Regions \label{tab:cont}}
\tablewidth{0pt}
\tablehead{
\colhead{} & \multicolumn{4}{c}{\underline{~~~~~~~~~~~~~~~~~R.A.~~~~~~~~~~~~~~~~~}} & 
\multicolumn{4}{c}{\underline{~~~~~~~~~~~~~~~~~Decl.~~~~~~~~~~~~~~~~~}} & 
\multicolumn{4}{c}{\underline{~~~~~~~~~~~~~~~Average~~~~~~~~~~~~~~~}} & \colhead{} \\ 
\colhead{} & \colhead{$T_{\rm C}$} & \colhead{$\sigma\,T_{\rm C}$} & 
\colhead{$\Theta$} & \colhead{$\sigma\,\Theta$} & 
\colhead{$T_{\rm C}$} & \colhead{$\sigma\,T_{\rm C}$} & 
\colhead{$\Theta$} & \colhead{$\sigma\,\Theta$} & 
\colhead{$T_{\rm C}$} & \colhead{$\sigma\,T_{\rm C}$} & 
\colhead{$\Theta$} & \colhead{$\sigma\,\Theta$} & \colhead{} \\ 
\colhead{Name} & \colhead{(mK)} & \colhead{(mK)} & \colhead{($\prime$)} & \colhead{($\prime$)} & 
\colhead{(mK)} & \colhead{(mK)} & \colhead{($\prime$)} & \colhead{($\prime$)} & 
\colhead{(mK)} & \colhead{(mK)} & \colhead{($\prime$)} & \colhead{($\prime$)} & \colhead{QF} 
} 
\startdata 
 G027.562+0.084 &  243.6 &   2.34 &    1.38 & 0.016 &  263.8 &   1.97 &    1.55 & 0.014 &  253.7 &   1.53 &    1.47 & 0.011 & C \\ 
 G031.274+0.485 &   96.9 &   2.95 &    3.31 & 0.116 &  100.2 &   4.22 &    3.27 & 0.159 &   98.5 &   2.57 &    3.29 & 0.099 & C \\ 
 G032.272-0.226 &  639.4 &   5.72 &    1.47 & 0.015 &  687.5 &   6.64 &    1.57 & 0.018 &  663.5 &   4.38 &    1.52 & 0.012 & C \\ 
 G032.733+0.209 &  126.8 &   3.47 &    3.23 & 0.103 &  149.6 &   1.60 &    2.02 & 0.043 &  138.2 &   1.91 &    2.56 & 0.049 & C \\ 
 G032.876-0.423 &  142.0 &   1.65 &    2.98 & 0.044 &  180.1 &   3.47 &    2.39 & 0.053 &  161.1 &   1.92 &    2.67 & 0.036 & C \\ 
 G032.928+0.607 &  649.2 &   2.53 &    1.44 & 0.007 &  655.5 &   3.94 &    1.45 & 0.010 &  652.4 &   2.34 &    1.45 & 0.006 & A \\ 
 G032.976-0.334 &  172.6 &   3.46 &    2.68 & 0.062 &  194.1 &   4.22 &    3.04 & 0.077 &  183.3 &   2.73 &    2.85 & 0.049 & B \\ 
 G034.041+0.053 &  264.5 &   2.01 &    1.86 & 0.021 &  261.1 &   1.48 &    2.00 & 0.017 &  262.8 &   1.25 &    1.93 & 0.014 & C \\
\enddata 
\end{deluxetable}

\begin{deluxetable}{lrrrrrrrrrrrrrc}
\tabletypesize{\scriptsize}
\tablecaption{\hii\ Region Properties \label{tab:prop}}
\tablewidth{0pt}
\setlength{\tabcolsep}{2.0pt}
\rotate
\tablehead{
\colhead{} &  \colhead{} &  \colhead{} & 
\multicolumn{3}{c}{\underline{~~~~~Kinematic-IAU~~~~~}} & 
\multicolumn{3}{c}{\underline{~~~~~Kinematic-Reid~~~~~}} & 
\multicolumn{3}{c}{\underline{~~~~~~~~~~Best~~~~~~~~~~}} & 
\colhead{} & \colhead{} & \colhead{} \\ 
\colhead{} & \colhead{$\ell$} & \colhead{$b$} & 
\colhead{$R_{\rm sun}$} & \colhead{$Az$} & \colhead{$R_{\rm gal}$} & 
\colhead{$R_{\rm sun}$} & \colhead{$Az$} & \colhead{$R_{\rm gal}$} & 
\colhead{$R_{\rm sun}$} & \colhead{$Az$} & \colhead{$R_{\rm gal}$} & 
\colhead{} & \colhead{$T_{\rm e}$} & \colhead{} \\ 
\colhead{Name} & \colhead{(deg)} & \colhead{(deg)} & 
\colhead{(kpc)} & \colhead{(deg)} & \colhead{(kpc)} & 
\colhead{(kpc)} & \colhead{(deg)} & \colhead{(kpc)} & 
\colhead{(kpc)} & \colhead{(deg)} & \colhead{(kpc)} & 
\colhead{y\tablenotemark{a}} & \colhead{(K)} & \colhead{Telescope} } 
\startdata 
   G0.605+0.325\tablenotemark{b} &   0.605 &  0.329 & $33.5 \pm 47.30$ & 179.2 & 25.0 & $33.3 \pm 47.30$ & 165.6 & 25.0 & $33.5 \pm 47.30$ & 179.2 & 25.0 & $  0.065 \pm  0.025$ & $6600 \pm 320$ & 140Foot \\ 
   G0.665$-$0.035 &   0.665 & $-$0.035 & $ 8.8 \pm 00.05$ & 160.0 &  0.3 & $ 8.6 \pm 00.05$ & 111.7 &  0.3 & $ 7.8 \pm 00.73$ &   6.7 &  0.8 & $  0.038 \pm  0.014$ & $8170 \pm 180$ & 140Foot \\ 
      G1.13$-$0.1\tablenotemark{b} &   1.119 & $-$0.098 & $ 8.5 \pm 54.11$ &   0.0 & 25.0 & $ 8.3 \pm 54.11$ &   0.0 & 25.0 & $ 8.5 \pm 54.11$ &   0.0 & 25.0 & $  0.065 \pm  0.008$ & $7130 \pm 70$ & 140Foot \\ 
      G2.90+0.0\tablenotemark{b} &   2.901 & $-$0.003 & $19.2 \pm 27.13$ & 174.8 & 10.7 & $18.2 \pm 27.13$ & 162.2 &  9.9 & $19.2 \pm 27.13$ & 174.8 & 10.7 & $  0.052 \pm  0.014$ & $4840 \pm 140$ & 140Foot \\ 
   G3.270$-$0.101 &   3.270 & $-$0.100 & $13.9 \pm 02.29$ & 171.3 &  5.4 & $13.7 \pm 02.29$ & 158.3 &  5.4 & $13.9 \pm 02.29$ & 171.3 &  5.4 & $  0.077 \pm  0.011$ & $7440 \pm 280$ & 140Foot \\ 
   G5.899$-$0.427 &   5.900 & $-$0.427 & $14.3 \pm 01.47$ & 165.6 &  5.9 & $14.1 \pm 01.47$ & 156.0 &  5.9 & $ 3.0 \pm 00.18$ &   3.2 &  5.5 & $  0.079 \pm  0.010$ & $11130 \pm 170$ & 140Foot \\ 
   G5.956$-$1.265 &   5.956 & $-$1.264 & $16.1 \pm 02.56$ & 167.2 &  7.7 & $15.7 \pm 02.56$ & 157.7 &  7.5 & $16.1 \pm 02.56$ & 167.2 &  7.7 & $  0.074 \pm  0.003$ & $3420 \pm 30$ & 140Foot \\ 
   G5.973$-$1.178 &   5.974 & $-$1.178 & $16.1 \pm 02.56$ & 167.2 &  7.7 & $15.7 \pm 02.56$ & 157.7 &  7.5 & $16.1 \pm 02.56$ & 167.2 &  7.7 & $  0.073 \pm  0.005$ & $8180 \pm 70$ & 140Foot \\
\enddata 
\tablenotetext{a}{For sources with no helium RRL detected, we assume
  $y = 0.080 \pm 0.000$.}  \tablenotetext{b}{Distances for these
  sources are unrealistic because they are located toward the Galactic
  Center.  Our Galactic rotation curve model assumes a maximum \rgal\
  value of 25\kpc.}
\end{deluxetable}

\begin{deluxetable}{lcc}
\tabletypesize{\scriptsize}
\tablecaption{Average LTE \te--\rgal\ Slope \label{tab:fit}}
\tablewidth{0pt}
\tablehead{
\colhead{Data Set} & \colhead{$Az =0$\degree$-60$\degree} & 
\colhead{$Az =90$\degree$-150$\degree}
}
\startdata 
Kinematic-IAU Data & $322 \pm\ 28$\Kkpc\  & $551 \pm\ 92$\Kkpc\ \\
Random \te\ Gradient & $386 \pm\ 15$\Kkpc\ & $384 \pm\ 34$\Kkpc\ \\
Monte Carlo Errors & $246 \pm\ 28$\Kkpc\ & $507 \pm\ 79$\Kkpc\ \\
\enddata 
\tablecomments{``Random \te\ Gradient'' corresponds to the average
  slope over 100 simulations assuming $T_{\rm e} = a + b\,R_{\rm gal}$
  (see text).  ``Monte Carlo Errors'' corresponds to the average slope
  over 100 Monte Carlo simulations assuming errors in \te\ and location.}
\end{deluxetable}

\end{document}